\documentclass[prb,reprint,a4paper,floatfix,superscriptaddress]{revtex4-1}

\usepackage{graphicx}
\usepackage{amsmath}
\usepackage{hyperref}
\usepackage{xfrac}
\usepackage{epstopdf}

\newcommand{\tc}{ T_{\mathrm{c}} }
\newcommand{\tn}{ T^{\mathrm{N}} }
\newcommand{\esdw}{ E_{\mathrm{SDW}} }

\newcommand{\nsc}{n_{ \mathrm{SC} }}

\newcommand{\bmu}{B_{\mu}}
\newcommand{\bmuvec}{\mathbf{B}_{\mu}}
\newcommand{\vmu}{\nu_{\mu}}
\newcommand{\pmu}{\mathbf{P}_{\mu}}
\newcommand{\mub}{\mu_{\mathrm{B}}}

\begin{document}

	\title{Muon spin rotation and infrared spectroscopy study of magnetism and superconductivity in Ba$ _{1-x} $K$ _{x} $Fe$ _{2} $As$ _{2} $}
	
	\author{B.P.P. Mallett}
	\email{benjamin.mallett@gmail.com}
	\affiliation{Robinson Research Institute, Victoria University, P.O. Box 600, Wellington, New Zealand}
	\affiliation{University of Fribourg, Department of Physics and Fribourg Center for Nanomaterials, Chemin du Mus\'{e}e 3, CH-1700 Fribourg, Switzerland}

	\author{C.N. Wang}
	\affiliation{University of Fribourg, Department of Physics and Fribourg Center for Nanomaterials, Chemin du Mus\'{e}e 3, CH-1700 Fribourg, Switzerland}
	\affiliation{Masaryk University, Department of Condensed Matter Physics and CEITEC - Central European Institute of Technology, Kotl\'{a}\v{r}sk\'{a} 2, 61137 Brno, Czech Republic}

	\author{P. Marsik}
	\affiliation{University of Fribourg, Department of Physics and Fribourg Center for Nanomaterials, Chemin du Mus\'{e}e 3, CH-1700 Fribourg, Switzerland}
	
	\author{E. Sheveleva}
	\affiliation{University of Fribourg, Department of Physics and Fribourg Center for Nanomaterials, Chemin du Mus\'{e}e 3, CH-1700 Fribourg, Switzerland}
	
	\author{M. Yazdi-Rizi}
	\affiliation{University of Fribourg, Department of Physics and Fribourg Center for Nanomaterials, Chemin du Mus\'{e}e 3, CH-1700 Fribourg, Switzerland}
		
	\author{J.L. Tallon}
	\affiliation{Robinson Research Institute, Victoria University, P.O. Box 600, Wellington, New Zealand}

	\author{P. Adelmann}
	\affiliation{Institute of Solid State Physics, Karlsruhe Institute of Technology, Postfach 3640, Karlsruhe 76021, Germany}
		
	\author{Th. Wolf}
	\affiliation{Institute of Solid State Physics, Karlsruhe Institute of Technology, Postfach 3640, Karlsruhe 76021, Germany}
		
	\author{C. Bernhard}
	\email{christian.bernhard@unifr.ch}
	\affiliation{University of Fribourg, Department of Physics and Fribourg Center for Nanomaterials, Chemin du Mus\'{e}e 3, CH-1700 Fribourg, Switzerland}

	\date{\today}
	
	\pacs{ 74.70.−b, 74.25.Gz, 78.30.−j}
	\keywords{}
	
	\begin{abstract}

Using muon spin rotation and infrared spectroscopy we study the relation between magnetism and superconductivity in Ba$ _{1-x} $K$ _{x} $Fe$ _{2} $As$ _{2} $ single crystals from the underdoped to the slightly overdoped regime. We find that the Fe magnetic moment is only moderately suppressed in most of the underdoped region where it decreases more slowly than the N\'{e}el-temperature, $ \tn $. This applies for both the total Fe moment obtained from muon spin rotation and for the itinerant component that is deduced from the spectral weight of the spin-density-wave pair breaking peak in the infrared response. In the moderately underdoped region, superconducting and static magnetic orders co-exist on the nano-scale and compete for the same electronic states. The static magnetic moment disappears rather sharply near optimal doping, however, in the slightly overdoped region there is still an enhancement or slowing down of spin fluctuations in the superconducting state. Similar to the gap magnitude reported from specific heat measurements, the superconducting condensate density is nearly constant in the optimally- and slightly overdoped region, but exhibits a rather pronounced decrease on the underdoped side. Several of these observations are similar to the phenomenology in the electron doped counterpart Ba(Fe$ _{1-y} $Co$ _{y} $)$ _{2} $As$ _{2} $. 

	\end{abstract}
	
	\maketitle

\section{Introduction}
\label{sec:introduction}
Of great current interest is the relationship between magnetism and superconductivity (SC). The majority of unconventional superconductors are found proximate to a magnetic state which is reached by controlling some tuning parameter such as electronic doping, magnetic field or pressure. It has also long been suspected that spin-fluctuations are involved in the formation of the high-temperature SC state, perhaps by mediating the pairing interaction\cite{miyake1986, scalapino1986, beal1986, bickers1987, wolf2004}. The Fe-based superconductors (FeSCs) have highlighted that this proximity of magnetic and SC states can be taken one step further with magnetism and SC co-existing on a nanoscopic scale in these materials\cite{marsik2010, lumsden2010, wiesenmayer2011,bernhard2012,dai2015}. This is observed in lightly underdoped FeSCs where the two phases compete for the same electronic states. With further doping, a SC ground state without static magnetic order is obtained. The FeSCs are thus especially interesting systems for studying the changing relationship between SC and magnetism. 

\begin{figure*}

	\includegraphics[width=0.75\textwidth]{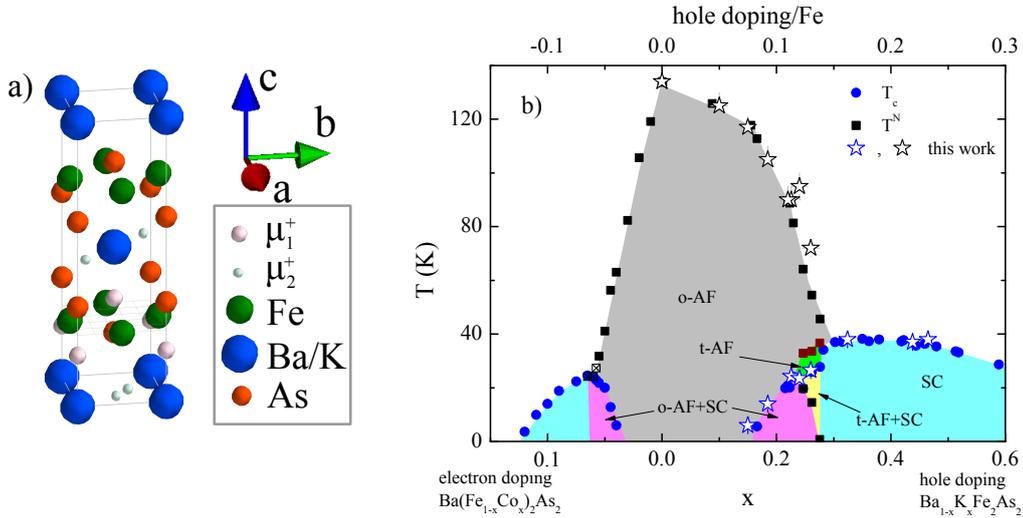}
	\caption{\label{fig:phasediagram}  
		a) Crystal structure of BKFA indicating the main muon site, $ \mu^{+}_{1} $, and secondary muon site, $ \mu^{+}_{2} $. b) Indicative phase diagram of BaFe$ _{2} $As$ _{2} $. `o-AF' is the single-$\mathbf{Q}$, stripe-like AF static order with orthorhombic unit-cell symmetry. `SC' annotates the superconducting regions. `t-AF' is the double-$ \mathbf{Q} $ AF static order with tetragonal unit-cell symmetry. Closed symbols are data from Refs.~\onlinecite{hardy2010, bernhard2012, bohmer2015}, open stars are the samples studied in this work.}
	
\end{figure*}

BaFe$ _{2} $As$ _{2} $ is a prototypical FeSC for such a study due to the high-quality, large single crystals that can now be synthesised. Its crystal structure is shown in Fig.~\ref{fig:phasediagram}a, along with calculated muon-stopping sites (see details in Sec.~\ref{subsec:zfmusr}), and an annotated phase diagram for the material is shown in Fig.~\ref{fig:phasediagram}b. The key structural component is the FeAs layer as the (five) electronic bands crossing the Fermi-level are predominantly of Fe-$ 3d $ character with a small admixture of As-$ 4p $ character\cite{paglione2010}. The undoped parent compound is metallic and paramagnetic at high temperature with a tetragonal $I4/mmm$ space group symmetry. Coincident with the N\'{e}el temperature, $ \tn $, of approximately 135~K, is an orthorhombic distortion of the lattice and an antiferromagnetic (AF) state, annotated `o-AF' in Fig.~\ref{fig:phasediagram}b. This AF state exhibits in-plane antiparallel spins along the $ (0,\pi) $ direction and parallel ones along $ (\pi,0) $ in a so-called single-$ \mathbf{Q} $ or stripe-like AF state\cite{dai2015}. The interactions underlying this magneto-structural transition are as yet unclear. Explanations range from itinerant models\cite{chubukov2012,mazin2008,fernandes2014}, in which Fermi-surface nesting governs the magnetic interactions, to local models which assume localized spins with exchange interactions determined by the orbital occupation\cite{lee2009, kruger2009}. It is probable that both itinerant and local models are partly applicable\cite{dai2015, mallett2015bkfaIR, mallett2015musr, pelliciari2016}. It has also been proposed that the spin-lattice coupling\cite{yildirim2009} and a near degeneracy of different spin states of the Fe ions\cite{chaloupka2013, gretarsson2013} play an important role.

From this starting point, a combined SC and magnetic ground state can be induced. This can be by electrical-doping with excess electrons by substitution of Ni or Co for Fe, with excess holes by substitution of Na or K for Ba, or by `chemical pressure' through the iso-valent substitution of P for As. Figure~\ref{fig:phasediagram}b shows the case for electron doping via Co substitution for Fe in Ba(Fe$ _{1-y} $Co$ _{y} $)$_{2}$As$ _{2} $ (BFCA), and for hole doping via K substitution for Ba in Ba$ _{1-x} $K$ _{x} $Fe$ _{2} $As$ _{2} $ (BKFA). Infra-red (IR) optical spectroscopy\cite{marsik2010,  mallett2015bkfaIR}, muon spin rotation ($\mu$SR) \cite{marsik2010, wiesenmayer2011, bernhard2012} and neutron scattering \cite{pratt2009, christianson2009} have shown that the magnetic and SC states compete for the same electronic states in the underdoped region. 

However, there are some signs of a qualitatively different relationship between magnetism and superconductivity in near optimally doped BFCA (open square symbols in Fig.~\ref{fig:phasediagram}b) \cite{bernhard2012}. In this regime magnetic order or magnetic fluctuations are seen to develop at the SC transition temperature, $ \tc $. A lingering issue with BFCA is the influence of the disorder on the Fe-site caused by Co substitution. This concerns, for example, pair-breaking effects in the SC state \cite{hardy2016} and the rather rapid suppression of the magnetic order parameter\cite{bernhard2012}. The influence of disorder is much reduced in BKFA where the dopant is outside of the FeAs layer, which provides confidence that observed behaviours are intrinsic rather than disorder effects. It is interesting to compare the behaviour of BFCA and BKFA in this regard to elucidate the importance or otherwise of disorder, doping and Fermi-surface shape. Initially, crystal quality was an issue for BKFA with phase separation being reported\cite{aczel2008,park2009}, presumably due to large distributions of $ x $ within a given crystal. With refinements to the growth processes however, high-quality BKFA crystals can now be made with narrow distributions of $ x $. 

Indeed, as the crystal quality improves for BKFA and BNFA, it is becoming apparent that near the termination of AF static order, novel magnetic states emerge \cite{bohmer2015, wang2016, hardy2016}. In particular, a double-$ \mathbf{Q} $ AF state with tetragonal-lattice symmetry and a spin-reorientation from in-plane to $ c $-axis alignment is now well documented \cite{hassinger2012, avci2014, wasser2015, allred2015xray, mallett2015musr}, and is annotated at `t-AF' in Fig.~\ref{fig:phasediagram}b. This state also competes with SC. Such novel-magnetic states offer new perspectives into the underlying interactions of the magnetic ordering and how they relate to SC \cite{fernandes2014, mallett2015bkfaIR}. They are evidently nearly degenerate in energy to the primary o-AF order, especially near the termination of static AF order, and thus are likely important considerations in the description of the spin fluctuations. 

In this work we track the evolution of the magnetic and SC properties of BKFA in undoped ($ x=0 $) to slightly-overdoped samples ($ x=0.47 $), shown by the open star symbols in Fig.~\ref{fig:phasediagram}, using $\mu$SR and infrared optical spectroscopy. $\mu$SR is a powerful technique for studying magnetism in the bulk. It is capable of measuring small magnetic fields, of order $ 0.1 $~mT, in true zero-external-field conditions. Since it is a probe of the local magnetic field at the muon site, rather than the volume-averaged magnetic field, it is possible to determine the volume fraction of a particular magnetic state. In turn, IR optical spectroscopy provides rich information about the bulk electronic properties. IR spectroscopy probes the energy, scattering and symmetry of the SC gap(s) and sum rules can be used to determine the SC condensate density. 
The IR response of underdoped FeSCs also has a prominent spin-density-wave pair-breaking feature \cite{hu2008} that is related to the ordered magnetic moment from the itinerant carriers and is thus also a probe of the magnetic state. Combined, these two techniques provide a rich picture of the magnetic and SC states and the phenomenological relationship between them. 

We find a co-existence and competition of static AF order and SC on a nanoscopic scale in single-crystalline BKFA for $ 0.19 \leq x \leq 0.26 $. The decrease in $ \tn $ in this region is more rapid than the decrease in the local-magnetic field, and the magnetic state is more ordered/uniform in BKFA than that of BFCA.  We find an absence of static magnetic order in $ x=0.43 $, but we do see an enhancement of spin-fluctuations below $ \tc $. In conjunction, the full SC volume fraction and in-plane London penetration depth, $ \lambda_{L} $, of approximately 200~nm at this doping level signals a robust SC state.

Our measurements complement previous optical studies on BKFA which have focused on the undoped \cite{hu2008, wang2012} or optimally doped \cite{li2008, yang2009, charnukha2011, charnukha2011prb,dai2013} compounds, with only a few studies of the underdoped compounds \cite{dai2012,mallett2015bkfaIR}. Similarly, previous $\mu$SR studies of BKFA have focused on near-optimally doped samples \cite{aczel2008, goko2009,park2009,khasanov2009,hiraishi2009} with just one study of polycrystalline, underdoped samples \cite{wiesenmayer2011}. On the other hand, BFCA has been more systematically studied by $\mu$SR \cite{williams2010, bernhard2012} and IR optical spectroscopy \cite{marsik2010, marsik2013}.

The paper is laid out as follows; experimental methods are discussed in Section \ref{sec:experiment}. We then present $\mu$SR results in Section \ref{sec:musr} which show the development of the magnetic state in underdoped BKFA. The optical response is described in the following Section \ref{sec:optics} followed by a summary of results in Section \ref{sec:summary}.
 
\section{Experiment}
\label{sec:experiment}

High-quality single crystals of Ba$ _{1-x} $K$ _{x} $Fe$ _{2} $As$ _{2} $ (BKFA) were grown in alumina crucibles using an FeAs flux as described in Ref.~\onlinecite{karkin2014}. The crystals were then characterised by x-ray diffraction refinement and electron dispersion spectroscopy. We did not detect impurity phases from these measurements. The K-content, $ x $,  determined by these measurements indicates a small distribution of $ x $ on the order of $\pm 0.02 $. The estimated uncertainties in $ x $ are indicated by the error bars in Fig.~\ref{fig:phasediagram} and following figures. Resistivity and magnetization data were obtained with a Quantum Design PPMS. 

The $\mu$SR measurements were performed at the GPS instrument of the $ \mu $M3 beamline at the Paul Scherrer Institute (PSI) in Villigen, Switzerland. Fully spin-polarized, positive muons with an energy of $ 4.2 $~MeV were implanted in the crystal (along the $ c $-axis of the crystal) where they rapidly thermalize and stop at interstitial lattice sites distributed over a depth of about 100~$ \mu $m. The muon spins precess in the magnetic field at the muon site, $ \bmuvec $, with a precession frequency, $ \vmu $, proportional to the magnitude of $ \bmuvec $, $\bmu$, as $ \vmu = \gamma_{\mu}\bmu/2\pi $, where $ \gamma_{\mu}=2\pi\times 135.5 $~MHz/T is the gyromagnetic ratio of the muon. The time evolution of the polarization of the muon spin ensemble, $ P(t) $, is detected via the asymmetry of the emission rate of the decay positrons as described in Refs.~\onlinecite{schenck1985,lee1999}. The zero field (ZF) and transverse field (TF) measurements were performed with the so-called up-down positron counters in spin-rotation mode for which the muon spin polarisation, $ \pmu $ is at about 54$ ^{\circ} $ with respect to the muon beam (pointing toward the upward counter). See Fig.~S7b of Ref.~\onlinecite{mallett2015musr} for an illustration of the experimental geometry. Longitudinal-field (LF) measurements were performed with $ \pmu $ anti-parallel to the muon momentum. 

The far-infrared optical reflectivity, $ R(\omega) $, was measured from $ 45-700 $~cm$ ^{-1} $ with a Bruker Vertex 70v FTIR spectrometer with an in-situ gold evaporation technique \cite{homes1993, kim2010infrared}. For the ellipsometry measurements we used a home-built rotating-analyzer setup attached to a Bruker 113v at $ 200-4500 $~cm$ ^{-1} $ \cite{bernhard2004thinfilms} and for the near-infrared to near-UV region a Woollam VASE ellipsometer at $ 4000-52000 $~cm$ ^{-1} $. We measured the in-plane component of the optical response representing an average of the $ a $- and $ b $-axis response \cite{schafgans2011, nakajima2011} since the twinning of the samples was not controlled. The combined ellipsometry and reflectivity data have been analyzed as described in Refs.~\onlinecite{kim2010infrared, kuzmenko2005} to obtain the complex optical response functions; the complex optical conductivity $ \sigma(\omega) = \sigma_{1}(\omega) + i\sigma_{2}(\omega) $ and the related complex dielectric function $ \epsilon(\omega) = \epsilon_{1}(\omega) + i\epsilon_{2}(\omega) = 1 + i 4\pi \sigma(\omega) / \omega $. The quantity $ \int_{\omega_{ \mathrm{min} }}^{\omega_{ \mathrm{max} }} {\sigma_{1}(\omega) d\omega}  $ is known as a spectral-weight (SW) and is related by a sum rule \cite{smith1998} to the density of electronic states within the energy integration window, $ [\omega_{\mathrm{min}}, \omega_{ \mathrm{max} }] $, by $ \int _{\omega_{ \mathrm{min} }} ^{\omega_{ \mathrm{max} }} \sigma_{1}(\omega) d\omega = \frac{\pi n e^{2}}{2m^{*}}$ where $ e $ is the electron charge, $ n $ is the density of electrons with energy within that window and $ m^{*} $ is the effective mass of those states. In the remainder of the manuscript we take the frequency dependence of these functions to be understood and drop the $ (\omega) $ notation. 
 
\section{$ \mu $SR}
\label{sec:musr}
\subsection{Zero-field $\mu$SR}
\label{subsec:zfmusr}
\begin{figure*}
	
	\includegraphics[width=1.0\textwidth]{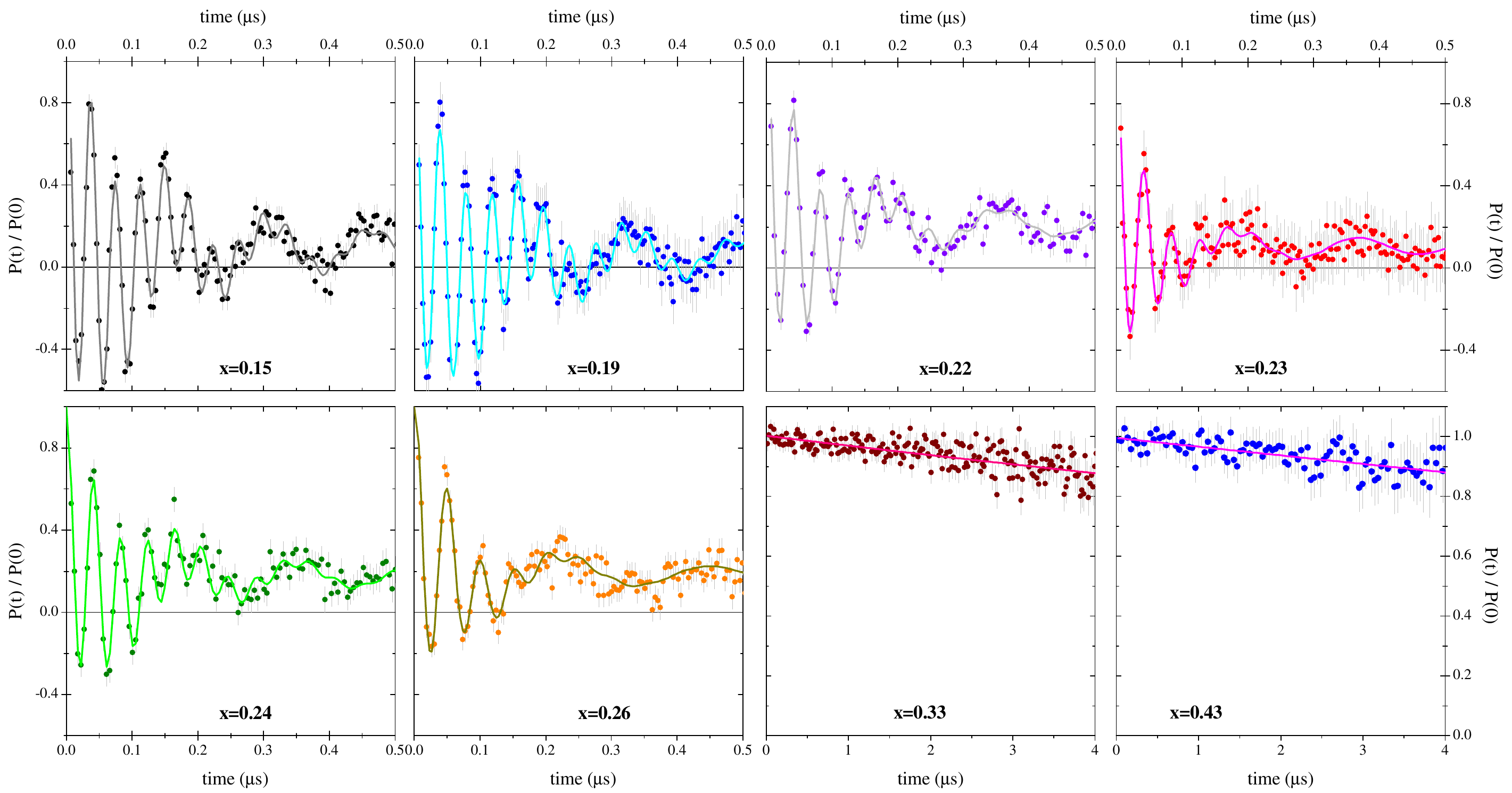}
	\caption{\label{fig:zfmusrraw}  
		Zero-field $\mu$SR spectra for BKFA single crystals at $ T=5 $~K showing the development of magnetic order in the underdoped samples and its absence in for the optimally- and overdoped samples with $ x \ge 0.33 $. Fits are shown as solid lines and are described in the text. The initial polarization is $ P(0)=0.21 $.  }
	
\end{figure*}

In our study of the magnetic and SC response of BKFA we begin with the ZF-$\mu$SR data where the magnetic field at the muon site is solely from the internal magnetic moments of the sample. An example of such data taken at low temperature, $ T=5 $~K, for various $ x $ is shown in Fig.~\ref{fig:zfmusrraw} as symbols with corresponding fits, described below, as solid lines. 

The data show an absence of static magnetic order or large magnetic moments for the $ x=0.43 $ sample due to the full polarisation at $ t=0 $~$ \mu $s and a very slow depolarization. This result is consistent with the findings of Ref.~\onlinecite{hiraishi2009}.  Similar behaviour was also observed at the lower doping of $ x=0.33 $. However, in this crystal batch there were some crystals that showed a partial magnetic fraction in ZF-$ \mu $SR measurements. Work is ongoing to understand these mixed phase crystals and will be published separately. Turning to the underdoped samples, $ x \leq 0.26 $, a clear oscillation can be seen which indicates well-defined local magnetic fields in the majority of the sample volume. At least two oscillation frequencies can in fact be easily resolved from these data and this is due to two inequivalent muon stopping sites, as was found in previous studies of undoped (Sr,Ba)Fe$ _{2} $As$ _{2} $ \cite{jesche2008,aczel2008,bernhard2012} and underdoped Ba$ _{1-x} $K$ _{x} $Fe$ _{2} $As$ _{2} $ \cite{wiesenmayer2011, mallett2015musr}. Calculations using a modified Thomas Fermi approach \cite{reznik1995} show that the majority muon site, which accounts for approximately 80\% of the muons and is due to the global minimum in the potential energy, is located at the coordinate $ (0, 0, 0.191) $ in the $ I4/mmm $ setting \cite{mallett2015musr}, i.e. on the line that connects the Ba and As ions along the $ c $-direction as shown in Fig.~\ref{fig:phasediagram}a. The oscillatory signal with the smaller amplitude and lower frequency originates from a secondary muon site due to a local potential minimum located at $ (0.4, 0.5, 0) $, slightly away from the line connecting the As ions along the c-direction, see Fig.~\ref{fig:phasediagram}a. It also has a high local symmetry with the same direction and qualitative change of the local magnetic field. 

With these considerations, we find that the data are well described by the fitting to the following function
\[
P(t) = P(0)\sum_{i=1}^{2}{A_{i}^{\mathrm{os}} \cos{(\gamma_{\mu}B_{\mu,i}t+\phi)e^{-\lambda_{i}t} }
	+ A_{3}^{\mathrm{no}}e^{-\lambda_{3}t} } 
\] 
\noindent where $ P $ describes the polarization of the muon spin ensemble, the sum is over the two muon sites, $ A_{i} $ is the relative amplitude of the signals (which relates to the volume fraction), $ B_{\mu,i} $ is the component of the local magnetic field at the muon site orthogonal to the muon spin, $ \phi $ is the initial phase of the muon spin and $ \lambda_{i} $ are the relaxation rates relating to a spread in $ B_{\mu,i} $. The non-precessing signal described by the third term arises from the non-orthogonal orientation of $ \pmu $ and $ \bmuvec $ and also from a small background due to muons that stop outside the sample. Fits to the data using this expression are shown as solid lines in Fig.~\ref{fig:zfmusrraw}. 

Without loss of generality in the conclusions we draw, we focus only on the majority muon site, $ \nu_{\mu,1} $ etc., and the change of its local field due to magnetic and superconducting transitions. 

\begin{figure}
	
	\includegraphics[width=1.0\columnwidth]{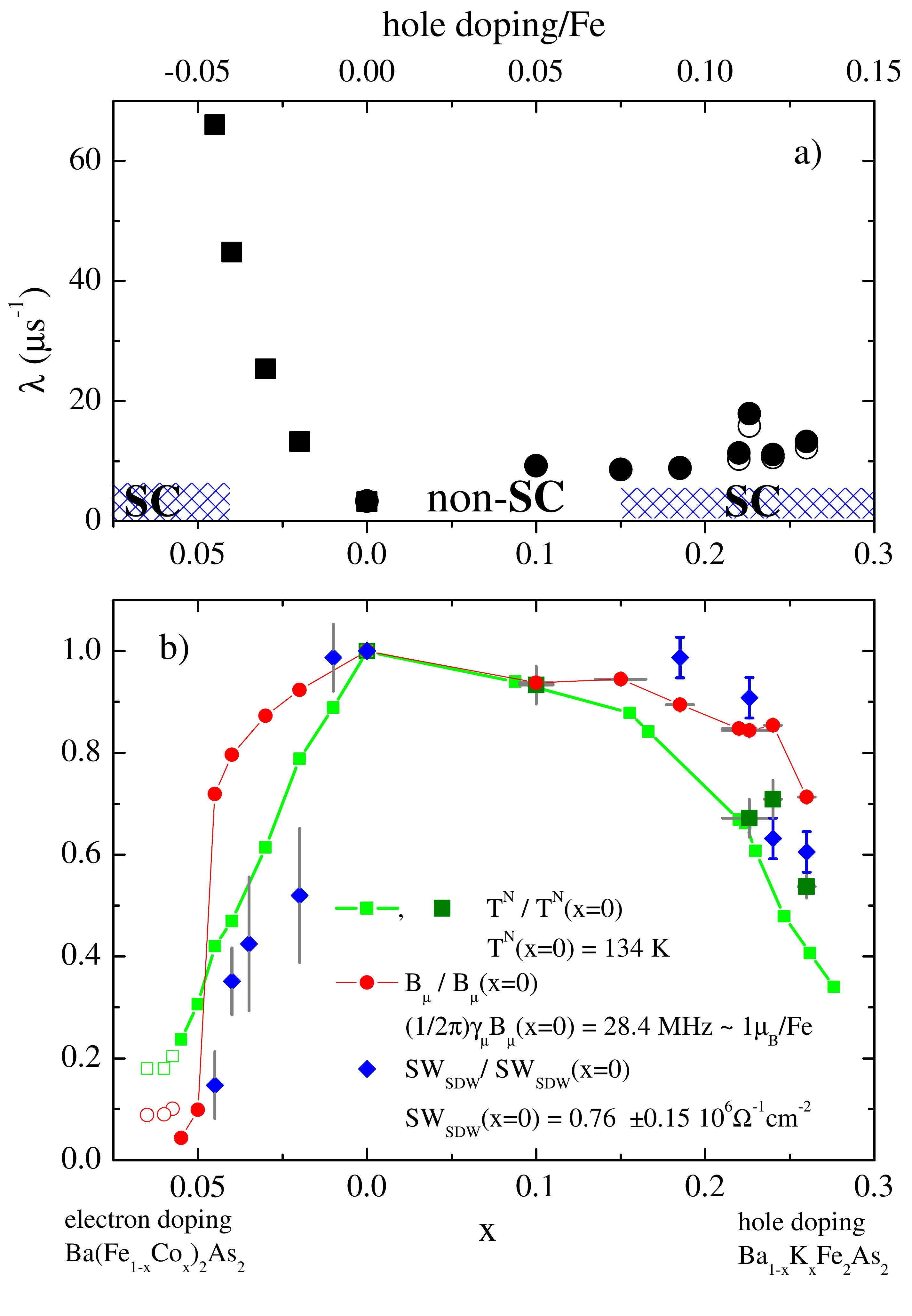}
	\caption{\label{fig:zfmusr}  
		Low-temperature ZF-$\mu$SR results comparing BKFA (this work) and BFCA \cite{bernhard2012}. a) Relaxation rates describing the distribution of $ \bmu $. Dashed blue areas represent doping-levels where superconductivity is observed. Open symbols show relaxation rates measured slightly above $ \tc $ (and in the o-AF magnetic phase for the $ x=0.26 $ sample). b) Magnetic ordering temperature, $ \tn $, (green squares) and the frequency of the muon rotation at the main muon site, $ \bmu $, (red circles) normalised to their values at $ x=0 $ as indicated in the legend. For BKFA, dark-green squares are from this work, connected light-green squares are literature values \cite{bohmer2015}. For comparison, the estimated low-temperature spectral weight of the SDW feature in the IR-optical response is also shown (blue diamonds), see Sec.~\ref{subsec:sdw} for details.}
	
\end{figure} 

The clear oscillation frequencies for doped BKFA in Fig.~\ref{fig:zfmusrraw} can be contrasted with BFCA where the magnetic signal is rapidly overdamped with Co-substitution\cite{bernhard2012}. This is probably due to lower disorder in BKFA since the dopant ion is not on the FeAs layer. To be quantitative, we show in Fig.~\ref{fig:zfmusr}a a comparison of the low temperature relaxation rates, $ \lambda $, for BKFA and BFCA. Whilst $ \lambda $ rapidly increases with Co doping from $3.3$~$ \mu $s for BFA to 50~$ \mu $s for Ba(Fe$ _{0.96} $Co$ _{0.04} $)$ _{2} $As$ _{2} $ where the first co-existence of SC and magnetism is seen, for BKFA $ \lambda $ remains a moderate $ \approx 10 $~$ \mu $s well into the doping regime where SC is established. We note that the $ x=0.23 $ sample has a slightly larger $ \lambda $, which may be due to a larger distribution of $ x $ values with respect to the other samples, or a larger degree of internal crystal strain. We also note that relaxation rates here are smaller than those of Ref.~\onlinecite{wiesenmayer2011}, possibly because they were measuring polycrystalline rather than single-crystal samples. 

Turning now to the low-temperature ZF-$\mu$SR oscillation frequency shown in Fig.~\ref{fig:zfmusr}b. With increasing doping, $ \tn $ (dark-green squares) varies little between $ x=0 $ and $ 0.19 $, but then falls rapidly at higher doping. Data from B\"{o}hmer \textit{et al.} \cite{bohmer2015} (connected light-green squares) are overlaid which indicate $ \tn $ falls to zero at $ x=0.3 $. In comparison, the low-temperature value of $ \bmu $ (red circles) initially decreases even more slowly than $ \tn $. A similar observation was made in BFCA. Barring any change in the magnetic order and/or orientation of the magnetic moment, $ \bmu $ is proportional to the magnetic moment on the Fe-site. For $ x=0 $, neutrons measure between 0.9 and 1 $ \mub $/Fe \cite{lumsden2010} and with ZF-$\mu$SR we measure $ \bmu =28.4$~MHz, which is about 10\% lower than expected from dipolar calculations based on a 0.9 $ \mub $/Fe moment \cite{mallett2015musr}. The relatively slow decrease of $ \bmu $ strongly suggests that the magnetic moment on the Fe site decreases more slowly than $ \tn $ with doping. We can discount that a re-orientation of the Fe-moment contributes to these changes in $ \bmu (x)$, since one can monitor the orientation of $ \bmu $ by inspecting the signal from the orthogonal ‘forward-backward’ detector set (see, for example, Fig.S7b of Ref.~\onlinecite{mallett2015musr} for an illustration). Doing so shows that $ \bmu $ remains parallel to the $ c $-axis (that means an in-plane Fe moment orientation), except in the t-AF phase of $ x=0.26 $ between 32 and 18~K. We note that the gradual reduction in the Fe moment with doping in BKFA that we report here is in good quantitative agreement with very recent first-principles calculations\cite{derondeau2016}. $ \tn $ may be significantly reduced below the highest temperature possible given by the energy of the order-parameter \cite{mazin2008, bernhard2012} and instead be tied to the near-concomitant structural transition discussed earlier. Additional information can be gleaned from measuring the spectral weight of the spin-density-wave pair-breaking peak in the optical spectra, SW$ _{\mathrm{SDW}} $, shown as blue diamonds in Fig.~\ref{fig:zfmusr}b. SW$ _{\mathrm{SDW}} $ is a measure of the itinerant magnetic moment and is discussed in more detail in Sec.~\ref{subsec:sdw}. It also falls more slowly than $ \tn $ with doping and for $ x<0.24 $ tracks $ \bmu $ within uncertainties suggesting that the ratio of itinerant to local magnetic contributions is similar there. There is somewhat of a drop for $ x \geq 0.24 $ which may suggest a larger relative contribution to $ \bmu $ from local moments in this doping range. On the electron doped side this effect seems to be even more enhanced.

\subsection{Magnetic volume fraction: phase separation?}
\label{subsec:magvolfrac}
\begin{figure}
	
	\includegraphics[width=1.0\columnwidth]{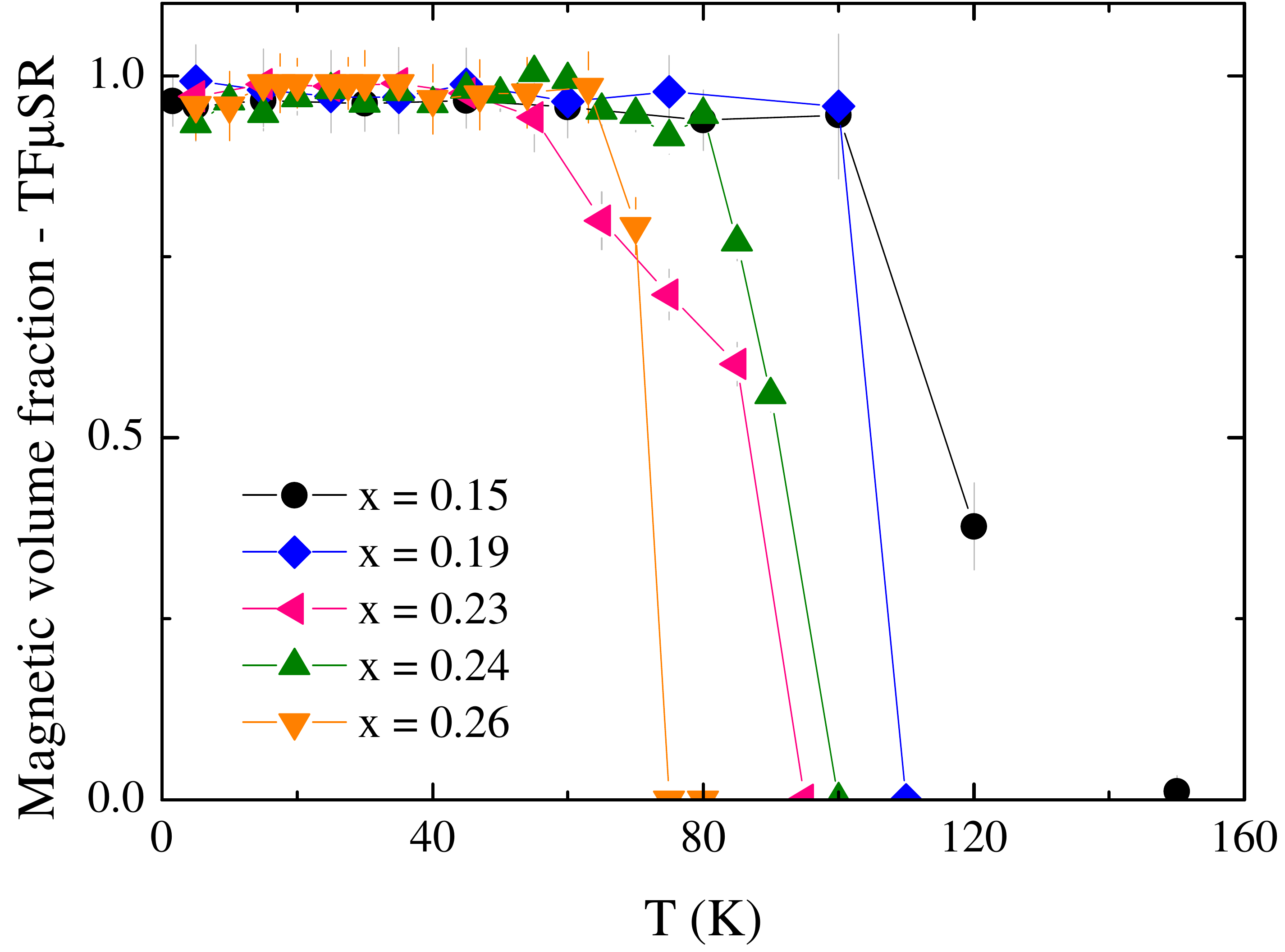}
	\caption{\label{fig:musrvolfrac}  
		The temperature dependence of the magnetic volume fraction determined from TF-$\mu$SR measurements for underdoped BKFA single crystals.}
	
\end{figure}

For samples $ x \leq 0.26 $, the volume fraction of the magnetic state is, with experimental uncertainty, 100\%. This was consistently determined from both ZF-$\mu$SR, where the fitted $ A_{i} $ give the magnetic volume fractions, and more directly from TF-$\mu$SR measurements at temperatures just above $ \tc $. The temperature dependence of the magnetic volume fraction for various $ x $ as determined by TF-$\mu$SR is shown in Fig.~\ref{fig:musrvolfrac}. In these measurements, if there were a non-magnetic fraction it would result in a signal with a small relaxation rate, $ \lambda \approx 0.2 $~$ \mu $s$ ^{-1} $, at the frequency of the external field (below $ \tc $, SC vortices broaden and shift $ \bmu $). A full magnetic volume fraction in this doping regime was also reported by the $\mu$SR study of Ref.~\onlinecite{wiesenmayer2011} on polycrystalline samples.  
  
This issue is of importance given earlier results which showed a real-space phase separation in BKFA single crystals into magnetic and non-magnetic regions \cite{aczel2008, goko2009, park2009, khasanov2009bkfa}. Goko \textit{et al.} reported a mixed volume fraction of magnetic and non-magnetic regions, with approximately equal volume fraction, below 70~K for single crystals of BKFA with $ x=0.45 $.\cite{goko2009} These samples have a somewhat broad magnetic transition which might suggest a large distribution of K content in their samples. In contrast, Hiraishi \textit{et al.} failed to see a magnetic volume fraction with ZF-$\mu$SR for their polycrystalline $ x=0.4 $ sample \cite{hiraishi2009}, in line with what we report here in our single-crystal samples. Park \textit{et al.} report mixed magnetic volume fraction for their slightly underdoped single crystals below 70~K and a sharp SC transition temperature at $ \tc =32 $~K \cite{park2009}. Given these values we estimate $ x=0.30 \pm 0.05 $. Based on the above, one might suspect that only single crystal samples show this phase-separation behaviour whereas polycrystalline samples at similar values of $ x $ do not. This may point to an important role of crystal strain. In this regard, it is important that we can show for our underdoped and the slightly overdoped $ x=0.43 $ single crystals an absence of significant real-space phase separation. We note however that somewhere between $ x=0.26 $ and $ x=0.43 $ there may be a doping state with intrinsic phase-separation and possibly a qualitatively different relation between magnetism and SC. Such a scenario has been reported in BFCA \cite{bernhard2012}. We mentioned earlier that some of our own $ x=0.33 $ crystals showed a partial magnetic volume fraction, and such a scenario appears to be consistent with the observations of Park \textit{et al.}.\cite{park2009} It is of great interest to study this region in more detail with $\mu$SR and fine control of sample quality will be key to this pursuit. This seems to mostly pertain to doping homogeneity, but there are possibly additional contributing factors such as crystal strain. 

\subsection{Coexistence of superconductivity and magnetism}
\label{subsec:scandmagmusr}
Although a SC transition is seen at $ \tc=6 $~K for the $ x=0.15 $ sample, our $\mu$SR data do not provide proof that SC is a bulk phenomenon at this doping since we do not see characteristic SC anomalies below $ \tc $. Similar observations were made by neutrons\cite{avci2011}. However, for all $ 0.19 \leq x \leq 0.26 $ there is clear evidence of a coexistence and competition of superconducting and magnetic order on a nanometer scale. 

\begin{figure}
	
	\includegraphics[width=0.9\columnwidth]{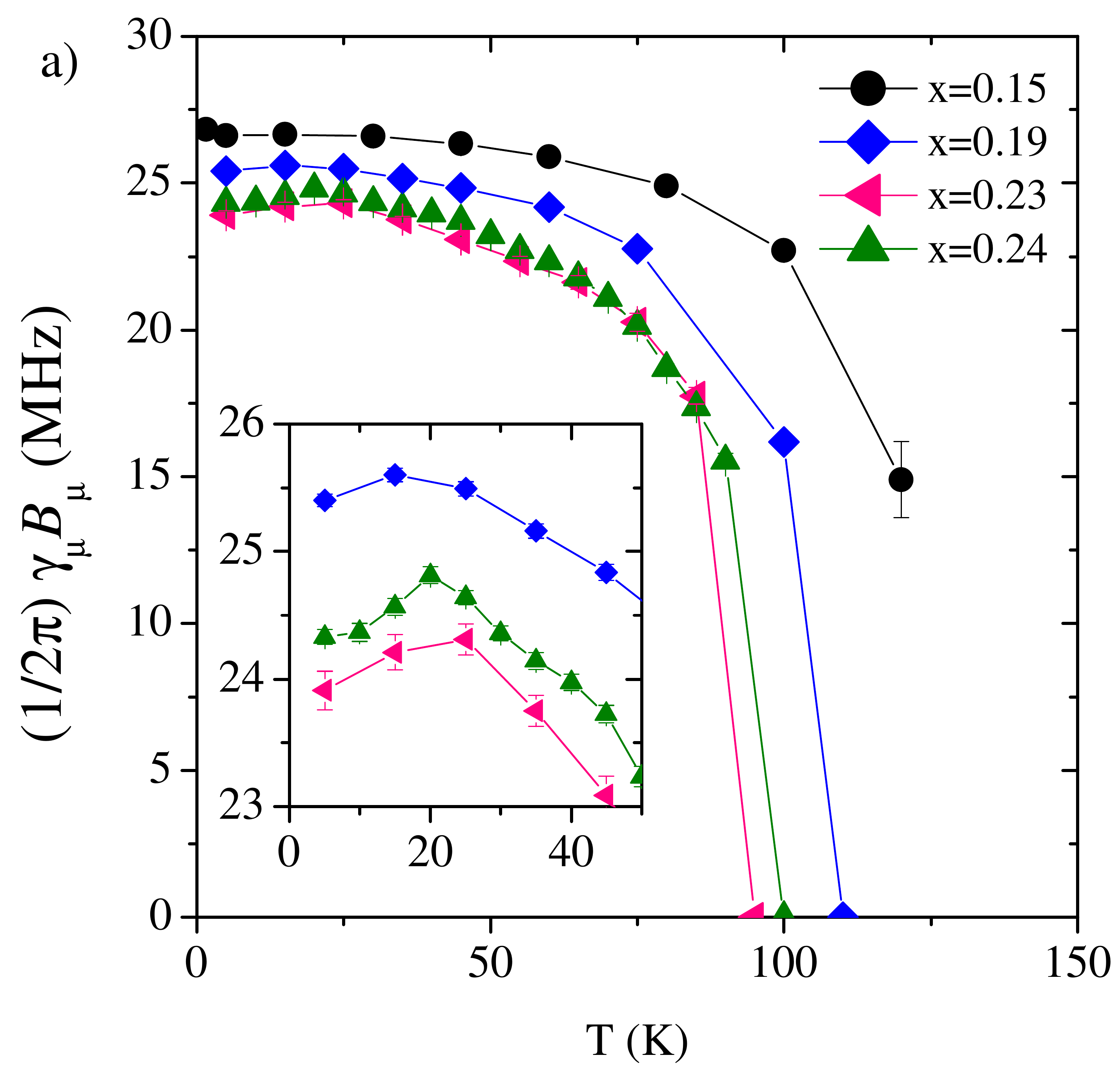}
	\includegraphics[width=1.0\columnwidth]{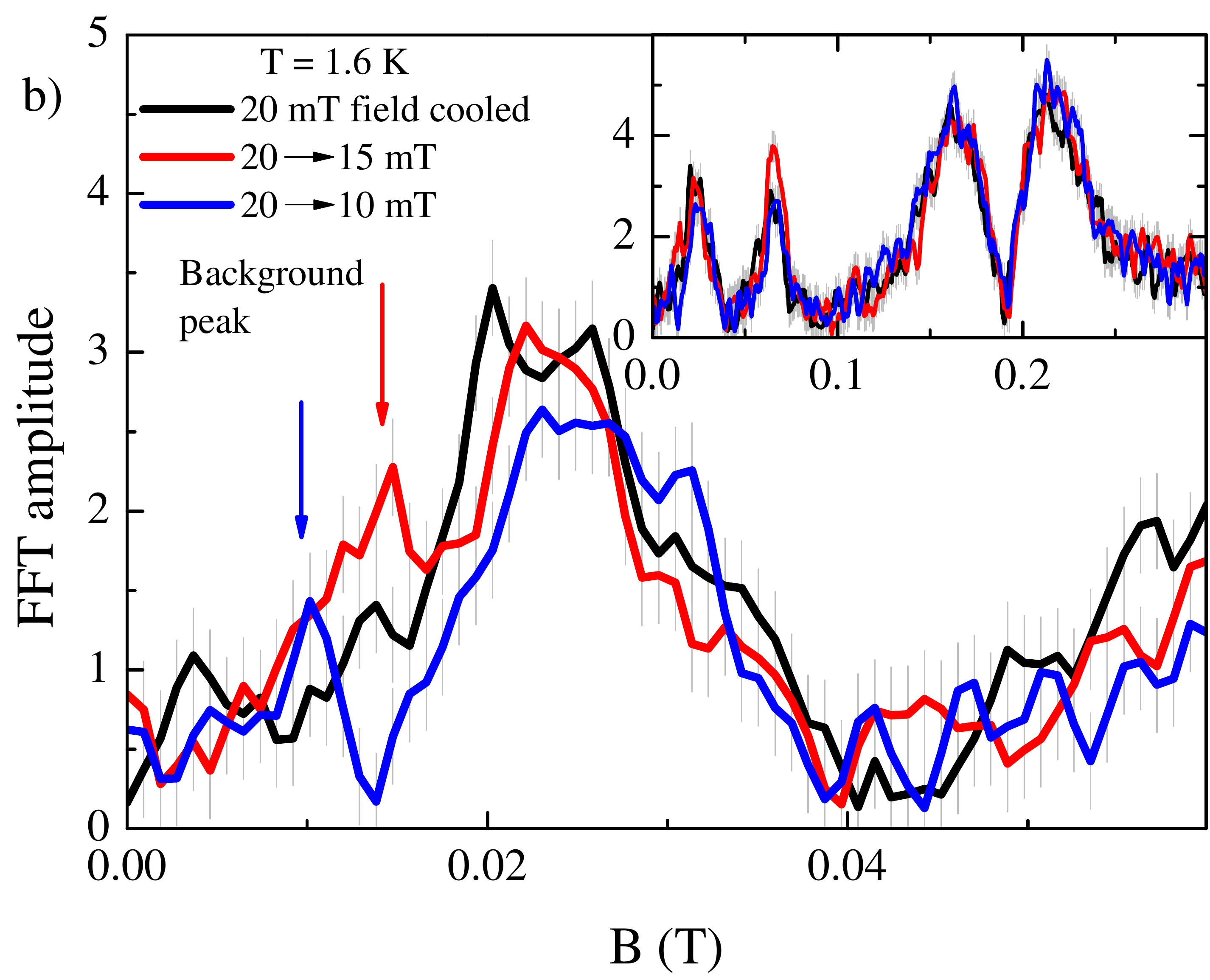}
	\caption{\label{fig:musrbmu}  
	Co-existence of superconductivity and magnetism. a) The temperature dependence of $ \bmu $ from ZF-$ \mu $SR for several of the measured samples. b) Results of a so-called pinning experiment for $ x=0.19 $ revealing a pinned vortex lattice. The sample was field cooled in 20~mT to $ T=1.6 $~K.}
	
\end{figure}

The first set of evidence from $\mu$SR for this concerns the decrease in ZF and TF $\mu$SR frequency below $ \tc $ as shown in the inset to Fig.~\ref{fig:musrbmu} for ZF-$\mu$SR. A decrease in the magnetic order parameter below $ \tc $ was also seen in polycrystalline BKFA with ZF-$\mu$SR for $ x=0.19$ and $0.23 $ \cite{wiesenmayer2011} and by neutron diffraction experiments for polycrystalline BKFA with $ x=0.21 $ \cite{avci2011}. The suppression of the magnetic order parameter concurrent with the onset of SC strongly suggests a co-existence and competition between the two orders. The second piece of evidence comes from so-called pinning experiments (Fig.~\ref{fig:musrbmu}b). In this experiment, the sample is cooled in a moderate field, 20~mT, followed by a high-statistics $\mu$SR spectrum collected at low temperature. Next, without changing the temperature, the applied field is changed. If the magnetic flux density inside the sample remains unchanged, despite the change in externally applied field, it shows the existence of pinned vortices and a bulk type-II SC state whose volume fraction can be estimated from the data. Whilst we did not observe a pinned vortex lattice for the $ x=0.15 $ doping state at $ 1.5 $~K, it was observed in the other two samples tested, $ x=0.19$ and $ 0.23 $, signifying a bulk SC state at these doping levels. Results of the pinning experiment for $ x=0.19 $ are shown in Fig.~\ref{fig:musrbmu}b. Here the sample was field cooled from $ T>\tn $ in 20~mT to $ T=1.6 $~K and a high-statistics spectrum collected. The four main peaks, shown in the inset to Fig.~\ref{fig:musrbmu}b, correspond to the zero-field $ B_{\mu} $ at the main and secondary muon site (which is parallel to the $ c $-axis\cite{mallett2015musr}) split by $ \pm 20$~mT, the applied field (that is also parallel to the $ c $-axis).  The field was then decreased by $ 5 $~mT and then a further 5~mT. A weak peak, indicated by the arrows and resulting from the small fraction of muons that stop outside of the sample, is seen to follow the external field, whilst the main features, due to the muons that stopped inside the sample, remain essentially unchanged. 

Combined with observation of a full magnetic-volume fraction shown in Fig.~\ref{fig:musrvolfrac}, these two observations show that SC and magnetism coexist on a nanoscopic scale and compete in underdoped BKFA. 

The behaviour for the $ x=0.43 $ sample is different. Firstly, as shown in Fig.~\ref{fig:zfmusrraw} there are no observable oscillations, no fast relaxation and/or missing volume fractions in the ZF-$\mu$SR data at $ T=2 $~K which indicates the absence of static magnetism. This is somewhat different to BFCA around optimal doping where a spatially inhomogeneous, static-magnetic state develops only below $ \tc $.\cite{bernhard2012} Instead, it is not until a more substantially overdoped Ba(Fe$ _{1.89} $Co$ _{0.11} $)$ _{2} $As$ _{2} $ that static magnetism is no longer observed at any temperature. However, even here, it was found that spin-fluctuations were enhanced below $ \tc $. Thus, to investigate the magnetic state of our $ x=0.43 $ BKFA sample further, we carried out a series of LF-$\mu$SR experiments for $ H=0.5$, 1 and 2~mT. In a weak LF, the small contribution to the relaxation rate from nuclear magnetic moments can be suppressed, so that a weak contribution of muon relaxation due to fluctuations can be resolved. 
The data were fitted to $ P(t)=P(0) g(\nu _0,\sigma,t) e{^{-\lambda^{\mathrm{LF}} t}}$ where $ e{^{-\lambda^{\mathrm{LF}} t}} $ is the contribution to the relaxation from spin fluctuations and $ g(\nu _0,\sigma,t) $ is a standard expression describing the contribution from nuclear magnetic moments in a weak LF.\cite{hayano1979} The full expression for $ g(\nu _0,\sigma,t) $ is
\begin{multline*}
g(\nu _0,\sigma,t) = 1 
- \frac{2\sigma ^ 2  }{ \nu _0^{2}  } \left[ 1 -     e^{ -\frac{1}{2} \sigma^{2} t ^{2}   } \cos(\nu _0 t   )  \right]\\ 
+ \frac{2\sigma^{4}  }{ \nu _0^{3} } \int_{0}^{t}   e^{ -\frac{1}{2} \sigma^{2} \tau ^{2} } \sin(\nu _0 \tau)  d\!\tau
\end{multline*}

\noindent where $ \sigma $ represents the distribution width of nuclear magnetic moment values and is approximately temperature independent. $ \sigma \approx 0.1 $~$ \mu $s$ ^{-1} $ was determined by fitting the highest-temperature data and was then kept fixed for lower-temperature-data fits.  $\nu _0 = \gamma_{\mu}\mu_{0}H_{0}/2\pi $ with $ H_{0} $ the applied magnetic field (a fixed parameter in the fitting).

\begin{figure}
	
	\includegraphics[width=1.0\columnwidth]{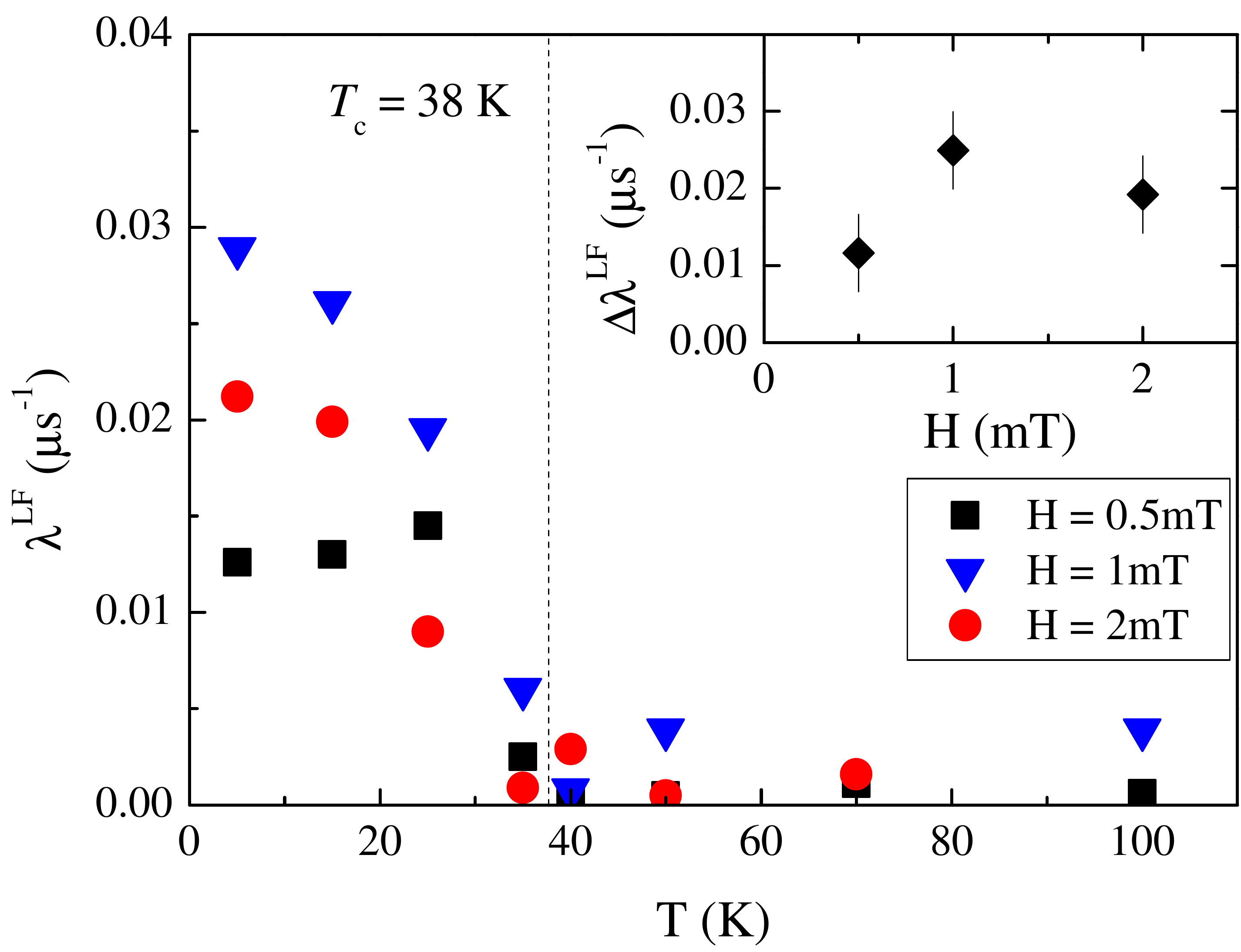}
	\caption{\label{fig:lfmusr}  
	The temperature dependence of the LF-$\mu$SR relaxation rate for $ x=0.43 $ showing the increase in magnetic fluctuations below $ \tc $. The inset shows the LF field-dependence of the change in the relaxation rate at low temperature with respect to $ T>\tc $, $ \Delta \lambda^{\mathrm{LF}} $. }
	
\end{figure}     

The results are shown in Fig.~\ref{fig:lfmusr} and reveal that $ \lambda^{\mathrm{LF}} $ exhibits an anomalous increase below $ \tc $. The effect is similar, though almost twice as large as that seen for the overdoped electron-doped sample with Ba(Fe$ _{1.89} $Co$ _{0.11} $)$ _{2} $As$ _{2} $ \cite{bernhard2012}. 
In our case however, the field dependence of the change in $ \lambda^{\mathrm{LF}} $ at low temperature with respect to the normal-state, $ \Delta \lambda^{\mathrm{LF}} $, shown in the inset does not follow a Redfield functional form. 
The increase may be due to a gapping of scattering channels for spins in the SC state that gives rise to observable spin fluctuations below $ \tc $, or a contribution from vortices that are not aligned with the external field (although in that case the size of the effect should scale with the vortex density and therefore the external field strength). On the other hand, the enhancement of spin fluctuations in the SC state might suggest a constructive rather than competitive relation between SC and magnetism for these doping states near the maximum of the $ \tc $ dome. There is also the possibility that the enhanced relaxation is due to the SC state itself, i.e. from an exotic SC order parameter with a spin-triplet component as seen in Sr$ _{2} $RuO$ _{4} $,\cite{luke1998} PrOs$ _{4} $Sb$ _{12} $,\cite{aoki2003} or LaNiC$ _{2} $.\cite{hillier2009}

Finally, for optimal- to overdoped samples we do not see the paramagnetic shift in $ \bmu $ below $ \tc $ in TF-$\mu$SR experiments that was reported before in the electron-doped 122 compounds \cite{khasanov2009, williams2010, sonier2011, bernhard2012}. Instead we see the more customary diamagnetic shift due to a screening of the magnetic field in the SC state (not shown). 
An estimate of the in-plane London penetration depth, $ \lambda_{L} $, was then calculated from the second-moment of the real-part of the Fourier transform of $ P(t) $ after, for example, Refs.~\onlinecite{brandt1988,pumpin1990}. We find $ \lambda_{L} \approx 200$~nm at $ T= 5$~K for both our optimally and over-doped samples. Residual magnetic order and/or vortex lattice disorder in these samples introduces a systematic uncertainty in such an estimate of $ \lambda_{L} $ based on TF-$ \mu $SR measurements with a possible underestimation of $\lambda_{L} $ to some degree.\cite{sonier2011} However, separate estimates of $ \lambda_{L} $ based on our IR spectroscopy data for these samples, shown in Sec.~\ref{subsec:scoptics}, also give a similar value of $ \approx 200 $~nm. 

The exact Fermi-surface shape, Hund's rule couplings and disorder in these compounds are undoubtedly very important factors for the manifest electronic and magnetic properties. However, the similarity between BFCA and BKFA phenomenology shown here suggest that the (i) co-existence and competition of SC and magnetism in underdoped compounds and (ii) SC-enhanced magnetism in near optimal-doped and overdoped samples are effects intrinsic to BFA that rather relate to the proximity of magnetic and SC orders. 
 
\section{Optical spectroscopy}
\label{sec:optics}
\begin{figure*}
	
	\includegraphics[width=1.0\textwidth]{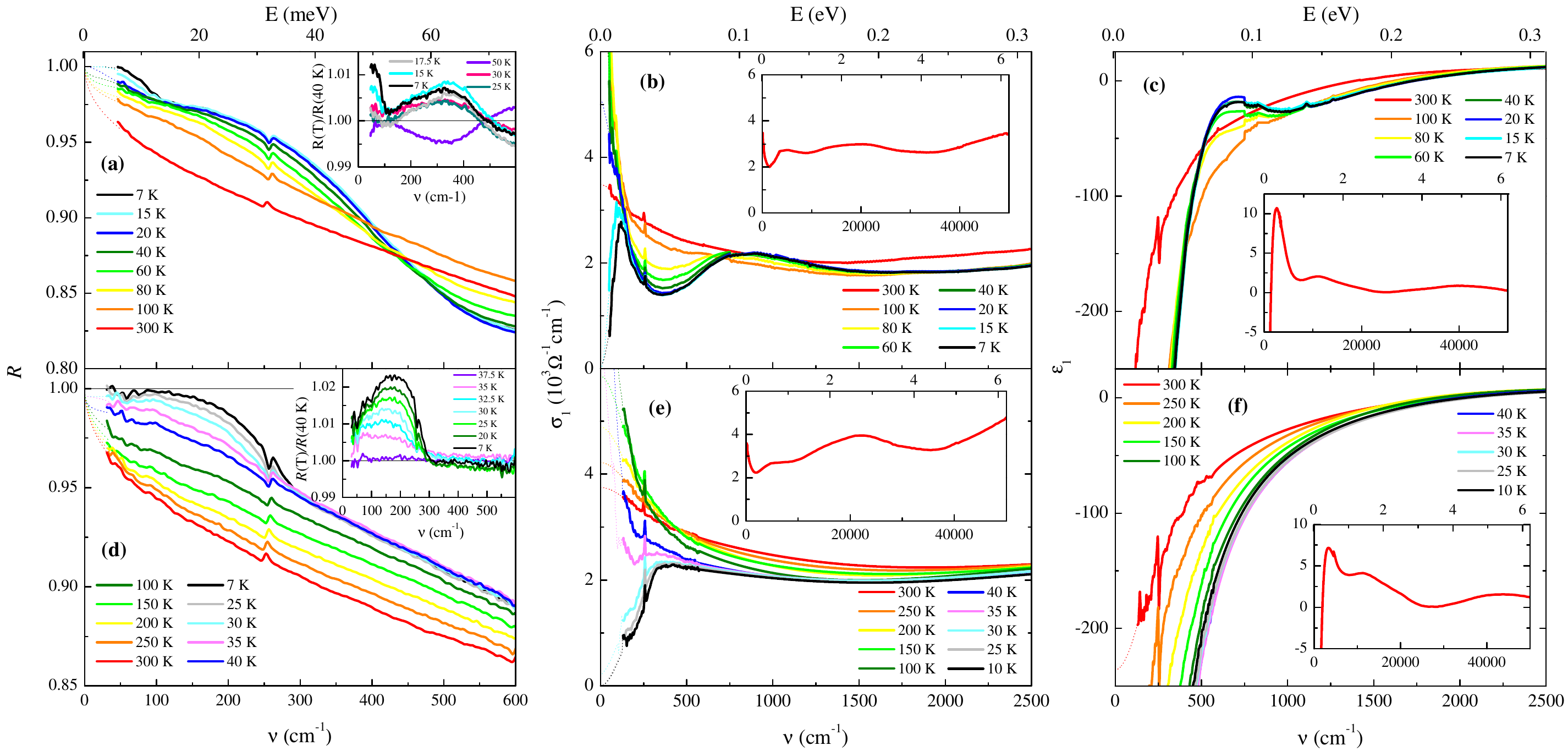}
	\caption{\label{fig:opticsraw}  
		Optical response of BKFA. (a) The far-infrared reflectivity for $ x=0.24 $ at selected temperatures. The inset shows the reflectivity at low temperatures divided by the spectrum with $ T\approx \tc $. Dotted lines in panel (a) show the low-frequency extrapolations used, in conjunction with mid-infrared ellipsometry, to determine the corresponding real-part of the optical conductivity $ \sigma_{1} $, shown in panel (b), and real-part of the dielectric function $ \epsilon_{1} $, shown in panel (c). The insets to panels (b) and (c) show the room temperature data to near-UV energies. Corresponding data is shown for $ x=0.43 $ in panels d, e and f. }
	
\end{figure*}

To complement the $\mu$SR results presented above, we have also studied several of the samples with IR optical spectroscopy. Representative spectra for $ x=0.24 $ of the reflectivity, $ R $, are shown in Fig.~\ref{fig:opticsraw}a, with the corresponding spectra of the real part of the optical conductivity, $ \sigma_{1} $, presented in Fig.~\ref{fig:opticsraw}b and the real part of the dielectric function, $ \epsilon_{1} $, in Fig.~\ref{fig:opticsraw}c. Similarly, we show the response for the slightly overdoped sample with $ x=0.43 $ in Figs.~\ref{fig:opticsraw}d, e and f. 

\subsection{Normal state response}
\label{subsec:nsresponse}
In the high-temperature paramagnetic state, $ T > \tn $, the spectra are similar to those previously reported in the literature \cite{hu2008, li2008, schafgans2011, nakajima2011, charnukha2011, wang2012, dai2012, dai2013, marsik2013, mallett2015bkfaIR}. They show a Drude response at low frequencies, arising from the itinerant carriers on the multiple bands crossing the Fermi-energy. From the extrapolation of the Drude response to zero frequency we estimate the DC resistivity, which we find decreases at room temperature from $\rho_{ab} \approx 400 $~$ \mu\Omega $cm for $ x=0 $ to $\rho_{ab} \approx 250 $~$ \mu\Omega $cm at higher $ x $. These values are consistent with those previously reported in the literature for BKFA ($\rho_{ab} \approx 250 $ to $400$~$ \mu\Omega $cm) \cite{hassinger2012, liu2014}. Looking to higher frequencies, the Drude response exhibits a pronounced tail that results from inelastic scattering of the itinerant carriers \cite{charnukhareview2014}. At these frequencies there are also contributions from the low-energy interband transitions that are, for example, responsible for the upturn in $ \sigma_{1} $ around 5000~cm$ ^{-1} $. With increasing $ x $, they are expected to move to lower energy leading to a significantly larger fraction of the low-energy spectral weight arising from interband transitions \cite{calderon2014}, and our ellipsometry data support this prediction.

With decreasing $ T $ the Drude response from at least one of the bands narrows and the spectral weight of the Drude terms decreases overall due to a transfer to high energies that originates from the strong Hund's-rule-coupling \cite{schafgans2011, wang2012, georges2013}. Changes in the response from the low-energy interband transitions are much less pronounced. In this temperature regime, we do not see evidence for a collective mode at $ \approx150 $~cm$ ^{-1} $ as was observed in Ref.~\onlinecite{charnukha2011prb}. A low-frequency feature at $ \approx 120 $~cm$ ^{-1} $ developing below a certain temperature $ T^* $ has also been reported in underdoped BKFA samples with $ x=0.20 $ ($ \tc = 19 $~K and $ \tn = 104 $~K) and, less prominently, $ x=0.12 $ ($ \tc = 11 $~K and $ \tn = 121 $~K) \cite{dai2012}. It is difficult to distinguish a similar feature in our raw data. We performed a similar spectral weight analysis to Ref.~\onlinecite{dai2012} in order to estimate $ T^* $ in our own samples. For $ x=0.23 $ and 0.24 the values of $ T^* $ we observe are approximately coincident with $ \tc $ making it difficult to distinguish from an effect relating to the formation of a SC gap. For $ x=0.19 $, $ T^* $ may be up to 40~K, which, while somewhat below the 80~K found in Ref.~\onlinecite{dai2012} for $ x=0.20 $, is still somewhat higher than $ \tc = 15 $~K for this sample. 

The IR-active phonon mode observable near 260~cm$ ^{-1} $ corresponds to the in-plane vibrations of Fe against As \cite{schafgans2011, litvinchuk2008, akrap2009} and generally displays only a weak temperature dependence and a weak dependence on $ x $. The second expected IR-active mode around 95~cm$ ^{-1} $, which is dominated by the displacement of Ba, is not resolved for the $ x>0 $ samples, most likely due to the disorder from the K substitution.

\subsection{Signatures of magnetic order in the IR response}
\label{subsec:sdw} 

\begin{figure}
	\includegraphics[width=1.0\columnwidth]{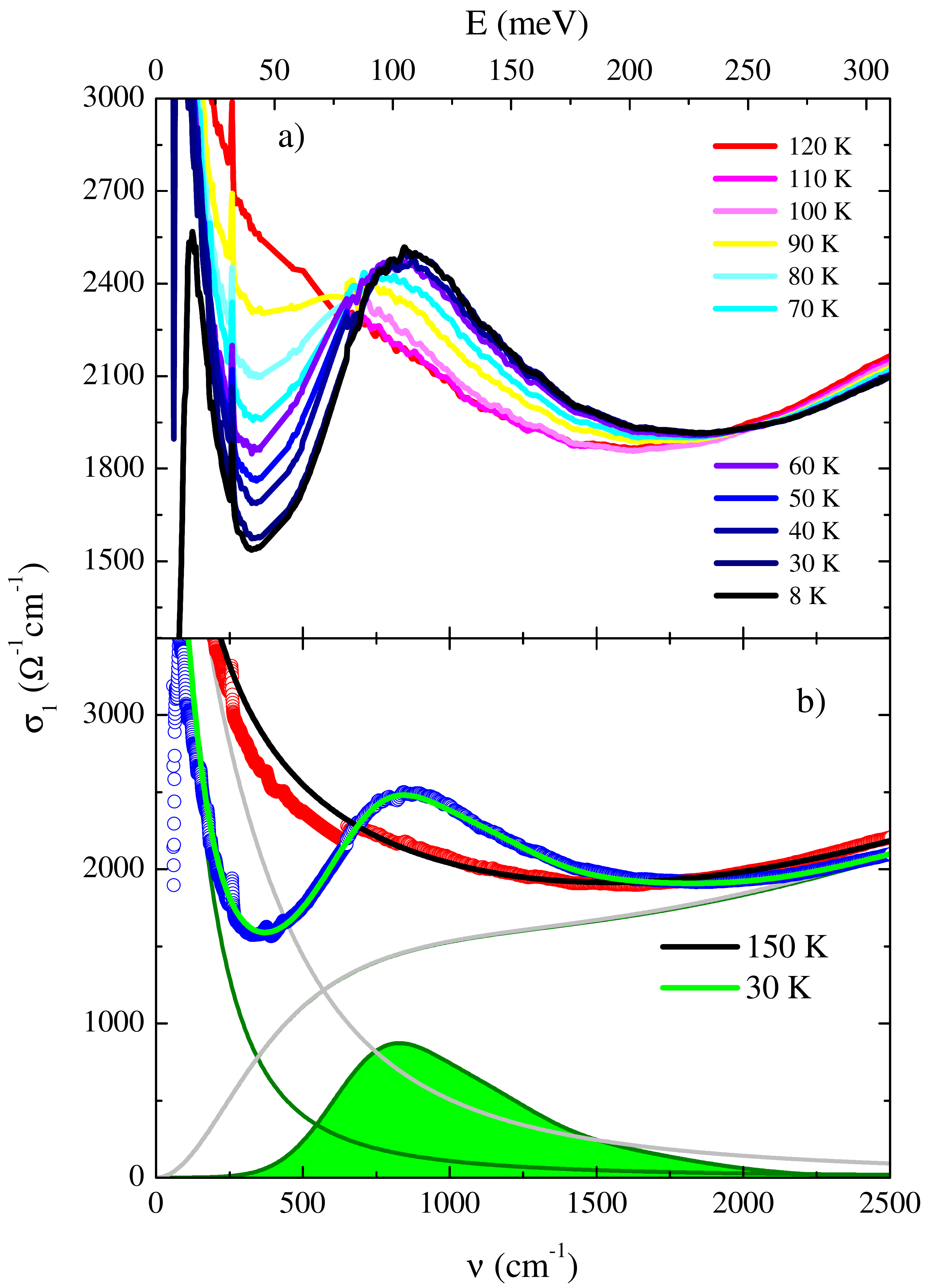}
		
	\vspace{0.3cm}
	\includegraphics[width=0.9\columnwidth]{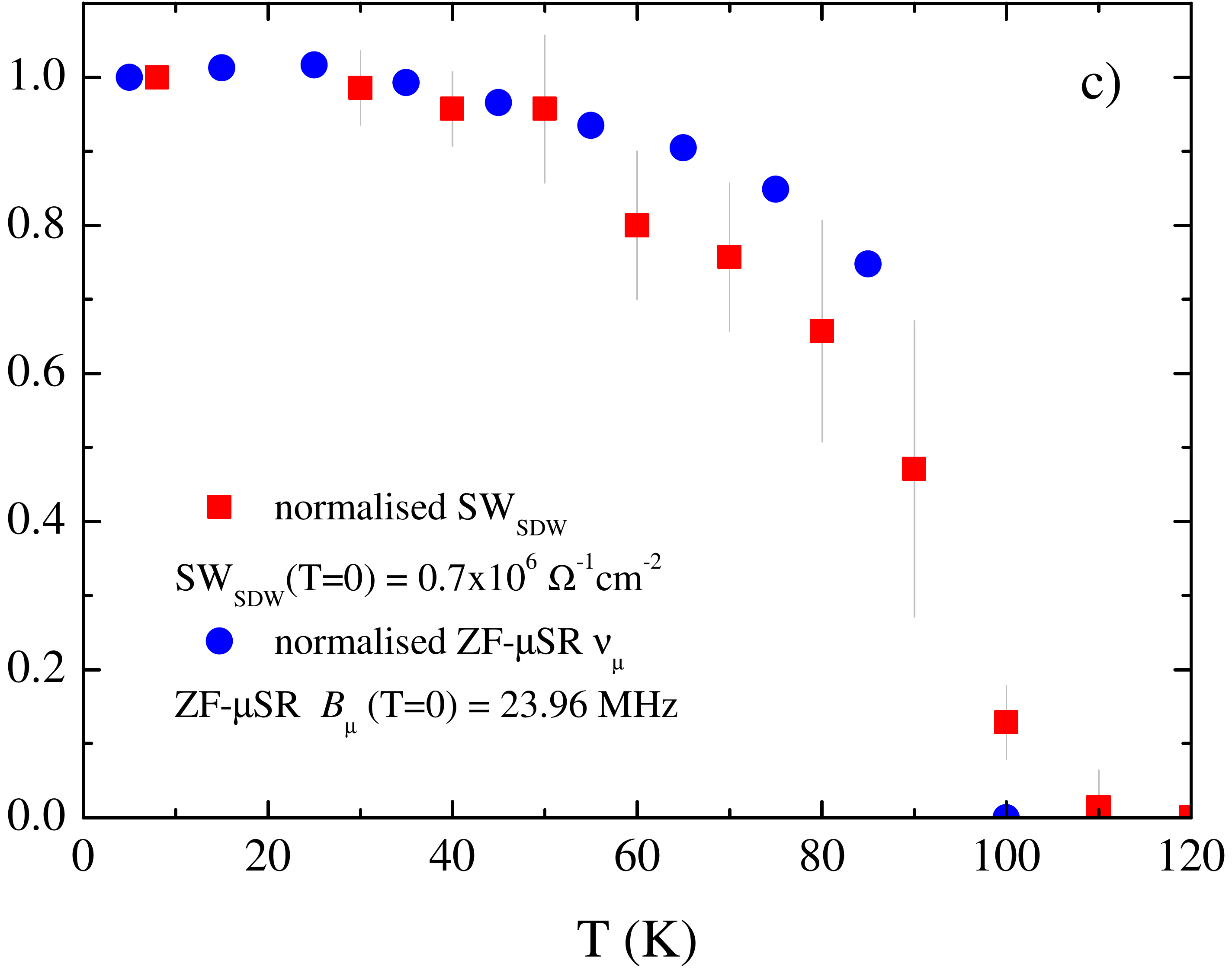}

	\caption{\label{fig:opticssdw}  
		Spin-density-wave (SDW) feature in the optical response of $ x=0.23 $. a) The temperature dependence of the SDW absorption as seen in $ \sigma_{1} $. b) The fitting procedure used to estimate the SW of the SDW corresponding to the filled green area. The red points show the 150~K data. The black solid line is the corresponding fit with the light grey lines showing the Drude and Lorentzian components of the fit. Similarly, the blue points are the 30~K data with the green lines the total fit and fit components. c) The estimated SW$ _{\mathrm{SDW}} $ as a function of temperature. Blue symbols show the magnetic field at the muon site, $ \bmu $, from ZF-$\mu$SR.}
	
\end{figure}

For the underdoped samples, in the magnetic state the optical response undergoes significant changes that stem from a partial gapping of the Fermi-surface due to the formation of a spin-density-wave (SDW). These are a peak around 800~cm$ ^{-1} $ in $ \sigma_{1} $ corresponding to the SDW pair-breaking peak, and a reduction of the spectral weight of the Drude-response \cite{hu2008, schafgans2011, nakajima2011, marsik2013}. Figure~\ref{fig:opticssdw}a highlights the temperature evolution of the SDW in $ \sigma_{1}(T) $ for $ x =0.23 $ as an example. The evolution is characterized by a progressive redistribution of spectral weight from lower energies, seen for example in Fig.~\ref{fig:opticsraw}b and c by the suppressed conductivity around 400~cm$ ^{-1} $ and corresponding peak in $ \epsilon_{1} $, to higher energies to form a characteristic pair-breaking peak in $ \sigma_{1} $. These features grow relatively rapidly below the $ \tn $ determined by $ \mu $SR (100~K in this case) and then display little change below 50~K. The peak centre in $ \sigma_{1} $ moves only slightly from $ \approx 700 $~cm$ ^{-1} $ just below $ \tn $ to $ \approx 850 $~cm$ ^{-1} $ at low temperature. 

The $ x=0 $ and $ x=0.26 $ samples have qualitatively different behaviour from that described above. For $ x = 0 $ this concerns an additional SDW feature peaked at a lower energy\cite{hu2008, marsik2013} of about 400~cm$ ^{-1} $. For $ x =0.26 $, in the t-AF phase a distinct SDW is observed in the temperature range 32~K$ \approx T^{ \mathrm{N2} } > T > T^{ \mathrm{N3} } \approx $18~K with a re-entrance of the o-AF state below $ T^{ \mathrm{N3} } $.\cite{bohmer2015, allred2015xray, mallett2015bkfaIR}

Signatures of a SDW state in the optical response, at any temperature, disappear between $ x=0.26 $ and $ x=0.30 $. This is consistent with the results from B\"{o}hmer \textit{et al.} that show no lattice distortions associated with magnetism above $ x \approx 0.28 $.\cite{bohmer2015}

The SW of the pair-breaking peak reflects the magnitude of the order parameter of the SDW and thus of the ordered magnetic moment due to the itinerant carriers. Our procedure for estimating the SW associated with the SDW is as follows; we firstly fit the `normal state' (i.e. $ T>\tn $) $ \sigma = \sigma_{1} + i\sigma_{2} $ data between 0 and 4000~cm$ ^{-1} $ using two Drude terms and three broad Lorentzian oscillators (representing the interband transitions). Such a fit is shown in Fig.~\ref{fig:opticssdw}b for $ x=0.23 $ at 150~K. Fitting the data below $ \tn $, we keep the Lorentzian oscillators fitted from the data above $ \tn $ essentially fixed (allowing only for small changes of the oscillator strength). The remainder of the spectrum is fitted by varying the Drude terms (which determine the low-frequency response) and introducing Gaussian oscillators to fit the pair-breaking peak.  Gaussian oscillators are used because they are more localized than Lorentz oscillators and still give a phenomenological description of the data. The SW of the pair-breaking peak is then determined from the fitted Gaussian oscillators. Whilst the absolute values obtained from this procedure are somewhat sensitive to the fitting model, they do provide a meaningful measure of the $ x $- and $ T $-dependence since the same fitting models are used for all $ x $.

By way of example, in Fig.~\ref{fig:opticssdw}c we show the $ T $-dependence of the spectral weight of the pair-breaking peak, SW$ _{\mathrm{SDW}} $, for $ x=0.23 $ estimated using the procedure above (red squares). Another measurable quantity related to the magnitude of the magnetic order parameter is the size of the magnetic moment, which is in-turn related to the magnitude of the magnetic field at the muon site, $ \bmu $. The temperature dependence of $ \bmu $ as measured by ZF-$\mu$SR for the main muon site is also plotted in Fig.~\ref{fig:opticssdw}c (blue circles) and shows a similar temperature evolution to the SW of the pair-breaking peak.

\begin{figure}
	
	\includegraphics[width=1.0\columnwidth]{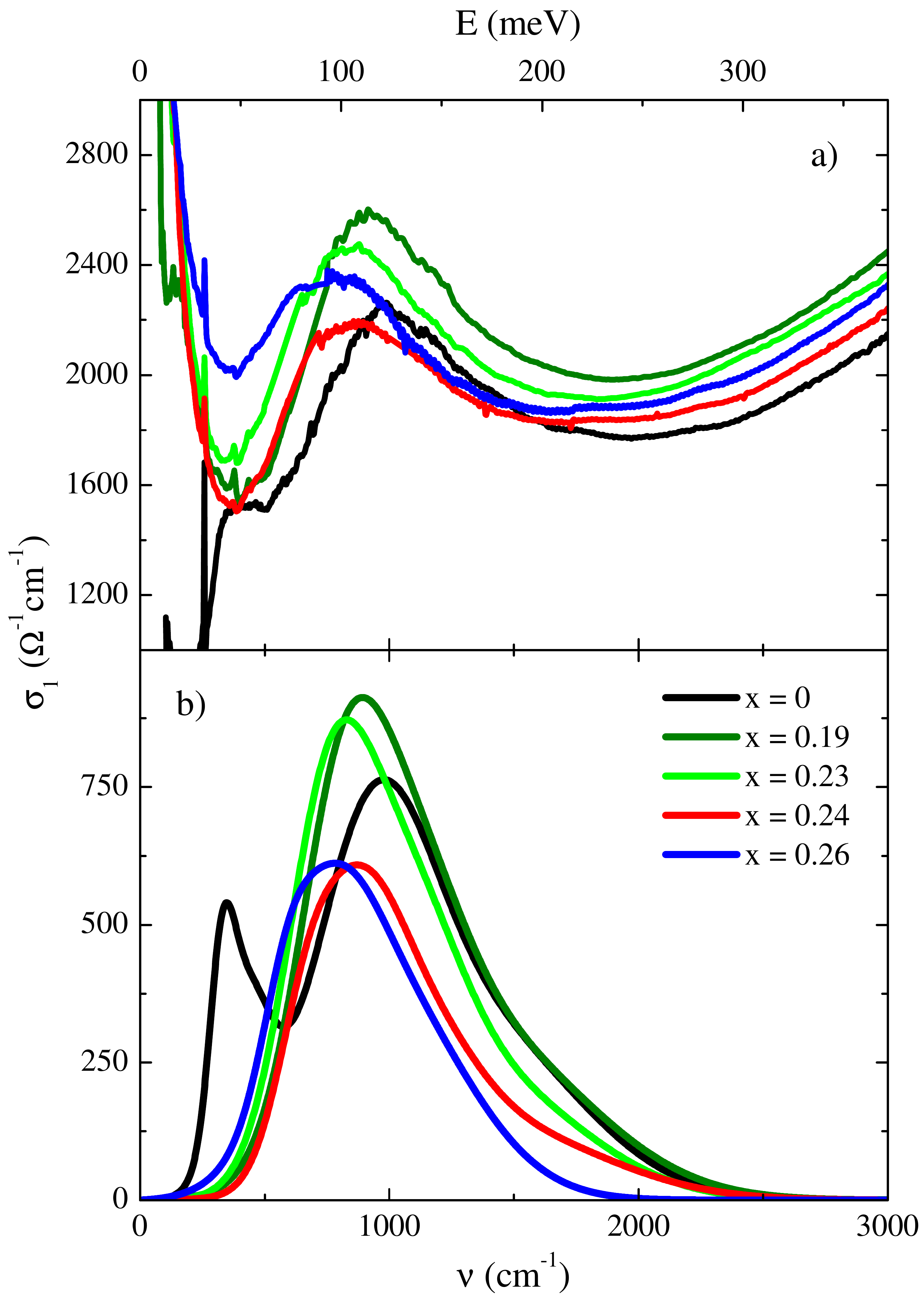}

	\caption{\label{fig:opticssdwvsx}  
		Doping dependence of the SDW feature. a) $ \sigma_{1} $ data at $ T = 30 $~K showing the evolution of the SDW with doping, $ x $. b) The fitted SDW from the data in panel a) using the procedure outlined in the text and Fig.~\ref{fig:opticssdw}b.}
	
\end{figure}

Figure~\ref{fig:opticssdwvsx}a and b shows raw data and fitted Gaussian oscillators for each $ x $ at low-temperature, $ \tc < T \ll \tn $, respectively. With increasing $ x $ the feature progressively moves to lower energy whilst its SW, a measure of the itinerant magnetic moment, decreases. Estimates of the SW are shown in Fig.~\ref{fig:zfmusr}b (blue diamonds) along with $ \vmu $ from ZF-$\mu$SR measurements. The reduction in the (itinerant) magnetic moment with increasing $ x $ is not monotonic with a steep decrease in the SW of the SDW feature between $ x=0.23 $ and $ x=0.26 $. In the $ x=0.26 $ sample (in the orthorhombic AF state rather than the lower-temperature tetragonal AF state \cite{mallett2015bkfaIR}) the SW is about 40\% lower than in the undoped parent compound \cite{hu2008,marsik2010}. 

By means of the above fitting procedure and from a spectral-weight analysis of the raw data, we also estimate a SDW-induced decrease of the Drude weight of about $ 65 \pm 10 $\% for the undoped $ x=0 $ sample to $ 35 \pm 10 $\% for $ x = 0.20 $ and falling to about $ 25 \pm 10 $\% for $ x=0.26 $ (in the o-AF state). These estimates are in reasonable agreement with the heat capacity data \cite{storey2013}. This reduction in the itinerant carrier density means that there are fewer states available for the SC state.

Meanwhile, the energy of the feature, $ \esdw $, that we estimate from the peak in $ \sigma_{1} $ in Fig.~\ref{fig:opticssdwvsx}b, moves slightly lower with $ x $, from $ \esdw \approx 950 $~cm$ ^{-1} $ (120~meV) for $ x=0 $ to about $ 800 $~cm$ ^{-1} $ at $ x=0.26 $. This suggests a lower-energy SDW-related gapping of the Fermi-surface. Within uncertainties however, the ratio $ 2\esdw / \tn $ remains $ 5.0 \pm 0.5 $ across the underdoped samples, where $ \tn $ has been taken from the $\mu$SR data. We note that signatures of the SDW state in the optical data are evident between 10 and 30~K higher than $ \tn $ across the underdoped series. This is probably related to predicted fluctuations of the SDW\cite{mazin2008} above the frequency probed by the experimental techniques (THz in IR spectroscopy compared with MHz for $\mu$SR).\cite{marsik2013, mallett2015bkfaIR} A very recent optical study of underdoped BKFA also notes signatures of the SDW above the $ \tn $ found from static probes.\cite{xu2016} Going further, at the fast, sub-fs time-scales probed with x-ray emission spectroscopy a significant Fe magnetic moment is observed at room temperature.\cite{pelliciari2016}

These observations paint the picture of a SDW of relatively constant energy scale, but which gaps a progressively smaller area of the Fermi-surface as $ x $ increases and which rather abruptly disappears above $ x \approx 0.28 $.

\subsection{Superconductivity}
\label{subsec:scoptics}
The pnictides are by now well-known for the microscopic coexistence of superconducting and magnetic phases that they exhibit and, as shown by the $\mu$SR data, underdoped BKFA is no exception. Optical spectroscopy was able to show early on that these two phases compete in the underdoped regime for the same electronic states\cite{marsik2010}, whereby the SW of the SDW decreased at the onset of SC. Although such a decrease was not clearly seen in our BKFA data, the optical data presented here do show a similar phenomenology for BKFA in that the SC gap energy and superfluid density are enhanced in the absence of a SDW state. 

At low energy and temperature, $ R=1 $ with experimental uncertainties for $ x \geq 0.19 $ implying a bulk, nodeless superconducting state. In conjunction with the $\mu$SR results on the same samples which showed a 100\% magnetic volume fraction, we can confirm the bulk co-existence of SC and magnetic states (in real space) in BKFA. It has been shown previously that more than one SC gaps are evident from the optical spectra of BKFA \cite{li2008, charnukha2011prb} and our data are consistent with this multi-gap scenario. A detailed analysis of the gap shape is a subject beyond the scope of this work and would likely provide more robust low-frequency extrapolations of the reflectivity data. 

\begin{figure}
	\begin{flushleft}
	\includegraphics[width=0.9\columnwidth]{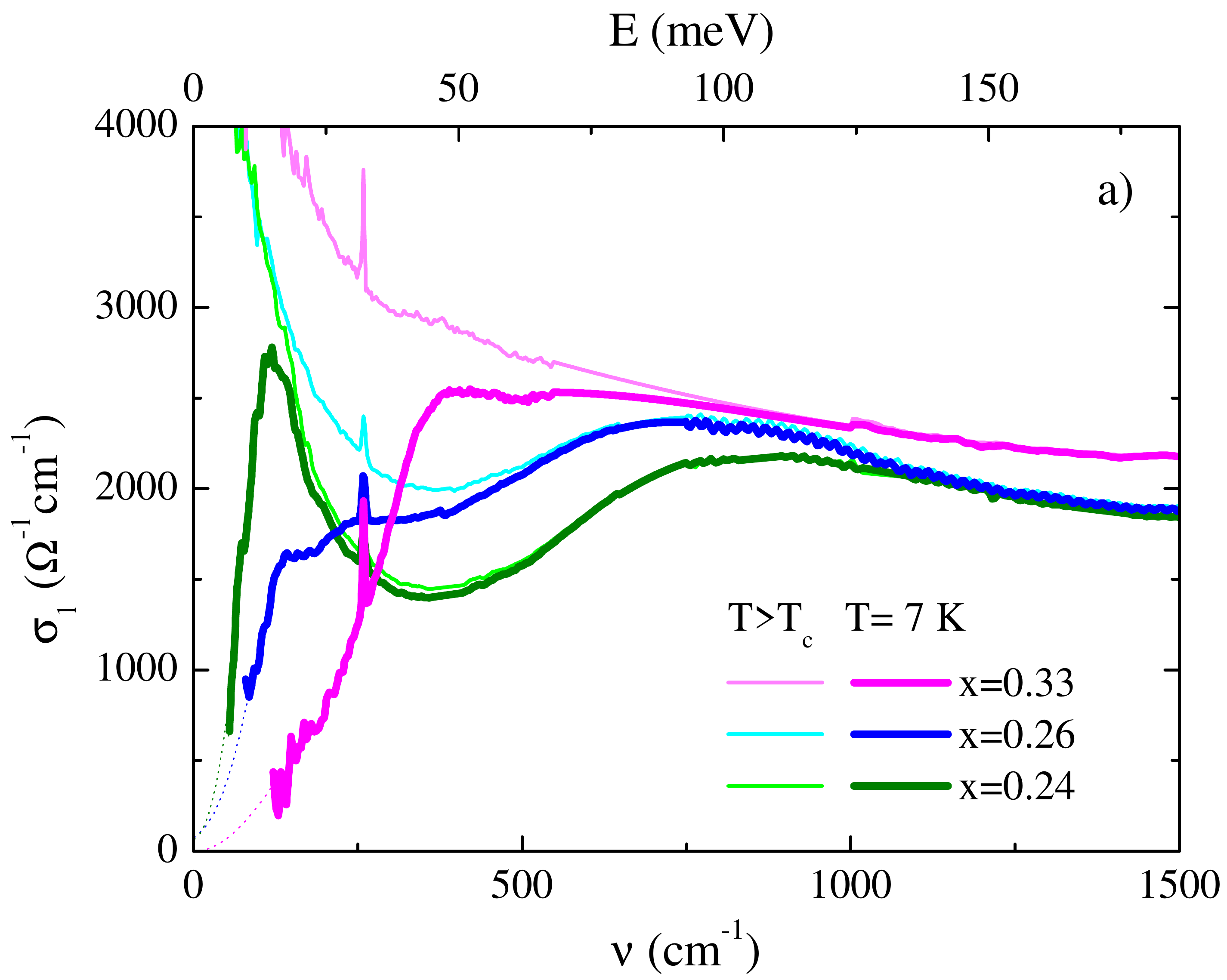}
	\end{flushleft}

	\vspace{0.25cm}
	\includegraphics[width=1.0\columnwidth]{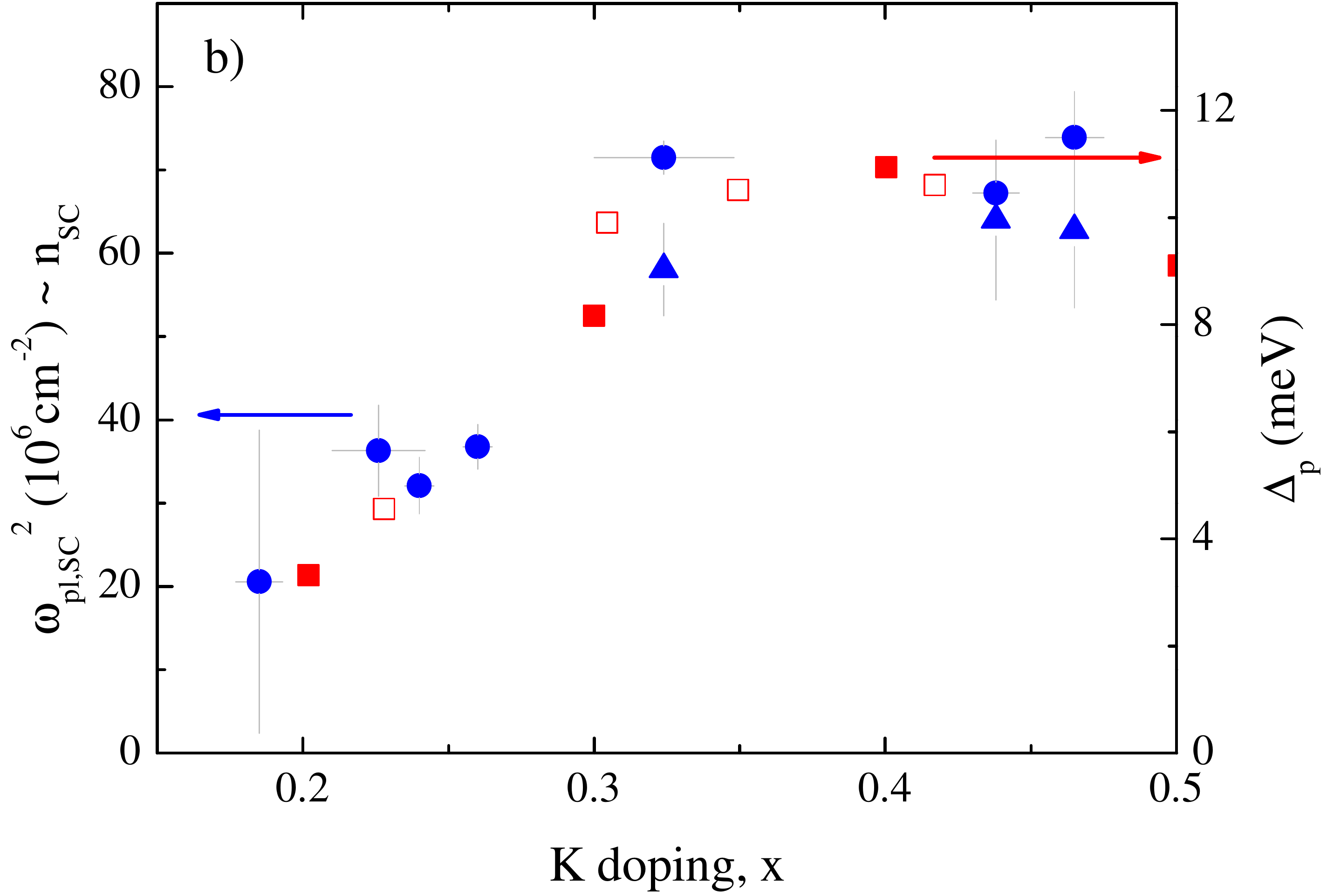}
	
	\caption{\label{fig:opticssc}  
	Superconducting response in the optical spectra. a) $ \sigma_{1} $ for $ T > \tc $ (thin lighter lines) and well below $ \tc $ (thick darker lines) from under- to optimally doped samples. b) The doping dependence of the SC superfluid density, $ \omega_{ \mathrm{pl, SC} }^{2} \propto n_{\mathrm{SC}}$, at $ T=7 $~K as determined from the optical data, blue circles, and $ \mu $SR data, blue triangles. The energy of the largest SC gap as determined from specific heat measurements is shown as closed red squares \cite{storey2013} and open red squares \cite{hardy2016}.}
	
\end{figure}

We find evidence that the shape of the SC gap appears to change between $ x=0.24 $ and $ x=0.33 $ as shown in Figs.~\ref{fig:opticsraw}a and d and in Fig.~\ref{fig:opticssc}.

The insets to Figs.~\ref{fig:opticsraw}a and d show the reflectivity ratio for a magnetic $ x=0.24 $ sample and a non-magnetic $ x=0.43 $ sample. The reflectivity edge in the SC state is clearly much more pronounced in the non-magnetic sample, occurs at higher energy and grows rapidly immediately below $ \tc $. Figure~\ref{fig:opticssc}a shows $ \sigma_{1} $ immediately above $ \tc $ and at $ T=7 $~K for three samples which span this transition. For $ x=0.24 $, a significant redistribution of SW from the normal state to the SC state, extends only up to 140~cm$ ^{-1} $ in energy. At the slightly higher doping of $ x=0.26 $ it is up to 440~cm$ ^{-1} $ and by $ x = 0.33 $ up to 725~cm$ ^{-1} $. Similarly, the steep decrease in $ \sigma_{1} $ at low-$ T $ due to SC occurs at 100~cm$ ^{-1} $ for $ x = 0.24 $ and increases to 345~cm$ ^{-1} $ for $ x = 0.33 $.

Using the methodology outlined in Ref.~\onlinecite{mallett2015bkfaIR} and references therein, we estimate the (in-plane) SC superfluid density, $ \nsc $, from $ \epsilon_{1} $ and from the ‘missing’ SW in $ \sigma_{1} $. Figure~\ref{fig:opticssc}b shows the doping dependence of the square of the SC condensate density plasma frequency, $ \omega_{ \mathrm{pl, SC} }^{2} $, which is proportional to $ \nsc $ and $ \lambda_{L}^{-2} $, as $ \lambda_{L} = 1.61\times10^{6} /  \omega_{ \mathrm{pl, SC} } $ (with $ \lambda_{L} $ in nm and $ \omega_{ \mathrm{pl, SC} } $ in ~cm$ ^{-1} $). There is a pronounced decrease in $ \nsc $ for $ x\leq 0.3 $, coincident with the appearance of the SDW signatures in the optical response - a phenomenology that is similar to what is found in the cuprates with the opening of the pseudogap. This decrease in $ \nsc $ also coincides with a decrease in the energy of the largest superconducting gap, $ \Delta_{ \mathrm{p} } $, as determined from specific heat measurements \cite{storey2013, hardy2016} and shown as red squares in Fig.~\ref{fig:opticssc}b. $ \nsc $ changes little between our $ x=0.33 $ and $ x=0.47 $ with values consistent with those previously reported \cite{li2008, charnukha2011prb, hiraishi2009}. 
The derived in-plane London penetration depth is $ \lambda_{L} \approx 200$~nm. This is in agreement with the values of $ \lambda_{L} $ obtained from the TF-$ \mu $SR data, which are shown as blue triangles in Fig.~\ref{fig:opticssc}b for optimally and slightly overdoped samples. Static magnetic order in the underdoped samples prevents us from estimating $ \lambda_{L} $ from the $ \mu $SR data there. 

A maximum in $ \nsc $ for $ x\approx0.4 $ has been reported from thermodynamic probes on BKFA \cite{storey2013, bohmer2015, hardy2016}. In BFCA \cite{hardy2016, bernhard2012} and BNFA \cite{wang2016} a maximum in $ \nsc $ has also been observed close to the point at which magnetism is no longer observable. For the case of BaFe$ _{2} $As$ _{2-z} $P$ _{z} $ (BFAP) there are reports of a different behaviour with a sharp minimum in the SC condensate density close to where magnetism disappears and the observation of nodes in the SC gap\cite{hashimoto2012, putzke2014}. This apparent qualitative difference between BFAP and the electronically doped BKFA (and BFCA) is intriguing, and is unlikely related to disorder since BKFA is a similarly `clean' from electronic disorder associated with the dopant. Measurements of $\lambda_{L}$ in BKFA between $ x=0.26 $ and $ x=0.5 $ in a finer mesh are desirable to test for any such increase in $ \lambda_{L} $ in a narrow range of $ x $. 

Without a full multi-band analysis of the gap shape it is difficult to estimate the ratio $ \Delta_{\mathrm{SC}}/k_{B}\tc $. Although, considering the similarity of our data to Ref.~\onlinecite{charnukha2011prb}, a ratio of about 6 for at least one of the bands is not inconsistent. 
 
 
\section{Summary}
\label{sec:summary}
In underdoped BKFA, we find a concomitant decrease in local magnetic field from the Fe moment as measured by $ \mu $SR, $ \bmu $, and the spectral weight of the spin density wave feature, SW$ _{\mathrm{SDW}} $, in the IR-optical response. On increasing the doping, $ \bmu $ initially decreases more slowly than the N\'{e}el temperature until it too falls rapidly close to optimal doping. A comparison of the evolution in doping of $ \bmu $ and SW$ _{\mathrm{SDW}} $ suggests that close to optimal doping, where the static magnetic order disappears sharply, the itinerant magnetic moment decreases even faster than the total magnetic moment. In the overdoped samples, there is no evidence of a static magnetic order. 

In the moderately underdoped region, we see a coexistence of static antiferromagnetic magnetic order and superconductivity on a nanoscopic scale. Here, the two orders compete for the same electronic states as shown by (i) the reduced magnetic order parameter below $ \tc $ from $\mu$SR, (ii) the optical conductivity showing a capture of spectral weight by the spin-density wave from the low-frequency Drude response - spectral weight that could otherwise contribute to the SC condensate - and (iii) the marked increased in the superconducting gap energy and condensate density between the slightly underdoped and optimally doped samples. In the slightly overdoped region however, where the optical response shows a strong superconducting state, our LF-$\mu$SR data suggest a qualitatively different relation between superconductivity and magnetism whereby spin fluctuations are enhanced below $ \tc $. A similar observation has been made in the electron doped system BFCA \cite{bernhard2012}.

We consider it possible that around optimal doping yet more novel electronic/magnetic properties might be discovered. This is partly suggested by the novel t-AF magnetic state in BKFA with a doping state of $ x=0.26 $\cite{hassinger2012, avci2014, wasser2015, allred2015xray, mallett2015musr} and by our observation here of spin-fluctuations below $ \tc $ in slightly overdoped samples. Specific-heat and thermal expansion on BKFA point to a rich competition of phases in this doping regime\cite{bohmer2015,hardy2016}, and it would be interesting to study in detail using the more sensitive magnetic probe of $\mu$SR. More recently, a similar specific-heat and thermal-expansion study of Ba$ _{1-x} $Na$ _{x} $Fe$ _{2} $As$ _{2} $ indeed showed novel magnetic states as the N\'{e}el temperature approaches the value of $ \tc $ \cite{wang2016}. BaFe$ _{2} $As$ _{2} $ is apparently host to several near-degenerate magnetic states whose energy depends on parameters such as bond-lengths, impurity scattering and Fermi-surface nesting. The similarities between the BFCA and BKFA phenomenology shown here may help to elucidate the role of each of these parameters. 

\begin{acknowledgments}
This work was supported by the Schweizerische Nationalfonds (SNF) through grant No. 200020-153660 and the Marsden Fund of New Zealand. Some measurements were performed at the IR beam line of the ANKA synchrotron at FZ Karlsruhe, where we acknowledge the support of Y.L. Mathis and M. S\"{u}pfle. We thank C. Neururer and B. Grobety for the electron dispersion spectroscopy measurements. The $ \mu $SR work has been performed at the Swiss Muon Source at the Paul Scherrer Institute, Switzerland.
	
\end{acknowledgments}


\begin{thebibliography}{79}%
	\makeatletter
	\providecommand \@ifxundefined [1]{%
		\@ifx{#1\undefined}
	}%
	\providecommand \@ifnum [1]{%
		\ifnum #1\expandafter \@firstoftwo
		\else \expandafter \@secondoftwo
		\fi
	}%
	\providecommand \@ifx [1]{%
		\ifx #1\expandafter \@firstoftwo
		\else \expandafter \@secondoftwo
		\fi
	}%
	\providecommand \natexlab [1]{#1}%
	\providecommand \enquote  [1]{``#1''}%
	\providecommand \bibnamefont  [1]{#1}%
	\providecommand \bibfnamefont [1]{#1}%
	\providecommand \citenamefont [1]{#1}%
	\providecommand \href@noop [0]{\@secondoftwo}%
	\providecommand \href [0]{\begingroup \@sanitize@url \@href}%
	\providecommand \@href[1]{\@@startlink{#1}\@@href}%
	\providecommand \@@href[1]{\endgroup#1\@@endlink}%
	\providecommand \@sanitize@url [0]{\catcode `\\12\catcode `\$12\catcode
		`\&12\catcode `\#12\catcode `\^12\catcode `\_12\catcode `\%12\relax}%
	\providecommand \@@startlink[1]{}%
	\providecommand \@@endlink[0]{}%
	\providecommand \url  [0]{\begingroup\@sanitize@url \@url }%
	\providecommand \@url [1]{\endgroup\@href {#1}{\urlprefix }}%
	\providecommand \urlprefix  [0]{URL }%
	\providecommand \Eprint [0]{\href }%
	\providecommand \doibase [0]{http://dx.doi.org/}%
	\providecommand \selectlanguage [0]{\@gobble}%
	\providecommand \bibinfo  [0]{\@secondoftwo}%
	\providecommand \bibfield  [0]{\@secondoftwo}%
	\providecommand \translation [1]{[#1]}%
	\providecommand \BibitemOpen [0]{}%
	\providecommand \bibitemStop [0]{}%
	\providecommand \bibitemNoStop [0]{.\EOS\space}%
	\providecommand \EOS [0]{\spacefactor3000\relax}%
	\providecommand \BibitemShut  [1]{\csname bibitem#1\endcsname}%
	\let\auto@bib@innerbib\@empty
	\bibitem [{\citenamefont {Miyake}\ \emph {et~al.}(1986)\citenamefont {Miyake},
		\citenamefont {Schmitt-Rink},\ and\ \citenamefont {Varma}}]{miyake1986}%
	\BibitemOpen
	\bibfield  {author} {\bibinfo {author} {\bibfnamefont {K.}~\bibnamefont
			{Miyake}}, \bibinfo {author} {\bibfnamefont {S.}~\bibnamefont
			{Schmitt-Rink}}, \ and\ \bibinfo {author} {\bibfnamefont {C.~M.}\
			\bibnamefont {Varma}},\ }\href {\doibase 10.1103/PhysRevB.34.6554} {\bibfield
		{journal} {\bibinfo  {journal} {Phys. Rev. B}\ }\textbf {\bibinfo {volume}
			{34}},\ \bibinfo {pages} {6554} (\bibinfo {year} {1986})}\BibitemShut
	{NoStop}%
	\bibitem [{\citenamefont {Scalapino}\ \emph {et~al.}(1986)\citenamefont
		{Scalapino}, \citenamefont {Loh},\ and\ \citenamefont
		{Hirsch}}]{scalapino1986}%
	\BibitemOpen
	\bibfield  {author} {\bibinfo {author} {\bibfnamefont {D.~J.}\ \bibnamefont
			{Scalapino}}, \bibinfo {author} {\bibfnamefont {E.}~\bibnamefont {Loh}}, \
		and\ \bibinfo {author} {\bibfnamefont {J.~E.}\ \bibnamefont {Hirsch}},\
	}\href {\doibase 10.1103/PhysRevB.34.8190} {\bibfield  {journal} {\bibinfo
		{journal} {Phys. Rev. B}\ }\textbf {\bibinfo {volume} {34}},\ \bibinfo
	{pages} {8190} (\bibinfo {year} {1986})}\BibitemShut {NoStop}%
\bibitem [{\citenamefont {B\'{e}al-Monod}\ \emph {et~al.}(1986)\citenamefont
	{B\'{e}al-Monod}, \citenamefont {Bourbonnais},\ and\ \citenamefont
	{Emery}}]{beal1986}%
\BibitemOpen
\bibfield  {author} {\bibinfo {author} {\bibfnamefont {M.~T.}\ \bibnamefont
		{B\'{e}al-Monod}}, \bibinfo {author} {\bibfnamefont {C.}~\bibnamefont
		{Bourbonnais}}, \ and\ \bibinfo {author} {\bibfnamefont {V.~J.}\ \bibnamefont
		{Emery}},\ }\href {\doibase 10.1103/PhysRevB.34.7716} {\bibfield  {journal}
	{\bibinfo  {journal} {Phys. Rev. B}\ }\textbf {\bibinfo {volume} {34}},\
	\bibinfo {pages} {7716} (\bibinfo {year} {1986})}\BibitemShut {NoStop}%
\bibitem [{\citenamefont {Bickers}\ \emph {et~al.}(1987)\citenamefont
	{Bickers}, \citenamefont {Scalapino},\ and\ \citenamefont
	{Scalettar}}]{bickers1987}%
\BibitemOpen
\bibfield  {author} {\bibinfo {author} {\bibfnamefont {N.}~\bibnamefont
		{Bickers}}, \bibinfo {author} {\bibfnamefont {D.}~\bibnamefont {Scalapino}},
	\ and\ \bibinfo {author} {\bibfnamefont {R.}~\bibnamefont {Scalettar}},\
}\href {\doibase 10.1142/S0217979287001079} {\bibfield  {journal} {\bibinfo
	{journal} {International Journal of Modern Physics B}\ }\textbf {\bibinfo
	{volume} {1}},\ \bibinfo {pages} {687} (\bibinfo {year} {1987})}\BibitemShut
{NoStop}%
\bibitem [{\citenamefont {Hott}\ \emph {et~al.}(2004)\citenamefont {Hott},
	\citenamefont {Kleiner}, \citenamefont {Wolf},\ and\ \citenamefont
	{Zwicknagl}}]{wolf2004}%
\BibitemOpen
\bibfield  {author} {\bibinfo {author} {\bibfnamefont {R.}~\bibnamefont
		{Hott}}, \bibinfo {author} {\bibfnamefont {R.}~\bibnamefont {Kleiner}},
	\bibinfo {author} {\bibfnamefont {T.}~\bibnamefont {Wolf}}, \ and\ \bibinfo
	{author} {\bibfnamefont {G.}~\bibnamefont {Zwicknagl}},\ }\href@noop {}
{\emph {\bibinfo {title} {Frontiers in Superconducting Materials;
			Superconducting Materials - a topical overview}}}\ (\bibinfo  {publisher}
{Springer Verlag},\ \bibinfo {address} {Berlin},\ \bibinfo {year}
{2004})\BibitemShut {NoStop}%
\bibitem [{\citenamefont {Marsik}\ \emph {et~al.}(2010)\citenamefont {Marsik},
	\citenamefont {Kim}, \citenamefont {Dubroka}, \citenamefont {R\"{o}ssle},
	\citenamefont {Malik}, \citenamefont {Schulz}, \citenamefont {Wang},
	\citenamefont {Niedermayer}, \citenamefont {Drew}, \citenamefont {Willis},
	\citenamefont {Wolf},\ and\ \citenamefont {Bernhard}}]{marsik2010}%
\BibitemOpen
\bibfield  {author} {\bibinfo {author} {\bibfnamefont {P.}~\bibnamefont
		{Marsik}}, \bibinfo {author} {\bibfnamefont {K.~W.}\ \bibnamefont {Kim}},
	\bibinfo {author} {\bibfnamefont {A.}~\bibnamefont {Dubroka}}, \bibinfo
	{author} {\bibfnamefont {M.}~\bibnamefont {R\"{o}ssle}}, \bibinfo {author}
	{\bibfnamefont {V.~K.}\ \bibnamefont {Malik}}, \bibinfo {author}
	{\bibfnamefont {L.}~\bibnamefont {Schulz}}, \bibinfo {author} {\bibfnamefont
		{C.~N.}\ \bibnamefont {Wang}}, \bibinfo {author} {\bibfnamefont
		{C.}~\bibnamefont {Niedermayer}}, \bibinfo {author} {\bibfnamefont {A.~J.}\
		\bibnamefont {Drew}}, \bibinfo {author} {\bibfnamefont {M.}~\bibnamefont
		{Willis}}, \bibinfo {author} {\bibfnamefont {T.}~\bibnamefont {Wolf}}, \ and\
	\bibinfo {author} {\bibfnamefont {C.}~\bibnamefont {Bernhard}},\ }\href
{\doibase 10.1103/PhysRevLett.105.057001} {\bibfield  {journal} {\bibinfo
		{journal} {Phys. Rev. Lett.}\ }\textbf {\bibinfo {volume} {105}},\ \bibinfo
	{pages} {057001} (\bibinfo {year} {2010})}\BibitemShut {NoStop}%
\bibitem [{\citenamefont {Lumsden}\ and\ \citenamefont
	{Christianson}(2010)}]{lumsden2010}%
\BibitemOpen
\bibfield  {author} {\bibinfo {author} {\bibfnamefont {M.~D.}\ \bibnamefont
		{Lumsden}}\ and\ \bibinfo {author} {\bibfnamefont {A.~D.}\ \bibnamefont
		{Christianson}},\ }\href {http://iopscience.iop.org/0953-8984/22/20/203203}
{\bibfield  {journal} {\bibinfo  {journal} {Journal of Physics: Condensed
			Matter}\ }\textbf {\bibinfo {volume} {22}},\ \bibinfo {pages} {203203}
	(\bibinfo {year} {2010})}\BibitemShut {NoStop}%
\bibitem [{\citenamefont {Wiesenmayer}\ \emph {et~al.}(2011)\citenamefont
	{Wiesenmayer}, \citenamefont {Luetkens}, \citenamefont {Pascua},
	\citenamefont {Khasanov}, \citenamefont {Amato}, \citenamefont {Potts},
	\citenamefont {Banusch}, \citenamefont {Klauss},\ and\ \citenamefont
	{Johrendt}}]{wiesenmayer2011}%
\BibitemOpen
\bibfield  {author} {\bibinfo {author} {\bibfnamefont {E.}~\bibnamefont
		{Wiesenmayer}}, \bibinfo {author} {\bibfnamefont {H.}~\bibnamefont
		{Luetkens}}, \bibinfo {author} {\bibfnamefont {G.}~\bibnamefont {Pascua}},
	\bibinfo {author} {\bibfnamefont {R.}~\bibnamefont {Khasanov}}, \bibinfo
	{author} {\bibfnamefont {A.}~\bibnamefont {Amato}}, \bibinfo {author}
	{\bibfnamefont {H.}~\bibnamefont {Potts}}, \bibinfo {author} {\bibfnamefont
		{B.}~\bibnamefont {Banusch}}, \bibinfo {author} {\bibfnamefont {H.-H.}\
		\bibnamefont {Klauss}}, \ and\ \bibinfo {author} {\bibfnamefont
		{D.}~\bibnamefont {Johrendt}},\ }\href {\doibase
	10.1103/PhysRevLett.107.237001} {\bibfield  {journal} {\bibinfo  {journal}
		{Phys. Rev. Lett.}\ }\textbf {\bibinfo {volume} {107}},\ \bibinfo {pages}
	{237001} (\bibinfo {year} {2011})}\BibitemShut {NoStop}%
\bibitem [{\citenamefont {Bernhard}\ \emph {et~al.}(2012)\citenamefont
	{Bernhard}, \citenamefont {Wang}, \citenamefont {Nuccio}, \citenamefont
	{Schulz}, \citenamefont {Zaharko}, \citenamefont {Larsen}, \citenamefont
	{Aristizabal}, \citenamefont {Willis}, \citenamefont {Drew}, \citenamefont
	{Varma}, \citenamefont {Wolf},\ and\ \citenamefont
	{Niedermayer}}]{bernhard2012}%
\BibitemOpen
\bibfield  {author} {\bibinfo {author} {\bibfnamefont {C.}~\bibnamefont
		{Bernhard}}, \bibinfo {author} {\bibfnamefont {C.~N.}\ \bibnamefont {Wang}},
	\bibinfo {author} {\bibfnamefont {L.}~\bibnamefont {Nuccio}}, \bibinfo
	{author} {\bibfnamefont {L.}~\bibnamefont {Schulz}}, \bibinfo {author}
	{\bibfnamefont {O.}~\bibnamefont {Zaharko}}, \bibinfo {author} {\bibfnamefont
		{J.}~\bibnamefont {Larsen}}, \bibinfo {author} {\bibfnamefont
		{C.}~\bibnamefont {Aristizabal}}, \bibinfo {author} {\bibfnamefont
		{M.}~\bibnamefont {Willis}}, \bibinfo {author} {\bibfnamefont {A.~J.}\
		\bibnamefont {Drew}}, \bibinfo {author} {\bibfnamefont {G.~D.}\ \bibnamefont
		{Varma}}, \bibinfo {author} {\bibfnamefont {T.}~\bibnamefont {Wolf}}, \ and\
	\bibinfo {author} {\bibfnamefont {C.}~\bibnamefont {Niedermayer}},\ }\href
{\doibase 10.1103/PhysRevB.86.184509} {\bibfield  {journal} {\bibinfo
		{journal} {Phys. Rev. B}\ }\textbf {\bibinfo {volume} {86}},\ \bibinfo
	{pages} {184509} (\bibinfo {year} {2012})}\BibitemShut {NoStop}%
\bibitem [{\citenamefont {Dai}(2015)}]{dai2015}%
\BibitemOpen
\bibfield  {author} {\bibinfo {author} {\bibfnamefont {P.}~\bibnamefont
		{Dai}},\ }\href {\doibase 10.1103/RevModPhys.87.855} {\bibfield  {journal}
	{\bibinfo  {journal} {Rev. Mod. Phys.}\ }\textbf {\bibinfo {volume} {87}},\
	\bibinfo {pages} {855} (\bibinfo {year} {2015})}\BibitemShut {NoStop}%
\bibitem [{\citenamefont {Paglione}\ and\ \citenamefont
	{Greene}(2010)}]{paglione2010}%
\BibitemOpen
\bibfield  {author} {\bibinfo {author} {\bibfnamefont {J.}~\bibnamefont
		{Paglione}}\ and\ \bibinfo {author} {\bibfnamefont {R.~L.}\ \bibnamefont
		{Greene}},\ }\href
{http://www.nature.com/nphys/journal/v6/n9/full/nphys1759.html} {\bibfield
	{journal} {\bibinfo  {journal} {Nature Physics}\ }\textbf {\bibinfo {volume}
		{6}},\ \bibinfo {pages} {645} (\bibinfo {year} {2010})}\BibitemShut {NoStop}%
\bibitem [{\citenamefont {Chubukov}(2012)}]{chubukov2012}%
\BibitemOpen
\bibfield  {author} {\bibinfo {author} {\bibfnamefont {A.}~\bibnamefont
		{Chubukov}},\ }\href {\doibase 10.1146/annurev-conmatphys-020911-125055}
{\bibfield  {journal} {\bibinfo  {journal} {Annual Review of Condensed Matter
			Physics}\ }\textbf {\bibinfo {volume} {3}},\ \bibinfo {pages} {57} (\bibinfo
	{year} {2012})}\BibitemShut {NoStop}%
\bibitem [{\citenamefont {Mazin}\ and\ \citenamefont
	{Johannes}(2008)}]{mazin2008}%
\BibitemOpen
\bibfield  {author} {\bibinfo {author} {\bibfnamefont {I.}~\bibnamefont
		{Mazin}}\ and\ \bibinfo {author} {\bibfnamefont {M.}~\bibnamefont
		{Johannes}},\ }\href
{http://www.nature.com/nphys/journal/v5/n2/full/nphys1160.html} {\bibfield
	{journal} {\bibinfo  {journal} {Nature Physics}\ }\textbf {\bibinfo {volume}
		{5}},\ \bibinfo {pages} {141} (\bibinfo {year} {2008})}\BibitemShut {NoStop}%
\bibitem [{\citenamefont {Fernandes}\ \emph {et~al.}(2014)\citenamefont
	{Fernandes}, \citenamefont {Chubukov},\ and\ \citenamefont
	{Schmalian}}]{fernandes2014}%
\BibitemOpen
\bibfield  {author} {\bibinfo {author} {\bibfnamefont {R.}~\bibnamefont
		{Fernandes}}, \bibinfo {author} {\bibfnamefont {A.}~\bibnamefont {Chubukov}},
	\ and\ \bibinfo {author} {\bibfnamefont {J.}~\bibnamefont {Schmalian}},\
}\href {http://www.nature.com/nphys/journal/v10/n2/full/nphys2877.html}
{\bibfield  {journal} {\bibinfo  {journal} {Nature Physics}\ }\textbf
	{\bibinfo {volume} {10}},\ \bibinfo {pages} {97} (\bibinfo {year}
	{2014})}\BibitemShut {NoStop}%
\bibitem [{\citenamefont {Lee}\ \emph {et~al.}(2009)\citenamefont {Lee},
	\citenamefont {Yin},\ and\ \citenamefont {Ku}}]{lee2009}%
\BibitemOpen
\bibfield  {author} {\bibinfo {author} {\bibfnamefont {C.-C.}\ \bibnamefont
		{Lee}}, \bibinfo {author} {\bibfnamefont {W.-G.}\ \bibnamefont {Yin}}, \ and\
	\bibinfo {author} {\bibfnamefont {W.}~\bibnamefont {Ku}},\ }\href {\doibase
	10.1103/PhysRevLett.103.267001} {\bibfield  {journal} {\bibinfo  {journal}
		{Phys. Rev. Lett.}\ }\textbf {\bibinfo {volume} {103}},\ \bibinfo {pages}
	{267001} (\bibinfo {year} {2009})}\BibitemShut {NoStop}%
\bibitem [{\citenamefont {Kr\"{u}ger}\ \emph {et~al.}(2009)\citenamefont
	{Kr\"{u}ger}, \citenamefont {Kumar}, \citenamefont {Zaanen},\ and\
	\citenamefont {van~den Brink}}]{kruger2009}%
\BibitemOpen
\bibfield  {author} {\bibinfo {author} {\bibfnamefont {F.}~\bibnamefont
		{Kr\"{u}ger}}, \bibinfo {author} {\bibfnamefont {S.}~\bibnamefont {Kumar}},
	\bibinfo {author} {\bibfnamefont {J.}~\bibnamefont {Zaanen}}, \ and\ \bibinfo
	{author} {\bibfnamefont {J.}~\bibnamefont {van~den Brink}},\ }\href {\doibase
	10.1103/PhysRevB.79.054504} {\bibfield  {journal} {\bibinfo  {journal} {Phys.
			Rev. B}\ }\textbf {\bibinfo {volume} {79}},\ \bibinfo {pages} {054504}
	(\bibinfo {year} {2009})}\BibitemShut {NoStop}%
\bibitem [{\citenamefont {Mallett}\ \emph
	{et~al.}(2015{\natexlab{a}})\citenamefont {Mallett}, \citenamefont {Marsik},
	\citenamefont {Yazdi-Rizi}, \citenamefont {Wolf}, \citenamefont {B\"ohmer},
	\citenamefont {Hardy}, \citenamefont {Meingast}, \citenamefont {Munzar},\
	and\ \citenamefont {Bernhard}}]{mallett2015bkfaIR}%
\BibitemOpen
\bibfield  {author} {\bibinfo {author} {\bibfnamefont {B.~P.~P.}\
		\bibnamefont {Mallett}}, \bibinfo {author} {\bibfnamefont {P.}~\bibnamefont
		{Marsik}}, \bibinfo {author} {\bibfnamefont {M.}~\bibnamefont {Yazdi-Rizi}},
	\bibinfo {author} {\bibfnamefont {T.}~\bibnamefont {Wolf}}, \bibinfo {author}
	{\bibfnamefont {A.~E.}\ \bibnamefont {B\"ohmer}}, \bibinfo {author}
	{\bibfnamefont {F.}~\bibnamefont {Hardy}}, \bibinfo {author} {\bibfnamefont
		{C.}~\bibnamefont {Meingast}}, \bibinfo {author} {\bibfnamefont
		{D.}~\bibnamefont {Munzar}}, \ and\ \bibinfo {author} {\bibfnamefont
		{C.}~\bibnamefont {Bernhard}},\ }\href {\doibase
	10.1103/PhysRevLett.115.027003} {\bibfield  {journal} {\bibinfo  {journal}
		{Phys. Rev. Lett.}\ }\textbf {\bibinfo {volume} {115}},\ \bibinfo {pages}
	{027003} (\bibinfo {year} {2015}{\natexlab{a}})}\BibitemShut {NoStop}%
\bibitem [{\citenamefont {Mallett}\ \emph
	{et~al.}(2015{\natexlab{b}})\citenamefont {Mallett}, \citenamefont
	{Pashkevich}, \citenamefont {Gusev}, \citenamefont {Wolf},\ and\
	\citenamefont {Bernhard}}]{mallett2015musr}%
\BibitemOpen
\bibfield  {author} {\bibinfo {author} {\bibfnamefont {B.~P.~P.}\
		\bibnamefont {Mallett}}, \bibinfo {author} {\bibfnamefont {Y.~G.}\
		\bibnamefont {Pashkevich}}, \bibinfo {author} {\bibfnamefont
		{A.}~\bibnamefont {Gusev}}, \bibinfo {author} {\bibfnamefont
		{T.}~\bibnamefont {Wolf}}, \ and\ \bibinfo {author} {\bibfnamefont
		{C.}~\bibnamefont {Bernhard}},\ }\href
{http://stacks.iop.org/0295-5075/111/i=5/a=57001} {\bibfield  {journal}
	{\bibinfo  {journal} {EPL (Europhysics Letters)}\ }\textbf {\bibinfo {volume}
		{111}},\ \bibinfo {pages} {57001} (\bibinfo {year}
	{2015}{\natexlab{b}})}\BibitemShut {NoStop}%
\bibitem [{\citenamefont {Pelliciari}\ \emph {et~al.}(2016)\citenamefont
	{Pelliciari}, \citenamefont {Huang}, \citenamefont {Ishii}, \citenamefont
	{Zhang}, \citenamefont {Dai}, \citenamefont {Chen}, \citenamefont {Xing},
	\citenamefont {Wang}, \citenamefont {Jin}, \citenamefont {Ding} \emph
	{et~al.}}]{pelliciari2016}%
\BibitemOpen
\bibfield  {author} {\bibinfo {author} {\bibfnamefont {J.}~\bibnamefont
		{Pelliciari}}, \bibinfo {author} {\bibfnamefont {Y.}~\bibnamefont {Huang}},
	\bibinfo {author} {\bibfnamefont {K.}~\bibnamefont {Ishii}}, \bibinfo
	{author} {\bibfnamefont {C.}~\bibnamefont {Zhang}}, \bibinfo {author}
	{\bibfnamefont {P.}~\bibnamefont {Dai}}, \bibinfo {author} {\bibfnamefont
		{G.~F.}\ \bibnamefont {Chen}}, \bibinfo {author} {\bibfnamefont
		{L.}~\bibnamefont {Xing}}, \bibinfo {author} {\bibfnamefont {X.}~\bibnamefont
		{Wang}}, \bibinfo {author} {\bibfnamefont {C.}~\bibnamefont {Jin}}, \bibinfo
	{author} {\bibfnamefont {H.}~\bibnamefont {Ding}},  \emph {et~al.},\ }\href
{http://arxiv.org/abs/1607.04038} {\bibfield  {journal} {\bibinfo  {journal}
		{arXiv preprint arXiv:1607.04038}\ } (\bibinfo {year} {2016})}\BibitemShut
{NoStop}%
\bibitem [{\citenamefont {Yildirim}(2009)}]{yildirim2009}%
\BibitemOpen
\bibfield  {author} {\bibinfo {author} {\bibfnamefont {T.}~\bibnamefont
		{Yildirim}},\ }\href
{http://www.sciencedirect.com/science/article/pii/S092145340900077X}
{\bibfield  {journal} {\bibinfo  {journal} {Physica C: Superconductivity}\
	}\textbf {\bibinfo {volume} {469}},\ \bibinfo {pages} {425} (\bibinfo {year}
	{2009})}\BibitemShut {NoStop}%
\bibitem [{\citenamefont {Chaloupka}\ and\ \citenamefont
	{Khaliullin}(2013)}]{chaloupka2013}%
\BibitemOpen
\bibfield  {author} {\bibinfo {author} {\bibfnamefont {J.~c.~v.}\
		\bibnamefont {Chaloupka}}\ and\ \bibinfo {author} {\bibfnamefont
		{G.}~\bibnamefont {Khaliullin}},\ }\href {\doibase
	10.1103/PhysRevLett.110.207205} {\bibfield  {journal} {\bibinfo  {journal}
		{Phys. Rev. Lett.}\ }\textbf {\bibinfo {volume} {110}},\ \bibinfo {pages}
	{207205} (\bibinfo {year} {2013})}\BibitemShut {NoStop}%
\bibitem [{\citenamefont {Gretarsson}\ \emph {et~al.}(2013)\citenamefont
	{Gretarsson}, \citenamefont {Saha}, \citenamefont {Drye}, \citenamefont
	{Paglione}, \citenamefont {Kim}, \citenamefont {Casa}, \citenamefont {Gog},
	\citenamefont {Wu}, \citenamefont {Julian},\ and\ \citenamefont
	{Kim}}]{gretarsson2013}%
\BibitemOpen
\bibfield  {author} {\bibinfo {author} {\bibfnamefont {H.}~\bibnamefont
		{Gretarsson}}, \bibinfo {author} {\bibfnamefont {S.~R.}\ \bibnamefont
		{Saha}}, \bibinfo {author} {\bibfnamefont {T.}~\bibnamefont {Drye}}, \bibinfo
	{author} {\bibfnamefont {J.}~\bibnamefont {Paglione}}, \bibinfo {author}
	{\bibfnamefont {J.}~\bibnamefont {Kim}}, \bibinfo {author} {\bibfnamefont
		{D.}~\bibnamefont {Casa}}, \bibinfo {author} {\bibfnamefont {T.}~\bibnamefont
		{Gog}}, \bibinfo {author} {\bibfnamefont {W.}~\bibnamefont {Wu}}, \bibinfo
	{author} {\bibfnamefont {S.~R.}\ \bibnamefont {Julian}}, \ and\ \bibinfo
	{author} {\bibfnamefont {Y.-J.}\ \bibnamefont {Kim}},\ }\href {\doibase
	10.1103/PhysRevLett.110.047003} {\bibfield  {journal} {\bibinfo  {journal}
		{Phys. Rev. Lett.}\ }\textbf {\bibinfo {volume} {110}},\ \bibinfo {pages}
	{047003} (\bibinfo {year} {2013})}\BibitemShut {NoStop}%
\bibitem [{\citenamefont {Pratt}\ \emph {et~al.}(2009)\citenamefont {Pratt},
	\citenamefont {Tian}, \citenamefont {Kreyssig}, \citenamefont {Zarestky},
	\citenamefont {Nandi}, \citenamefont {Ni}, \citenamefont {Bud'ko},
	\citenamefont {Canfield}, \citenamefont {Goldman},\ and\ \citenamefont
	{McQueeney}}]{pratt2009}%
\BibitemOpen
\bibfield  {author} {\bibinfo {author} {\bibfnamefont {D.~K.}\ \bibnamefont
		{Pratt}}, \bibinfo {author} {\bibfnamefont {W.}~\bibnamefont {Tian}},
	\bibinfo {author} {\bibfnamefont {A.}~\bibnamefont {Kreyssig}}, \bibinfo
	{author} {\bibfnamefont {J.~L.}\ \bibnamefont {Zarestky}}, \bibinfo {author}
	{\bibfnamefont {S.}~\bibnamefont {Nandi}}, \bibinfo {author} {\bibfnamefont
		{N.}~\bibnamefont {Ni}}, \bibinfo {author} {\bibfnamefont {S.~L.}\
		\bibnamefont {Bud'ko}}, \bibinfo {author} {\bibfnamefont {P.~C.}\
		\bibnamefont {Canfield}}, \bibinfo {author} {\bibfnamefont {A.~I.}\
		\bibnamefont {Goldman}}, \ and\ \bibinfo {author} {\bibfnamefont {R.~J.}\
		\bibnamefont {McQueeney}},\ }\href {\doibase 10.1103/PhysRevLett.103.087001}
{\bibfield  {journal} {\bibinfo  {journal} {Phys. Rev. Lett.}\ }\textbf
	{\bibinfo {volume} {103}},\ \bibinfo {pages} {087001} (\bibinfo {year}
	{2009})}\BibitemShut {NoStop}%
\bibitem [{\citenamefont {Christianson}\ \emph {et~al.}(2009)\citenamefont
	{Christianson}, \citenamefont {Lumsden}, \citenamefont {Nagler},
	\citenamefont {MacDougall}, \citenamefont {McGuire}, \citenamefont {Sefat},
	\citenamefont {Jin}, \citenamefont {Sales},\ and\ \citenamefont
	{Mandrus}}]{christianson2009}%
\BibitemOpen
\bibfield  {author} {\bibinfo {author} {\bibfnamefont {A.~D.}\ \bibnamefont
		{Christianson}}, \bibinfo {author} {\bibfnamefont {M.~D.}\ \bibnamefont
		{Lumsden}}, \bibinfo {author} {\bibfnamefont {S.~E.}\ \bibnamefont {Nagler}},
	\bibinfo {author} {\bibfnamefont {G.~J.}\ \bibnamefont {MacDougall}},
	\bibinfo {author} {\bibfnamefont {M.~A.}\ \bibnamefont {McGuire}}, \bibinfo
	{author} {\bibfnamefont {A.~S.}\ \bibnamefont {Sefat}}, \bibinfo {author}
	{\bibfnamefont {R.}~\bibnamefont {Jin}}, \bibinfo {author} {\bibfnamefont
		{B.~C.}\ \bibnamefont {Sales}}, \ and\ \bibinfo {author} {\bibfnamefont
		{D.}~\bibnamefont {Mandrus}},\ }\href {\doibase
	10.1103/PhysRevLett.103.087002} {\bibfield  {journal} {\bibinfo  {journal}
		{Phys. Rev. Lett.}\ }\textbf {\bibinfo {volume} {103}},\ \bibinfo {pages}
	{087002} (\bibinfo {year} {2009})}\BibitemShut {NoStop}%
\bibitem [{\citenamefont {Hardy}\ \emph {et~al.}(2016)\citenamefont {Hardy},
	\citenamefont {B{\"o}hmer}, \citenamefont {deMedici}, \citenamefont {Capone},
	\citenamefont {Giovannetti}, \citenamefont {Eder}, \citenamefont {Wang},
	\citenamefont {He}, \citenamefont {Wolf}, \citenamefont {Schweiss} \emph
	{et~al.}}]{hardy2016}%
\BibitemOpen
\bibfield  {author} {\bibinfo {author} {\bibfnamefont {F.}~\bibnamefont
		{Hardy}}, \bibinfo {author} {\bibfnamefont {A.}~\bibnamefont {B{\"o}hmer}},
	\bibinfo {author} {\bibfnamefont {L.}~\bibnamefont {deMedici}}, \bibinfo
	{author} {\bibfnamefont {M.}~\bibnamefont {Capone}}, \bibinfo {author}
	{\bibfnamefont {G.}~\bibnamefont {Giovannetti}}, \bibinfo {author}
	{\bibfnamefont {R.}~\bibnamefont {Eder}}, \bibinfo {author} {\bibfnamefont
		{L.}~\bibnamefont {Wang}}, \bibinfo {author} {\bibfnamefont {M.}~\bibnamefont
		{He}}, \bibinfo {author} {\bibfnamefont {T.}~\bibnamefont {Wolf}}, \bibinfo
	{author} {\bibfnamefont {P.}~\bibnamefont {Schweiss}},  \emph {et~al.},\
}\href {http://arxiv.org/abs/1605.05485} {\bibfield  {journal} {\bibinfo
	{journal} {arXiv:1605.05485}\ } (\bibinfo {year} {2016})}\BibitemShut
{NoStop}%
\bibitem [{\citenamefont {Aczel}\ \emph {et~al.}(2008)\citenamefont {Aczel},
	\citenamefont {Baggio-Saitovitch}, \citenamefont {Budko}, \citenamefont
	{Canfield}, \citenamefont {Carlo}, \citenamefont {Chen}, \citenamefont {Dai},
	\citenamefont {Goko}, \citenamefont {Hu}, \citenamefont {Luke}, \citenamefont
	{Luo}, \citenamefont {Ni}, \citenamefont {Sanchez-Candela}, \citenamefont
	{Tafti}, \citenamefont {Wang}, \citenamefont {Williams}, \citenamefont {Yu},\
	and\ \citenamefont {Uemura}}]{aczel2008}%
\BibitemOpen
\bibfield  {author} {\bibinfo {author} {\bibfnamefont {A.~A.}\ \bibnamefont
		{Aczel}}, \bibinfo {author} {\bibfnamefont {E.}~\bibnamefont
		{Baggio-Saitovitch}}, \bibinfo {author} {\bibfnamefont {S.~L.}\ \bibnamefont
		{Budko}}, \bibinfo {author} {\bibfnamefont {P.~C.}\ \bibnamefont {Canfield}},
	\bibinfo {author} {\bibfnamefont {J.~P.}\ \bibnamefont {Carlo}}, \bibinfo
	{author} {\bibfnamefont {G.~F.}\ \bibnamefont {Chen}}, \bibinfo {author}
	{\bibfnamefont {P.}~\bibnamefont {Dai}}, \bibinfo {author} {\bibfnamefont
		{T.}~\bibnamefont {Goko}}, \bibinfo {author} {\bibfnamefont {W.~Z.}\
		\bibnamefont {Hu}}, \bibinfo {author} {\bibfnamefont {G.~M.}\ \bibnamefont
		{Luke}}, \bibinfo {author} {\bibfnamefont {J.~L.}\ \bibnamefont {Luo}},
	\bibinfo {author} {\bibfnamefont {N.}~\bibnamefont {Ni}}, \bibinfo {author}
	{\bibfnamefont {D.~R.}\ \bibnamefont {Sanchez-Candela}}, \bibinfo {author}
	{\bibfnamefont {F.~F.}\ \bibnamefont {Tafti}}, \bibinfo {author}
	{\bibfnamefont {N.~L.}\ \bibnamefont {Wang}}, \bibinfo {author}
	{\bibfnamefont {T.~J.}\ \bibnamefont {Williams}}, \bibinfo {author}
	{\bibfnamefont {W.}~\bibnamefont {Yu}}, \ and\ \bibinfo {author}
	{\bibfnamefont {Y.~J.}\ \bibnamefont {Uemura}},\ }\href {\doibase
	10.1103/PhysRevB.78.214503} {\bibfield  {journal} {\bibinfo  {journal} {Phys.
			Rev. B}\ }\textbf {\bibinfo {volume} {78}},\ \bibinfo {pages} {214503}
	(\bibinfo {year} {2008})}\BibitemShut {NoStop}%
\bibitem [{\citenamefont {Park}\ \emph {et~al.}(2009)\citenamefont {Park},
	\citenamefont {Inosov}, \citenamefont {Niedermayer}, \citenamefont {Sun},
	\citenamefont {Haug}, \citenamefont {Christensen}, \citenamefont {Dinnebier},
	\citenamefont {Boris}, \citenamefont {Drew}, \citenamefont {Schulz},
	\citenamefont {Shapoval}, \citenamefont {Wolff}, \citenamefont {Neu},
	\citenamefont {Yang}, \citenamefont {Lin}, \citenamefont {Keimer},\ and\
	\citenamefont {Hinkov}}]{park2009}%
\BibitemOpen
\bibfield  {author} {\bibinfo {author} {\bibfnamefont {J.~T.}\ \bibnamefont
		{Park}}, \bibinfo {author} {\bibfnamefont {D.~S.}\ \bibnamefont {Inosov}},
	\bibinfo {author} {\bibfnamefont {C.}~\bibnamefont {Niedermayer}}, \bibinfo
	{author} {\bibfnamefont {G.~L.}\ \bibnamefont {Sun}}, \bibinfo {author}
	{\bibfnamefont {D.}~\bibnamefont {Haug}}, \bibinfo {author} {\bibfnamefont
		{N.~B.}\ \bibnamefont {Christensen}}, \bibinfo {author} {\bibfnamefont
		{R.}~\bibnamefont {Dinnebier}}, \bibinfo {author} {\bibfnamefont {A.~V.}\
		\bibnamefont {Boris}}, \bibinfo {author} {\bibfnamefont {A.~J.}\ \bibnamefont
		{Drew}}, \bibinfo {author} {\bibfnamefont {L.}~\bibnamefont {Schulz}},
	\bibinfo {author} {\bibfnamefont {T.}~\bibnamefont {Shapoval}}, \bibinfo
	{author} {\bibfnamefont {U.}~\bibnamefont {Wolff}}, \bibinfo {author}
	{\bibfnamefont {V.}~\bibnamefont {Neu}}, \bibinfo {author} {\bibfnamefont
		{X.}~\bibnamefont {Yang}}, \bibinfo {author} {\bibfnamefont {C.~T.}\
		\bibnamefont {Lin}}, \bibinfo {author} {\bibfnamefont {B.}~\bibnamefont
		{Keimer}}, \ and\ \bibinfo {author} {\bibfnamefont {V.}~\bibnamefont
		{Hinkov}},\ }\href {\doibase 10.1103/PhysRevLett.102.117006} {\bibfield
	{journal} {\bibinfo  {journal} {Phys. Rev. Lett.}\ }\textbf {\bibinfo
		{volume} {102}},\ \bibinfo {pages} {117006} (\bibinfo {year}
	{2009})}\BibitemShut {NoStop}%
\bibitem [{\citenamefont {B{\"o}hmer}\ \emph {et~al.}(2015)\citenamefont
	{B{\"o}hmer}, \citenamefont {Hardy}, \citenamefont {Wang}, \citenamefont
	{Wolf}, \citenamefont {Schweiss},\ and\ \citenamefont
	{Meingast}}]{bohmer2015}%
\BibitemOpen
\bibfield  {author} {\bibinfo {author} {\bibfnamefont {A.}~\bibnamefont
		{B{\"o}hmer}}, \bibinfo {author} {\bibfnamefont {F.}~\bibnamefont {Hardy}},
	\bibinfo {author} {\bibfnamefont {L.}~\bibnamefont {Wang}}, \bibinfo {author}
	{\bibfnamefont {T.}~\bibnamefont {Wolf}}, \bibinfo {author} {\bibfnamefont
		{P.}~\bibnamefont {Schweiss}}, \ and\ \bibinfo {author} {\bibfnamefont
		{C.}~\bibnamefont {Meingast}},\ }\href {\doibase 10.1038/ncomms8911}
{\bibfield  {journal} {\bibinfo  {journal} {Nature communications}\ }\textbf
	{\bibinfo {volume} {6}} (\bibinfo {year} {2015}),\
	10.1038/ncomms8911}\BibitemShut {NoStop}%
\bibitem [{\citenamefont {Wang}\ \emph {et~al.}(2016)\citenamefont {Wang},
	\citenamefont {Hardy}, \citenamefont {B\"ohmer}, \citenamefont {Wolf},
	\citenamefont {Schweiss},\ and\ \citenamefont {Meingast}}]{wang2016}%
\BibitemOpen
\bibfield  {author} {\bibinfo {author} {\bibfnamefont {L.}~\bibnamefont
		{Wang}}, \bibinfo {author} {\bibfnamefont {F.}~\bibnamefont {Hardy}},
	\bibinfo {author} {\bibfnamefont {A.~E.}\ \bibnamefont {B\"ohmer}}, \bibinfo
	{author} {\bibfnamefont {T.}~\bibnamefont {Wolf}}, \bibinfo {author}
	{\bibfnamefont {P.}~\bibnamefont {Schweiss}}, \ and\ \bibinfo {author}
	{\bibfnamefont {C.}~\bibnamefont {Meingast}},\ }\href {\doibase
	10.1103/PhysRevB.93.014514} {\bibfield  {journal} {\bibinfo  {journal} {Phys.
			Rev. B}\ }\textbf {\bibinfo {volume} {93}},\ \bibinfo {pages} {014514}
	(\bibinfo {year} {2016})}\BibitemShut {NoStop}%
\bibitem [{\citenamefont {Hassinger}\ \emph {et~al.}(2012)\citenamefont
	{Hassinger}, \citenamefont {Gredat}, \citenamefont {Valade}, \citenamefont
	{de~Cotret}, \citenamefont {Juneau-Fecteau}, \citenamefont {Reid},
	\citenamefont {Kim}, \citenamefont {Tanatar}, \citenamefont {Prozorov},
	\citenamefont {Shen}, \citenamefont {Wen}, \citenamefont {Doiron-Leyraud},\
	and\ \citenamefont {Taillefer}}]{hassinger2012}%
\BibitemOpen
\bibfield  {author} {\bibinfo {author} {\bibfnamefont {E.}~\bibnamefont
		{Hassinger}}, \bibinfo {author} {\bibfnamefont {G.}~\bibnamefont {Gredat}},
	\bibinfo {author} {\bibfnamefont {F.}~\bibnamefont {Valade}}, \bibinfo
	{author} {\bibfnamefont {S.~R.}\ \bibnamefont {de~Cotret}}, \bibinfo {author}
	{\bibfnamefont {A.}~\bibnamefont {Juneau-Fecteau}}, \bibinfo {author}
	{\bibfnamefont {J.-P.}\ \bibnamefont {Reid}}, \bibinfo {author}
	{\bibfnamefont {H.}~\bibnamefont {Kim}}, \bibinfo {author} {\bibfnamefont
		{M.~A.}\ \bibnamefont {Tanatar}}, \bibinfo {author} {\bibfnamefont
		{R.}~\bibnamefont {Prozorov}}, \bibinfo {author} {\bibfnamefont
		{B.}~\bibnamefont {Shen}}, \bibinfo {author} {\bibfnamefont {H.-H.}\
		\bibnamefont {Wen}}, \bibinfo {author} {\bibfnamefont {N.}~\bibnamefont
		{Doiron-Leyraud}}, \ and\ \bibinfo {author} {\bibfnamefont {L.}~\bibnamefont
		{Taillefer}},\ }\href {\doibase 10.1103/PhysRevB.86.140502} {\bibfield
	{journal} {\bibinfo  {journal} {Phys. Rev. B}\ }\textbf {\bibinfo {volume}
		{86}},\ \bibinfo {pages} {140502} (\bibinfo {year} {2012})}\BibitemShut
{NoStop}%
\bibitem [{\citenamefont {Avci}\ \emph {et~al.}(2014)\citenamefont {Avci},
	\citenamefont {Chmaissem}, \citenamefont {Allred}, \citenamefont
	{Rosenkranz}, \citenamefont {Eremin}, \citenamefont {Chubukov}, \citenamefont
	{Bugaris}, \citenamefont {Chung}, \citenamefont {Kanatzidis}, \citenamefont
	{Castellan} \emph {et~al.}}]{avci2014}%
\BibitemOpen
\bibfield  {author} {\bibinfo {author} {\bibfnamefont {S.}~\bibnamefont
		{Avci}}, \bibinfo {author} {\bibfnamefont {O.}~\bibnamefont {Chmaissem}},
	\bibinfo {author} {\bibfnamefont {J.}~\bibnamefont {Allred}}, \bibinfo
	{author} {\bibfnamefont {S.}~\bibnamefont {Rosenkranz}}, \bibinfo {author}
	{\bibfnamefont {I.}~\bibnamefont {Eremin}}, \bibinfo {author} {\bibfnamefont
		{A.}~\bibnamefont {Chubukov}}, \bibinfo {author} {\bibfnamefont
		{D.}~\bibnamefont {Bugaris}}, \bibinfo {author} {\bibfnamefont
		{D.}~\bibnamefont {Chung}}, \bibinfo {author} {\bibfnamefont
		{M.}~\bibnamefont {Kanatzidis}}, \bibinfo {author} {\bibfnamefont {J.-P.}\
		\bibnamefont {Castellan}},  \emph {et~al.},\ }\href {\doibase
	10.1038/ncomms4845} {\bibfield  {journal} {\bibinfo  {journal} {Nature
			Communications}\ }\textbf {\bibinfo {volume} {5}},\ \bibinfo {pages} {4845}
	(\bibinfo {year} {2014})}\BibitemShut {NoStop}%
\bibitem [{\citenamefont {Wa\ss{}er}\ \emph {et~al.}(2015)\citenamefont
	{Wa\ss{}er}, \citenamefont {Schneidewind}, \citenamefont {Sidis},
	\citenamefont {Wurmehl}, \citenamefont {Aswartham}, \citenamefont
	{B\"uchner},\ and\ \citenamefont {Braden}}]{wasser2015}%
\BibitemOpen
\bibfield  {author} {\bibinfo {author} {\bibfnamefont {F.}~\bibnamefont
		{Wa\ss{}er}}, \bibinfo {author} {\bibfnamefont {A.}~\bibnamefont
		{Schneidewind}}, \bibinfo {author} {\bibfnamefont {Y.}~\bibnamefont {Sidis}},
	\bibinfo {author} {\bibfnamefont {S.}~\bibnamefont {Wurmehl}}, \bibinfo
	{author} {\bibfnamefont {S.}~\bibnamefont {Aswartham}}, \bibinfo {author}
	{\bibfnamefont {B.}~\bibnamefont {B\"uchner}}, \ and\ \bibinfo {author}
	{\bibfnamefont {M.}~\bibnamefont {Braden}},\ }\href {\doibase
	10.1103/PhysRevB.91.060505} {\bibfield  {journal} {\bibinfo  {journal} {Phys.
			Rev. B}\ }\textbf {\bibinfo {volume} {91}},\ \bibinfo {pages} {060505}
	(\bibinfo {year} {2015})}\BibitemShut {NoStop}%
\bibitem [{\citenamefont {Allred}\ \emph {et~al.}(2015)\citenamefont {Allred},
	\citenamefont {Avci}, \citenamefont {Chung}, \citenamefont {Claus},
	\citenamefont {Khalyavin}, \citenamefont {Manuel}, \citenamefont {Taddei},
	\citenamefont {Kanatzidis}, \citenamefont {Rosenkranz}, \citenamefont
	{Osborn} \emph {et~al.}}]{allred2015xray}%
\BibitemOpen
\bibfield  {author} {\bibinfo {author} {\bibfnamefont {J.~M.}\ \bibnamefont
		{Allred}}, \bibinfo {author} {\bibfnamefont {S.}~\bibnamefont {Avci}},
	\bibinfo {author} {\bibfnamefont {D.~Y.}\ \bibnamefont {Chung}}, \bibinfo
	{author} {\bibfnamefont {H.}~\bibnamefont {Claus}}, \bibinfo {author}
	{\bibfnamefont {D.~D.}\ \bibnamefont {Khalyavin}}, \bibinfo {author}
	{\bibfnamefont {P.}~\bibnamefont {Manuel}}, \bibinfo {author} {\bibfnamefont
		{K.~M.}\ \bibnamefont {Taddei}}, \bibinfo {author} {\bibfnamefont {M.~G.}\
		\bibnamefont {Kanatzidis}}, \bibinfo {author} {\bibfnamefont
		{S.}~\bibnamefont {Rosenkranz}}, \bibinfo {author} {\bibfnamefont
		{R.}~\bibnamefont {Osborn}},  \emph {et~al.},\ }\href
{http://arxiv.org/abs/1505.01433} {\bibfield  {journal} {\bibinfo  {journal}
		{arXiv:1505.01433}\ } (\bibinfo {year} {2015})}\BibitemShut {NoStop}%
\bibitem [{\citenamefont {Hu}\ \emph {et~al.}(2008)\citenamefont {Hu},
	\citenamefont {Dong}, \citenamefont {Li}, \citenamefont {Li}, \citenamefont
	{Zheng}, \citenamefont {Chen}, \citenamefont {Luo},\ and\ \citenamefont
	{Wang}}]{hu2008}%
\BibitemOpen
\bibfield  {author} {\bibinfo {author} {\bibfnamefont {W.~Z.}\ \bibnamefont
		{Hu}}, \bibinfo {author} {\bibfnamefont {J.}~\bibnamefont {Dong}}, \bibinfo
	{author} {\bibfnamefont {G.}~\bibnamefont {Li}}, \bibinfo {author}
	{\bibfnamefont {Z.}~\bibnamefont {Li}}, \bibinfo {author} {\bibfnamefont
		{P.}~\bibnamefont {Zheng}}, \bibinfo {author} {\bibfnamefont {G.~F.}\
		\bibnamefont {Chen}}, \bibinfo {author} {\bibfnamefont {J.~L.}\ \bibnamefont
		{Luo}}, \ and\ \bibinfo {author} {\bibfnamefont {N.~L.}\ \bibnamefont
		{Wang}},\ }\href {\doibase 10.1103/PhysRevLett.101.257005} {\bibfield
	{journal} {\bibinfo  {journal} {Phys. Rev. Lett.}\ }\textbf {\bibinfo
		{volume} {101}},\ \bibinfo {pages} {257005} (\bibinfo {year}
	{2008})}\BibitemShut {NoStop}%
\bibitem [{\citenamefont {Wang}\ \emph {et~al.}(2012)\citenamefont {Wang},
	\citenamefont {Hu}, \citenamefont {Chen}, \citenamefont {Yuan}, \citenamefont
	{Li}, \citenamefont {Chen},\ and\ \citenamefont {Xiang}}]{wang2012}%
\BibitemOpen
\bibfield  {author} {\bibinfo {author} {\bibfnamefont {N.}~\bibnamefont
		{Wang}}, \bibinfo {author} {\bibfnamefont {W.}~\bibnamefont {Hu}}, \bibinfo
	{author} {\bibfnamefont {Z.}~\bibnamefont {Chen}}, \bibinfo {author}
	{\bibfnamefont {R.}~\bibnamefont {Yuan}}, \bibinfo {author} {\bibfnamefont
		{G.}~\bibnamefont {Li}}, \bibinfo {author} {\bibfnamefont {G.}~\bibnamefont
		{Chen}}, \ and\ \bibinfo {author} {\bibfnamefont {T.}~\bibnamefont {Xiang}},\
}\href {http://stacks.iop.org/0953-8984/24/i=29/a=294202} {\bibfield
{journal} {\bibinfo  {journal} {Journal of Physics: Condensed Matter}\
}\textbf {\bibinfo {volume} {24}},\ \bibinfo {pages} {294202} (\bibinfo
{year} {2012})}\BibitemShut {NoStop}%
\bibitem [{\citenamefont {Li}\ \emph {et~al.}(2008)\citenamefont {Li},
	\citenamefont {Hu}, \citenamefont {Dong}, \citenamefont {Li}, \citenamefont
	{Zheng}, \citenamefont {Chen}, \citenamefont {Luo},\ and\ \citenamefont
	{Wang}}]{li2008}%
\BibitemOpen
\bibfield  {author} {\bibinfo {author} {\bibfnamefont {G.}~\bibnamefont
		{Li}}, \bibinfo {author} {\bibfnamefont {W.~Z.}\ \bibnamefont {Hu}}, \bibinfo
	{author} {\bibfnamefont {J.}~\bibnamefont {Dong}}, \bibinfo {author}
	{\bibfnamefont {Z.}~\bibnamefont {Li}}, \bibinfo {author} {\bibfnamefont
		{P.}~\bibnamefont {Zheng}}, \bibinfo {author} {\bibfnamefont {G.~F.}\
		\bibnamefont {Chen}}, \bibinfo {author} {\bibfnamefont {J.~L.}\ \bibnamefont
		{Luo}}, \ and\ \bibinfo {author} {\bibfnamefont {N.~L.}\ \bibnamefont
		{Wang}},\ }\href {\doibase 10.1103/PhysRevLett.101.107004} {\bibfield
	{journal} {\bibinfo  {journal} {Phys. Rev. Lett.}\ }\textbf {\bibinfo
		{volume} {101}},\ \bibinfo {pages} {107004} (\bibinfo {year}
	{2008})}\BibitemShut {NoStop}%
\bibitem [{\citenamefont {Yang}\ \emph {et~al.}(2009)\citenamefont {Yang},
	\citenamefont {H\"uvonen}, \citenamefont {Nagel}, \citenamefont {R\~o\ om},
	\citenamefont {Ni}, \citenamefont {Canfield}, \citenamefont {Bud'ko},
	\citenamefont {Carbotte},\ and\ \citenamefont {Timusk}}]{yang2009}%
\BibitemOpen
\bibfield  {author} {\bibinfo {author} {\bibfnamefont {J.}~\bibnamefont
		{Yang}}, \bibinfo {author} {\bibfnamefont {D.}~\bibnamefont {H\"uvonen}},
	\bibinfo {author} {\bibfnamefont {U.}~\bibnamefont {Nagel}}, \bibinfo
	{author} {\bibfnamefont {T.}~\bibnamefont {R\~o\ om}}, \bibinfo {author}
	{\bibfnamefont {N.}~\bibnamefont {Ni}}, \bibinfo {author} {\bibfnamefont
		{P.~C.}\ \bibnamefont {Canfield}}, \bibinfo {author} {\bibfnamefont {S.~L.}\
		\bibnamefont {Bud'ko}}, \bibinfo {author} {\bibfnamefont {J.~P.}\
		\bibnamefont {Carbotte}}, \ and\ \bibinfo {author} {\bibfnamefont
		{T.}~\bibnamefont {Timusk}},\ }\href {\doibase
	10.1103/PhysRevLett.102.187003} {\bibfield  {journal} {\bibinfo  {journal}
		{Phys. Rev. Lett.}\ }\textbf {\bibinfo {volume} {102}},\ \bibinfo {pages}
	{187003} (\bibinfo {year} {2009})}\BibitemShut {NoStop}%
\bibitem [{\citenamefont {Charnukha}\ \emph
	{et~al.}(2011{\natexlab{a}})\citenamefont {Charnukha}, \citenamefont
	{Popovich}, \citenamefont {Matiks}, \citenamefont {Sun}, \citenamefont {Lin},
	\citenamefont {Yaresko}, \citenamefont {Keimer},\ and\ \citenamefont
	{Boris}}]{charnukha2011}%
\BibitemOpen
\bibfield  {author} {\bibinfo {author} {\bibfnamefont {A.}~\bibnamefont
		{Charnukha}}, \bibinfo {author} {\bibfnamefont {P.}~\bibnamefont {Popovich}},
	\bibinfo {author} {\bibfnamefont {Y.}~\bibnamefont {Matiks}}, \bibinfo
	{author} {\bibfnamefont {D.}~\bibnamefont {Sun}}, \bibinfo {author}
	{\bibfnamefont {C.}~\bibnamefont {Lin}}, \bibinfo {author} {\bibfnamefont
		{A.}~\bibnamefont {Yaresko}}, \bibinfo {author} {\bibfnamefont
		{B.}~\bibnamefont {Keimer}}, \ and\ \bibinfo {author} {\bibfnamefont
		{A.}~\bibnamefont {Boris}},\ }\href {\doibase 10.1038/ncomms1223} {\bibfield
	{journal} {\bibinfo  {journal} {Nature Commun.}\ }\textbf {\bibinfo {volume}
		{2}},\ \bibinfo {pages} {219} (\bibinfo {year}
	{2011}{\natexlab{a}})}\BibitemShut {NoStop}%
\bibitem [{\citenamefont {Charnukha}\ \emph
	{et~al.}(2011{\natexlab{b}})\citenamefont {Charnukha}, \citenamefont
	{Dolgov}, \citenamefont {Golubov}, \citenamefont {Matiks}, \citenamefont
	{Sun}, \citenamefont {Lin}, \citenamefont {Keimer},\ and\ \citenamefont
	{Boris}}]{charnukha2011prb}%
\BibitemOpen
\bibfield  {author} {\bibinfo {author} {\bibfnamefont {A.}~\bibnamefont
		{Charnukha}}, \bibinfo {author} {\bibfnamefont {O.~V.}\ \bibnamefont
		{Dolgov}}, \bibinfo {author} {\bibfnamefont {A.~A.}\ \bibnamefont {Golubov}},
	\bibinfo {author} {\bibfnamefont {Y.}~\bibnamefont {Matiks}}, \bibinfo
	{author} {\bibfnamefont {D.~L.}\ \bibnamefont {Sun}}, \bibinfo {author}
	{\bibfnamefont {C.~T.}\ \bibnamefont {Lin}}, \bibinfo {author} {\bibfnamefont
		{B.}~\bibnamefont {Keimer}}, \ and\ \bibinfo {author} {\bibfnamefont {A.~V.}\
		\bibnamefont {Boris}},\ }\href {\doibase 10.1103/PhysRevB.84.174511}
{\bibfield  {journal} {\bibinfo  {journal} {Phys. Rev. B}\ }\textbf {\bibinfo
		{volume} {84}},\ \bibinfo {pages} {174511} (\bibinfo {year}
	{2011}{\natexlab{b}})}\BibitemShut {NoStop}%
\bibitem [{\citenamefont {Dai}\ \emph {et~al.}(2013)\citenamefont {Dai},
	\citenamefont {Xu}, \citenamefont {Shen}, \citenamefont {Wen}, \citenamefont
	{Qiu},\ and\ \citenamefont {Lobo}}]{dai2013}%
\BibitemOpen
\bibfield  {author} {\bibinfo {author} {\bibfnamefont {Y.~M.}\ \bibnamefont
		{Dai}}, \bibinfo {author} {\bibfnamefont {B.}~\bibnamefont {Xu}}, \bibinfo
	{author} {\bibfnamefont {B.}~\bibnamefont {Shen}}, \bibinfo {author}
	{\bibfnamefont {H.~H.}\ \bibnamefont {Wen}}, \bibinfo {author} {\bibfnamefont
		{X.~G.}\ \bibnamefont {Qiu}}, \ and\ \bibinfo {author} {\bibfnamefont {R.~P.
			S.~M.}\ \bibnamefont {Lobo}},\ }\href
{http://stacks.iop.org/0295-5075/104/i=4/a=47006} {\bibfield  {journal}
	{\bibinfo  {journal} {EPL (Europhysics Letters)}\ }\textbf {\bibinfo {volume}
		{104}},\ \bibinfo {pages} {47006} (\bibinfo {year} {2013})}\BibitemShut
{NoStop}%
\bibitem [{\citenamefont {Dai}\ \emph {et~al.}(2012)\citenamefont {Dai},
	\citenamefont {Xu}, \citenamefont {Shen}, \citenamefont {Wen}, \citenamefont
	{Hu}, \citenamefont {Qiu},\ and\ \citenamefont {Lobo}}]{dai2012}%
\BibitemOpen
\bibfield  {author} {\bibinfo {author} {\bibfnamefont {Y.~M.}\ \bibnamefont
		{Dai}}, \bibinfo {author} {\bibfnamefont {B.}~\bibnamefont {Xu}}, \bibinfo
	{author} {\bibfnamefont {B.}~\bibnamefont {Shen}}, \bibinfo {author}
	{\bibfnamefont {H.~H.}\ \bibnamefont {Wen}}, \bibinfo {author} {\bibfnamefont
		{J.~P.}\ \bibnamefont {Hu}}, \bibinfo {author} {\bibfnamefont {X.~G.}\
		\bibnamefont {Qiu}}, \ and\ \bibinfo {author} {\bibfnamefont {R.~P. S.~M.}\
		\bibnamefont {Lobo}},\ }\href {\doibase 10.1103/PhysRevB.86.100501}
{\bibfield  {journal} {\bibinfo  {journal} {Phys. Rev. B}\ }\textbf {\bibinfo
		{volume} {86}},\ \bibinfo {pages} {100501} (\bibinfo {year}
	{2012})}\BibitemShut {NoStop}%
\bibitem [{\citenamefont {Goko}\ \emph {et~al.}(2009)\citenamefont {Goko},
	\citenamefont {Aczel}, \citenamefont {Baggio-Saitovitch}, \citenamefont
	{Bud'ko}, \citenamefont {Canfield}, \citenamefont {Carlo}, \citenamefont
	{Chen}, \citenamefont {Dai}, \citenamefont {Hamann}, \citenamefont {Hu},
	\citenamefont {Kageyama}, \citenamefont {Luke}, \citenamefont {Luo},
	\citenamefont {Nachumi}, \citenamefont {Ni}, \citenamefont {Reznik},
	\citenamefont {Sanchez-Candela}, \citenamefont {Savici}, \citenamefont
	{Sikes}, \citenamefont {Wang}, \citenamefont {Wiebe}, \citenamefont
	{Williams}, \citenamefont {Yamamoto}, \citenamefont {Yu},\ and\ \citenamefont
	{Uemura}}]{goko2009}%
\BibitemOpen
\bibfield  {author} {\bibinfo {author} {\bibfnamefont {T.}~\bibnamefont
		{Goko}}, \bibinfo {author} {\bibfnamefont {A.~A.}\ \bibnamefont {Aczel}},
	\bibinfo {author} {\bibfnamefont {E.}~\bibnamefont {Baggio-Saitovitch}},
	\bibinfo {author} {\bibfnamefont {S.~L.}\ \bibnamefont {Bud'ko}}, \bibinfo
	{author} {\bibfnamefont {P.~C.}\ \bibnamefont {Canfield}}, \bibinfo {author}
	{\bibfnamefont {J.~P.}\ \bibnamefont {Carlo}}, \bibinfo {author}
	{\bibfnamefont {G.~F.}\ \bibnamefont {Chen}}, \bibinfo {author}
	{\bibfnamefont {P.}~\bibnamefont {Dai}}, \bibinfo {author} {\bibfnamefont
		{A.~C.}\ \bibnamefont {Hamann}}, \bibinfo {author} {\bibfnamefont {W.~Z.}\
		\bibnamefont {Hu}}, \bibinfo {author} {\bibfnamefont {H.}~\bibnamefont
		{Kageyama}}, \bibinfo {author} {\bibfnamefont {G.~M.}\ \bibnamefont {Luke}},
	\bibinfo {author} {\bibfnamefont {J.~L.}\ \bibnamefont {Luo}}, \bibinfo
	{author} {\bibfnamefont {B.}~\bibnamefont {Nachumi}}, \bibinfo {author}
	{\bibfnamefont {N.}~\bibnamefont {Ni}}, \bibinfo {author} {\bibfnamefont
		{D.}~\bibnamefont {Reznik}}, \bibinfo {author} {\bibfnamefont {D.~R.}\
		\bibnamefont {Sanchez-Candela}}, \bibinfo {author} {\bibfnamefont {A.~T.}\
		\bibnamefont {Savici}}, \bibinfo {author} {\bibfnamefont {K.~J.}\
		\bibnamefont {Sikes}}, \bibinfo {author} {\bibfnamefont {N.~L.}\ \bibnamefont
		{Wang}}, \bibinfo {author} {\bibfnamefont {C.~R.}\ \bibnamefont {Wiebe}},
	\bibinfo {author} {\bibfnamefont {T.~J.}\ \bibnamefont {Williams}}, \bibinfo
	{author} {\bibfnamefont {T.}~\bibnamefont {Yamamoto}}, \bibinfo {author}
	{\bibfnamefont {W.}~\bibnamefont {Yu}}, \ and\ \bibinfo {author}
	{\bibfnamefont {Y.~J.}\ \bibnamefont {Uemura}},\ }\href {\doibase
	10.1103/PhysRevB.80.024508} {\bibfield  {journal} {\bibinfo  {journal} {Phys.
			Rev. B}\ }\textbf {\bibinfo {volume} {80}},\ \bibinfo {pages} {024508}
	(\bibinfo {year} {2009})}\BibitemShut {NoStop}%
\bibitem [{\citenamefont {Khasanov}\ \emph
	{et~al.}(2009{\natexlab{a}})\citenamefont {Khasanov}, \citenamefont
	{Maisuradze}, \citenamefont {Maeter}, \citenamefont {Kwadrin}, \citenamefont
	{Luetkens}, \citenamefont {Amato}, \citenamefont {Schnelle}, \citenamefont
	{Rosner}, \citenamefont {Leithe-Jasper},\ and\ \citenamefont
	{Klauss}}]{khasanov2009}%
\BibitemOpen
\bibfield  {author} {\bibinfo {author} {\bibfnamefont {R.}~\bibnamefont
		{Khasanov}}, \bibinfo {author} {\bibfnamefont {A.}~\bibnamefont
		{Maisuradze}}, \bibinfo {author} {\bibfnamefont {H.}~\bibnamefont {Maeter}},
	\bibinfo {author} {\bibfnamefont {A.}~\bibnamefont {Kwadrin}}, \bibinfo
	{author} {\bibfnamefont {H.}~\bibnamefont {Luetkens}}, \bibinfo {author}
	{\bibfnamefont {A.}~\bibnamefont {Amato}}, \bibinfo {author} {\bibfnamefont
		{W.}~\bibnamefont {Schnelle}}, \bibinfo {author} {\bibfnamefont
		{H.}~\bibnamefont {Rosner}}, \bibinfo {author} {\bibfnamefont
		{A.}~\bibnamefont {Leithe-Jasper}}, \ and\ \bibinfo {author} {\bibfnamefont
		{H.-H.}\ \bibnamefont {Klauss}},\ }\href {\doibase
	10.1103/PhysRevLett.103.067010} {\bibfield  {journal} {\bibinfo  {journal}
		{Phys. Rev. Lett.}\ }\textbf {\bibinfo {volume} {103}},\ \bibinfo {pages}
	{067010} (\bibinfo {year} {2009}{\natexlab{a}})}\BibitemShut {NoStop}%
\bibitem [{\citenamefont {Hiraishi}\ \emph {et~al.}(2009)\citenamefont
	{Hiraishi}, \citenamefont {Kadono}, \citenamefont {Takeshita}, \citenamefont
	{Miyazaki}, \citenamefont {Koda}, \citenamefont {Okabe},\ and\ \citenamefont
	{Akimitsu}}]{hiraishi2009}%
\BibitemOpen
\bibfield  {author} {\bibinfo {author} {\bibfnamefont {M.}~\bibnamefont
		{Hiraishi}}, \bibinfo {author} {\bibfnamefont {R.}~\bibnamefont {Kadono}},
	\bibinfo {author} {\bibfnamefont {S.}~\bibnamefont {Takeshita}}, \bibinfo
	{author} {\bibfnamefont {M.}~\bibnamefont {Miyazaki}}, \bibinfo {author}
	{\bibfnamefont {A.}~\bibnamefont {Koda}}, \bibinfo {author} {\bibfnamefont
		{H.}~\bibnamefont {Okabe}}, \ and\ \bibinfo {author} {\bibfnamefont
		{J.}~\bibnamefont {Akimitsu}},\ }\href {\doibase 10.1143/JPSJ.78.023710}
{\bibfield  {journal} {\bibinfo  {journal} {Journal of the Physical Society
			of Japan}\ }\textbf {\bibinfo {volume} {78}},\ \bibinfo {pages} {023710}
	(\bibinfo {year} {2009})}\BibitemShut {NoStop}%
\bibitem [{\citenamefont {Williams}\ \emph {et~al.}(2010)\citenamefont
	{Williams}, \citenamefont {Aczel}, \citenamefont {Baggio-Saitovitch},
	\citenamefont {Bud'ko}, \citenamefont {Canfield}, \citenamefont {Carlo},
	\citenamefont {Goko}, \citenamefont {Kageyama}, \citenamefont {Kitada},
	\citenamefont {Munevar}, \citenamefont {Ni}, \citenamefont {Saha},
	\citenamefont {Kirschenbaum}, \citenamefont {Paglione}, \citenamefont
	{Sanchez-Candela}, \citenamefont {Uemura},\ and\ \citenamefont
	{Luke}}]{williams2010}%
\BibitemOpen
\bibfield  {author} {\bibinfo {author} {\bibfnamefont {T.~J.}\ \bibnamefont
		{Williams}}, \bibinfo {author} {\bibfnamefont {A.~A.}\ \bibnamefont {Aczel}},
	\bibinfo {author} {\bibfnamefont {E.}~\bibnamefont {Baggio-Saitovitch}},
	\bibinfo {author} {\bibfnamefont {S.~L.}\ \bibnamefont {Bud'ko}}, \bibinfo
	{author} {\bibfnamefont {P.~C.}\ \bibnamefont {Canfield}}, \bibinfo {author}
	{\bibfnamefont {J.~P.}\ \bibnamefont {Carlo}}, \bibinfo {author}
	{\bibfnamefont {T.}~\bibnamefont {Goko}}, \bibinfo {author} {\bibfnamefont
		{H.}~\bibnamefont {Kageyama}}, \bibinfo {author} {\bibfnamefont
		{A.}~\bibnamefont {Kitada}}, \bibinfo {author} {\bibfnamefont
		{J.}~\bibnamefont {Munevar}}, \bibinfo {author} {\bibfnamefont
		{N.}~\bibnamefont {Ni}}, \bibinfo {author} {\bibfnamefont {S.~R.}\
		\bibnamefont {Saha}}, \bibinfo {author} {\bibfnamefont {K.}~\bibnamefont
		{Kirschenbaum}}, \bibinfo {author} {\bibfnamefont {J.}~\bibnamefont
		{Paglione}}, \bibinfo {author} {\bibfnamefont {D.~R.}\ \bibnamefont
		{Sanchez-Candela}}, \bibinfo {author} {\bibfnamefont {Y.~J.}\ \bibnamefont
		{Uemura}}, \ and\ \bibinfo {author} {\bibfnamefont {G.~M.}\ \bibnamefont
		{Luke}},\ }\href {\doibase 10.1103/PhysRevB.82.094512} {\bibfield  {journal}
	{\bibinfo  {journal} {Phys. Rev. B}\ }\textbf {\bibinfo {volume} {82}},\
	\bibinfo {pages} {094512} (\bibinfo {year} {2010})}\BibitemShut {NoStop}%
\bibitem [{\citenamefont {Marsik}\ \emph {et~al.}(2013)\citenamefont {Marsik},
	\citenamefont {Wang}, \citenamefont {R\"ossle}, \citenamefont {Yazdi-Rizi},
	\citenamefont {Schuster}, \citenamefont {Kim}, \citenamefont {Dubroka},
	\citenamefont {Munzar}, \citenamefont {Wolf}, \citenamefont {Chen},\ and\
	\citenamefont {Bernhard}}]{marsik2013}%
\BibitemOpen
\bibfield  {author} {\bibinfo {author} {\bibfnamefont {P.}~\bibnamefont
		{Marsik}}, \bibinfo {author} {\bibfnamefont {C.~N.}\ \bibnamefont {Wang}},
	\bibinfo {author} {\bibfnamefont {M.}~\bibnamefont {R\"ossle}}, \bibinfo
	{author} {\bibfnamefont {M.}~\bibnamefont {Yazdi-Rizi}}, \bibinfo {author}
	{\bibfnamefont {R.}~\bibnamefont {Schuster}}, \bibinfo {author}
	{\bibfnamefont {K.~W.}\ \bibnamefont {Kim}}, \bibinfo {author} {\bibfnamefont
		{A.}~\bibnamefont {Dubroka}}, \bibinfo {author} {\bibfnamefont
		{D.}~\bibnamefont {Munzar}}, \bibinfo {author} {\bibfnamefont
		{T.}~\bibnamefont {Wolf}}, \bibinfo {author} {\bibfnamefont {X.~H.}\
		\bibnamefont {Chen}}, \ and\ \bibinfo {author} {\bibfnamefont
		{C.}~\bibnamefont {Bernhard}},\ }\href {\doibase 10.1103/PhysRevB.88.180508}
{\bibfield  {journal} {\bibinfo  {journal} {Phys. Rev. B}\ }\textbf {\bibinfo
		{volume} {88}},\ \bibinfo {pages} {180508 (R)} (\bibinfo {year}
	{2013})}\BibitemShut {NoStop}%
\bibitem [{\citenamefont {Hardy}\ \emph {et~al.}(2010)\citenamefont {Hardy},
	\citenamefont {Burger}, \citenamefont {Wolf}, \citenamefont {Fisher},
	\citenamefont {Schweiss}, \citenamefont {Adelmann}, \citenamefont {Heid},
	\citenamefont {Fromknecht}, \citenamefont {Eder}, \citenamefont {Ernst},
	\citenamefont {v.~L\"{o}hneysen},\ and\ \citenamefont
	{Meingast}}]{hardy2010}%
\BibitemOpen
\bibfield  {author} {\bibinfo {author} {\bibfnamefont {F.}~\bibnamefont
		{Hardy}}, \bibinfo {author} {\bibfnamefont {P.}~\bibnamefont {Burger}},
	\bibinfo {author} {\bibfnamefont {T.}~\bibnamefont {Wolf}}, \bibinfo {author}
	{\bibfnamefont {R.~A.}\ \bibnamefont {Fisher}}, \bibinfo {author}
	{\bibfnamefont {P.}~\bibnamefont {Schweiss}}, \bibinfo {author}
	{\bibfnamefont {P.}~\bibnamefont {Adelmann}}, \bibinfo {author}
	{\bibfnamefont {R.}~\bibnamefont {Heid}}, \bibinfo {author} {\bibfnamefont
		{R.}~\bibnamefont {Fromknecht}}, \bibinfo {author} {\bibfnamefont
		{R.}~\bibnamefont {Eder}}, \bibinfo {author} {\bibfnamefont {D.}~\bibnamefont
		{Ernst}}, \bibinfo {author} {\bibfnamefont {H.}~\bibnamefont
		{v.~L\"{o}hneysen}}, \ and\ \bibinfo {author} {\bibfnamefont
		{C.}~\bibnamefont {Meingast}},\ }\href {\doibase 10.1209/0295-5075/91/47008}
{\bibfield  {journal} {\bibinfo  {journal} {EPL (Europhysics Letters)}\
	}\textbf {\bibinfo {volume} {91}},\ \bibinfo {pages} {47008} (\bibinfo {year}
	{2010})}\BibitemShut {NoStop}%
\bibitem [{\citenamefont {Karkin}\ \emph {et~al.}(2014)\citenamefont {Karkin},
	\citenamefont {Wolf},\ and\ \citenamefont {Goshchitskii}}]{karkin2014}%
\BibitemOpen
\bibfield  {author} {\bibinfo {author} {\bibfnamefont {A.}~\bibnamefont
		{Karkin}}, \bibinfo {author} {\bibfnamefont {T.}~\bibnamefont {Wolf}}, \ and\
	\bibinfo {author} {\bibfnamefont {B.}~\bibnamefont {Goshchitskii}},\ }\href
{http://stacks.iop.org/0953-8984/26/i=27/a=275702} {\bibfield  {journal}
	{\bibinfo  {journal} {Journal of Physics: Condensed Matter}\ }\textbf
	{\bibinfo {volume} {26}},\ \bibinfo {pages} {275702} (\bibinfo {year}
	{2014})}\BibitemShut {NoStop}%
\bibitem [{\citenamefont {Schenck}(1985)}]{schenck1985}%
\BibitemOpen
\bibfield  {author} {\bibinfo {author} {\bibfnamefont {A.}~\bibnamefont
		{Schenck}},\ }\href@noop {} {\emph {\bibinfo {title} {{Muon Spin Rotation
				Spectroscopy: Principles and Applications in Solid State Physics}}}}\
(\bibinfo  {publisher} {Adam Hilger Ltd.},\ \bibinfo {year}
{1985})\BibitemShut {NoStop}%
\bibitem [{\citenamefont {Lee}\ \emph {et~al.}(1999)\citenamefont {Lee},
	\citenamefont {Cywinski},\ and\ \citenamefont {Kilcoyne}}]{lee1999}%
\BibitemOpen
\bibinfo {editor} {\bibfnamefont {S.}~\bibnamefont {Lee}}, \bibinfo {editor}
{\bibfnamefont {R.}~\bibnamefont {Cywinski}}, \ and\ \bibinfo {editor}
{\bibfnamefont {S.~H.}\ \bibnamefont {Kilcoyne}},\ eds.,\ \href@noop {}
{\emph {\bibinfo {title} {{Muon Science: Proceedings of the 51st Scottish
				Universities Summer School in Physics: NATO Advanced Study Institute on Muon
				Science, 17-28 August 1998}}}}\ (\bibinfo  {publisher} {Institute of Physics,
	Bristol, UK},\ \bibinfo {year} {1999})\BibitemShut {NoStop}%
\bibitem [{\citenamefont {Homes}\ \emph {et~al.}(1993)\citenamefont {Homes},
	\citenamefont {Reedyk}, \citenamefont {Cradles},\ and\ \citenamefont
	{Timusk}}]{homes1993}%
\BibitemOpen
\bibfield  {author} {\bibinfo {author} {\bibfnamefont {C.~C.}\ \bibnamefont
		{Homes}}, \bibinfo {author} {\bibfnamefont {M.}~\bibnamefont {Reedyk}},
	\bibinfo {author} {\bibfnamefont {D.}~\bibnamefont {Cradles}}, \ and\
	\bibinfo {author} {\bibfnamefont {T.}~\bibnamefont {Timusk}},\ }\href
{https://www.osapublishing.org/ao/fulltext.cfm?uri=ao-32-16-2976&id=40940}
{\bibfield  {journal} {\bibinfo  {journal} {Applied Optics}\ }\textbf
	{\bibinfo {volume} {32}},\ \bibinfo {pages} {2976} (\bibinfo {year}
	{1993})}\BibitemShut {NoStop}%
\bibitem [{\citenamefont {Kim}\ \emph {et~al.}(2010)\citenamefont {Kim},
	\citenamefont {R\"{o}ssle}, \citenamefont {Dubroka}, \citenamefont {Malik},
	\citenamefont {Wolf},\ and\ \citenamefont {Bernhard}}]{kim2010infrared}%
\BibitemOpen
\bibfield  {author} {\bibinfo {author} {\bibfnamefont {K.~W.}\ \bibnamefont
		{Kim}}, \bibinfo {author} {\bibfnamefont {M.}~\bibnamefont {R\"{o}ssle}},
	\bibinfo {author} {\bibfnamefont {A.}~\bibnamefont {Dubroka}}, \bibinfo
	{author} {\bibfnamefont {V.~K.}\ \bibnamefont {Malik}}, \bibinfo {author}
	{\bibfnamefont {T.}~\bibnamefont {Wolf}}, \ and\ \bibinfo {author}
	{\bibfnamefont {C.}~\bibnamefont {Bernhard}},\ }\href {\doibase
	10.1103/PhysRevB.81.214508} {\bibfield  {journal} {\bibinfo  {journal} {Phys.
			Rev. B}\ }\textbf {\bibinfo {volume} {81}},\ \bibinfo {pages} {214508}
	(\bibinfo {year} {2010})}\BibitemShut {NoStop}%
\bibitem [{\citenamefont {Bernhard}\ \emph {et~al.}(2004)\citenamefont
	{Bernhard}, \citenamefont {Humlicek},\ and\ \citenamefont
	{Keimer}}]{bernhard2004thinfilms}%
\BibitemOpen
\bibfield  {author} {\bibinfo {author} {\bibfnamefont {C.}~\bibnamefont
		{Bernhard}}, \bibinfo {author} {\bibfnamefont {J.}~\bibnamefont {Humlicek}},
	\ and\ \bibinfo {author} {\bibfnamefont {B.}~\bibnamefont {Keimer}},\ }\href
{http://www.sciencedirect.com/science/article/pii/S0040609004000033?np=y}
{\bibfield  {journal} {\bibinfo  {journal} {Thin Solid Films}\ }\textbf
	{\bibinfo {volume} {455}},\ \bibinfo {pages} {143} (\bibinfo {year}
	{2004})}\BibitemShut {NoStop}%
\bibitem [{\citenamefont {Schafgans}\ \emph {et~al.}(2011)\citenamefont
	{Schafgans}, \citenamefont {Pursley}, \citenamefont {LaForge}, \citenamefont
	{Sefat}, \citenamefont {Mandrus},\ and\ \citenamefont
	{Basov}}]{schafgans2011}%
\BibitemOpen
\bibfield  {author} {\bibinfo {author} {\bibfnamefont {A.~A.}\ \bibnamefont
		{Schafgans}}, \bibinfo {author} {\bibfnamefont {B.~C.}\ \bibnamefont
		{Pursley}}, \bibinfo {author} {\bibfnamefont {A.~D.}\ \bibnamefont
		{LaForge}}, \bibinfo {author} {\bibfnamefont {A.~S.}\ \bibnamefont {Sefat}},
	\bibinfo {author} {\bibfnamefont {D.}~\bibnamefont {Mandrus}}, \ and\
	\bibinfo {author} {\bibfnamefont {D.~N.}\ \bibnamefont {Basov}},\ }\href
{\doibase 10.1103/PhysRevB.84.052501} {\bibfield  {journal} {\bibinfo
		{journal} {Phys. Rev. B}\ }\textbf {\bibinfo {volume} {84}},\ \bibinfo
	{pages} {052501} (\bibinfo {year} {2011})}\BibitemShut {NoStop}%
\bibitem [{\citenamefont {Nakajima}\ \emph {et~al.}(2011)\citenamefont
	{Nakajima}, \citenamefont {Liang}, \citenamefont {Ishida}, \citenamefont
	{Tomioka}, \citenamefont {Kihou}, \citenamefont {Lee}, \citenamefont {Iyo},
	\citenamefont {Eisaki}, \citenamefont {Kakeshita}, \citenamefont {Ito} \emph
	{et~al.}}]{nakajima2011}%
\BibitemOpen
\bibfield  {author} {\bibinfo {author} {\bibfnamefont {M.}~\bibnamefont
		{Nakajima}}, \bibinfo {author} {\bibfnamefont {T.}~\bibnamefont {Liang}},
	\bibinfo {author} {\bibfnamefont {S.}~\bibnamefont {Ishida}}, \bibinfo
	{author} {\bibfnamefont {Y.}~\bibnamefont {Tomioka}}, \bibinfo {author}
	{\bibfnamefont {K.}~\bibnamefont {Kihou}}, \bibinfo {author} {\bibfnamefont
		{C.}~\bibnamefont {Lee}}, \bibinfo {author} {\bibfnamefont {A.}~\bibnamefont
		{Iyo}}, \bibinfo {author} {\bibfnamefont {H.}~\bibnamefont {Eisaki}},
	\bibinfo {author} {\bibfnamefont {T.}~\bibnamefont {Kakeshita}}, \bibinfo
	{author} {\bibfnamefont {T.}~\bibnamefont {Ito}},  \emph {et~al.},\ }\href
{http://www.pnas.org/content/108/30/12238.full} {\bibfield  {journal}
	{\bibinfo  {journal} {Proceedings of the National Academy of Sciences}\
	}\textbf {\bibinfo {volume} {108}},\ \bibinfo {pages} {12238} (\bibinfo
	{year} {2011})}\BibitemShut {NoStop}%
\bibitem [{\citenamefont {Kuzmenko}(2005)}]{kuzmenko2005}%
\BibitemOpen
\bibfield  {author} {\bibinfo {author} {\bibfnamefont {A.}~\bibnamefont
		{Kuzmenko}},\ }\href
{http://scitation.aip.org/content/aip/journal/rsi/76/8/10.1063/1.1979470}
{\bibfield  {journal} {\bibinfo  {journal} {Review of Scientific
			Instruments}\ }\textbf {\bibinfo {volume} {76}},\ \bibinfo {pages} {083108}
	(\bibinfo {year} {2005})}\BibitemShut {NoStop}%
\bibitem [{\citenamefont {Smith}(1998)}]{smith1998}%
\BibitemOpen
\bibfield  {author} {\bibinfo {author} {\bibfnamefont {D.}~\bibnamefont
		{Smith}},\ }\href {http://store.elsevier.com/product.jsp?isbn=9780125444231}
{\emph {\bibinfo {title} {Handbook of optical constants of solids}}},\ edited
by\ \bibinfo {editor} {\bibfnamefont {E.~D.}\ \bibnamefont {Palik}},\
Vol.~\bibinfo {volume} {3}\ (\bibinfo  {publisher} {Academic press},\
\bibinfo {year} {1998})\ p.~\bibinfo {pages} {35}\BibitemShut {NoStop}%
\bibitem [{\citenamefont {Jesche}\ \emph {et~al.}(2008)\citenamefont {Jesche},
	\citenamefont {Caroca-Canales}, \citenamefont {Rosner}, \citenamefont
	{Borrmann}, \citenamefont {Ormeci}, \citenamefont {Kasinathan}, \citenamefont
	{Klauss}, \citenamefont {Luetkens}, \citenamefont {Khasanov}, \citenamefont
	{Amato}, \citenamefont {Hoser}, \citenamefont {Kaneko}, \citenamefont
	{Krellner},\ and\ \citenamefont {Geibel}}]{jesche2008}%
\BibitemOpen
\bibfield  {author} {\bibinfo {author} {\bibfnamefont {A.}~\bibnamefont
		{Jesche}}, \bibinfo {author} {\bibfnamefont {N.}~\bibnamefont
		{Caroca-Canales}}, \bibinfo {author} {\bibfnamefont {H.}~\bibnamefont
		{Rosner}}, \bibinfo {author} {\bibfnamefont {H.}~\bibnamefont {Borrmann}},
	\bibinfo {author} {\bibfnamefont {A.}~\bibnamefont {Ormeci}}, \bibinfo
	{author} {\bibfnamefont {D.}~\bibnamefont {Kasinathan}}, \bibinfo {author}
	{\bibfnamefont {H.~H.}\ \bibnamefont {Klauss}}, \bibinfo {author}
	{\bibfnamefont {H.}~\bibnamefont {Luetkens}}, \bibinfo {author}
	{\bibfnamefont {R.}~\bibnamefont {Khasanov}}, \bibinfo {author}
	{\bibfnamefont {A.}~\bibnamefont {Amato}}, \bibinfo {author} {\bibfnamefont
		{A.}~\bibnamefont {Hoser}}, \bibinfo {author} {\bibfnamefont
		{K.}~\bibnamefont {Kaneko}}, \bibinfo {author} {\bibfnamefont
		{C.}~\bibnamefont {Krellner}}, \ and\ \bibinfo {author} {\bibfnamefont
		{C.}~\bibnamefont {Geibel}},\ }\href {\doibase 10.1103/PhysRevB.78.180504}
{\bibfield  {journal} {\bibinfo  {journal} {Phys. Rev. B}\ }\textbf {\bibinfo
		{volume} {78}},\ \bibinfo {pages} {180504} (\bibinfo {year}
	{2008})}\BibitemShut {NoStop}%
\bibitem [{\citenamefont {Reznik}\ \emph {et~al.}(1995)\citenamefont {Reznik},
	\citenamefont {Vagizov},\ and\ \citenamefont {Tro\ifmmode~\acute{c}\else
		\'{c}\fi{}}}]{reznik1995}%
\BibitemOpen
\bibfield  {author} {\bibinfo {author} {\bibfnamefont {I.~M.}\ \bibnamefont
		{Reznik}}, \bibinfo {author} {\bibfnamefont {F.~G.}\ \bibnamefont {Vagizov}},
	\ and\ \bibinfo {author} {\bibfnamefont {R.}~\bibnamefont
		{Tro\ifmmode~\acute{c}\else \'{c}\fi{}}},\ }\href {\doibase
	10.1103/PhysRevB.51.3013} {\bibfield  {journal} {\bibinfo  {journal} {Phys.
			Rev. B}\ }\textbf {\bibinfo {volume} {51}},\ \bibinfo {pages} {3013}
	(\bibinfo {year} {1995})}\BibitemShut {NoStop}%
\bibitem [{\citenamefont {Derondeau}\ \emph {et~al.}(2016)\citenamefont
	{Derondeau}, \citenamefont {Min\'{a}r}, \citenamefont {Wimmer},\ and\
	\citenamefont {Ebert}}]{derondeau2016}%
\BibitemOpen
\bibfield  {author} {\bibinfo {author} {\bibfnamefont {G.}~\bibnamefont
		{Derondeau}}, \bibinfo {author} {\bibfnamefont {J.}~\bibnamefont
		{Min\'{a}r}}, \bibinfo {author} {\bibfnamefont {S.}~\bibnamefont {Wimmer}}, \
	and\ \bibinfo {author} {\bibfnamefont {H.}~\bibnamefont {Ebert}},\
}\href@noop {} {\bibfield  {journal} {\bibinfo  {journal} {arXiv:1608.08077}\
} (\bibinfo {year} {2016})}\BibitemShut {NoStop}%
\bibitem [{\citenamefont {Khasanov}\ \emph
	{et~al.}(2009{\natexlab{b}})\citenamefont {Khasanov}, \citenamefont
	{Evtushinsky}, \citenamefont {Amato}, \citenamefont {Klauss}, \citenamefont
	{Luetkens}, \citenamefont {Niedermayer}, \citenamefont {B\"uchner},
	\citenamefont {Sun}, \citenamefont {Lin}, \citenamefont {Park}, \citenamefont
	{Inosov},\ and\ \citenamefont {Hinkov}}]{khasanov2009bkfa}%
\BibitemOpen
\bibfield  {author} {\bibinfo {author} {\bibfnamefont {R.}~\bibnamefont
		{Khasanov}}, \bibinfo {author} {\bibfnamefont {D.~V.}\ \bibnamefont
		{Evtushinsky}}, \bibinfo {author} {\bibfnamefont {A.}~\bibnamefont {Amato}},
	\bibinfo {author} {\bibfnamefont {H.-H.}\ \bibnamefont {Klauss}}, \bibinfo
	{author} {\bibfnamefont {H.}~\bibnamefont {Luetkens}}, \bibinfo {author}
	{\bibfnamefont {C.}~\bibnamefont {Niedermayer}}, \bibinfo {author}
	{\bibfnamefont {B.}~\bibnamefont {B\"uchner}}, \bibinfo {author}
	{\bibfnamefont {G.~L.}\ \bibnamefont {Sun}}, \bibinfo {author} {\bibfnamefont
		{C.~T.}\ \bibnamefont {Lin}}, \bibinfo {author} {\bibfnamefont {J.~T.}\
		\bibnamefont {Park}}, \bibinfo {author} {\bibfnamefont {D.~S.}\ \bibnamefont
		{Inosov}}, \ and\ \bibinfo {author} {\bibfnamefont {V.}~\bibnamefont
		{Hinkov}},\ }\href {\doibase 10.1103/PhysRevLett.102.187005} {\bibfield
	{journal} {\bibinfo  {journal} {Phys. Rev. Lett.}\ }\textbf {\bibinfo
		{volume} {102}},\ \bibinfo {pages} {187005} (\bibinfo {year}
	{2009}{\natexlab{b}})}\BibitemShut {NoStop}%
\bibitem [{\citenamefont {Avci}\ \emph {et~al.}(2011)\citenamefont {Avci},
	\citenamefont {Chmaissem}, \citenamefont {Goremychkin}, \citenamefont
	{Rosenkranz}, \citenamefont {Castellan}, \citenamefont {Chung}, \citenamefont
	{Todorov}, \citenamefont {Schlueter}, \citenamefont {Claus}, \citenamefont
	{Kanatzidis}, \citenamefont {Daoud-Aladine}, \citenamefont {Khalyavin},\ and\
	\citenamefont {Osborn}}]{avci2011}%
\BibitemOpen
\bibfield  {author} {\bibinfo {author} {\bibfnamefont {S.}~\bibnamefont
		{Avci}}, \bibinfo {author} {\bibfnamefont {O.}~\bibnamefont {Chmaissem}},
	\bibinfo {author} {\bibfnamefont {E.~A.}\ \bibnamefont {Goremychkin}},
	\bibinfo {author} {\bibfnamefont {S.}~\bibnamefont {Rosenkranz}}, \bibinfo
	{author} {\bibfnamefont {J.-P.}\ \bibnamefont {Castellan}}, \bibinfo {author}
	{\bibfnamefont {D.~Y.}\ \bibnamefont {Chung}}, \bibinfo {author}
	{\bibfnamefont {I.~S.}\ \bibnamefont {Todorov}}, \bibinfo {author}
	{\bibfnamefont {J.~A.}\ \bibnamefont {Schlueter}}, \bibinfo {author}
	{\bibfnamefont {H.}~\bibnamefont {Claus}}, \bibinfo {author} {\bibfnamefont
		{M.~G.}\ \bibnamefont {Kanatzidis}}, \bibinfo {author} {\bibfnamefont
		{A.}~\bibnamefont {Daoud-Aladine}}, \bibinfo {author} {\bibfnamefont
		{D.}~\bibnamefont {Khalyavin}}, \ and\ \bibinfo {author} {\bibfnamefont
		{R.}~\bibnamefont {Osborn}},\ }\href {\doibase 10.1103/PhysRevB.83.172503}
{\bibfield  {journal} {\bibinfo  {journal} {Phys. Rev. B}\ }\textbf {\bibinfo
		{volume} {83}},\ \bibinfo {pages} {172503} (\bibinfo {year}
	{2011})}\BibitemShut {NoStop}%
\bibitem [{\citenamefont {Hayano}\ \emph {et~al.}(1979)\citenamefont {Hayano},
	\citenamefont {Uemura}, \citenamefont {Imazato}, \citenamefont {Nishida},
	\citenamefont {Yamazaki},\ and\ \citenamefont {Kubo}}]{hayano1979}%
\BibitemOpen
\bibfield  {author} {\bibinfo {author} {\bibfnamefont {R.~S.}\ \bibnamefont
		{Hayano}}, \bibinfo {author} {\bibfnamefont {Y.~J.}\ \bibnamefont {Uemura}},
	\bibinfo {author} {\bibfnamefont {J.}~\bibnamefont {Imazato}}, \bibinfo
	{author} {\bibfnamefont {N.}~\bibnamefont {Nishida}}, \bibinfo {author}
	{\bibfnamefont {T.}~\bibnamefont {Yamazaki}}, \ and\ \bibinfo {author}
	{\bibfnamefont {R.}~\bibnamefont {Kubo}},\ }\href {\doibase
	10.1103/PhysRevB.20.850} {\bibfield  {journal} {\bibinfo  {journal} {Phys.
			Rev. B}\ }\textbf {\bibinfo {volume} {20}},\ \bibinfo {pages} {850} (\bibinfo
	{year} {1979})}\BibitemShut {NoStop}%
\bibitem [{\citenamefont {Luke}\ \emph {et~al.}(1998)\citenamefont {Luke},
	\citenamefont {Fudamoto}, \citenamefont {Kojima}, \citenamefont {Larkin},
	\citenamefont {Merrin}, \citenamefont {Nachumi}, \citenamefont {Uemura},
	\citenamefont {Maeno}, \citenamefont {Mao}, \citenamefont {Mori} \emph
	{et~al.}}]{luke1998}%
\BibitemOpen
\bibfield  {author} {\bibinfo {author} {\bibfnamefont {G.}~\bibnamefont
		{Luke}}, \bibinfo {author} {\bibfnamefont {Y.}~\bibnamefont {Fudamoto}},
	\bibinfo {author} {\bibfnamefont {K.}~\bibnamefont {Kojima}}, \bibinfo
	{author} {\bibfnamefont {M.}~\bibnamefont {Larkin}}, \bibinfo {author}
	{\bibfnamefont {J.}~\bibnamefont {Merrin}}, \bibinfo {author} {\bibfnamefont
		{B.}~\bibnamefont {Nachumi}}, \bibinfo {author} {\bibfnamefont
		{Y.}~\bibnamefont {Uemura}}, \bibinfo {author} {\bibfnamefont
		{Y.}~\bibnamefont {Maeno}}, \bibinfo {author} {\bibfnamefont
		{Z.}~\bibnamefont {Mao}}, \bibinfo {author} {\bibfnamefont {Y.}~\bibnamefont
		{Mori}},  \emph {et~al.},\ }\href
{http://www.nature.com/nature/journal/v394/n6693/full/394558a0.html}
{\bibfield  {journal} {\bibinfo  {journal} {Nature}\ }\textbf {\bibinfo
		{volume} {394}},\ \bibinfo {pages} {558} (\bibinfo {year}
	{1998})}\BibitemShut {NoStop}%
\bibitem [{\citenamefont {Aoki}\ \emph {et~al.}(2003)\citenamefont {Aoki},
	\citenamefont {Tsuchiya}, \citenamefont {Kanayama}, \citenamefont {Saha},
	\citenamefont {Sugawara}, \citenamefont {Sato}, \citenamefont {Higemoto},
	\citenamefont {Koda}, \citenamefont {Ohishi}, \citenamefont {Nishiyama},\
	and\ \citenamefont {Kadono}}]{aoki2003}%
\BibitemOpen
\bibfield  {author} {\bibinfo {author} {\bibfnamefont {Y.}~\bibnamefont
		{Aoki}}, \bibinfo {author} {\bibfnamefont {A.}~\bibnamefont {Tsuchiya}},
	\bibinfo {author} {\bibfnamefont {T.}~\bibnamefont {Kanayama}}, \bibinfo
	{author} {\bibfnamefont {S.~R.}\ \bibnamefont {Saha}}, \bibinfo {author}
	{\bibfnamefont {H.}~\bibnamefont {Sugawara}}, \bibinfo {author}
	{\bibfnamefont {H.}~\bibnamefont {Sato}}, \bibinfo {author} {\bibfnamefont
		{W.}~\bibnamefont {Higemoto}}, \bibinfo {author} {\bibfnamefont
		{A.}~\bibnamefont {Koda}}, \bibinfo {author} {\bibfnamefont {K.}~\bibnamefont
		{Ohishi}}, \bibinfo {author} {\bibfnamefont {K.}~\bibnamefont {Nishiyama}}, \
	and\ \bibinfo {author} {\bibfnamefont {R.}~\bibnamefont {Kadono}},\ }\href
{\doibase 10.1103/PhysRevLett.91.067003} {\bibfield  {journal} {\bibinfo
		{journal} {Phys. Rev. Lett.}\ }\textbf {\bibinfo {volume} {91}},\ \bibinfo
	{pages} {067003} (\bibinfo {year} {2003})}\BibitemShut {NoStop}%
\bibitem [{\citenamefont {Hillier}\ \emph {et~al.}(2009)\citenamefont
	{Hillier}, \citenamefont {Quintanilla},\ and\ \citenamefont
	{Cywinski}}]{hillier2009}%
\BibitemOpen
\bibfield  {author} {\bibinfo {author} {\bibfnamefont {A.~D.}\ \bibnamefont
		{Hillier}}, \bibinfo {author} {\bibfnamefont {J.}~\bibnamefont
		{Quintanilla}}, \ and\ \bibinfo {author} {\bibfnamefont {R.}~\bibnamefont
		{Cywinski}},\ }\href {\doibase 10.1103/PhysRevLett.102.117007} {\bibfield
	{journal} {\bibinfo  {journal} {Phys. Rev. Lett.}\ }\textbf {\bibinfo
		{volume} {102}},\ \bibinfo {pages} {117007} (\bibinfo {year}
	{2009})}\BibitemShut {NoStop}%
\bibitem [{\citenamefont {Sonier}\ \emph {et~al.}(2011)\citenamefont {Sonier},
	\citenamefont {Huang}, \citenamefont {Kaiser}, \citenamefont {Cochrane},
	\citenamefont {Pacradouni}, \citenamefont {Sabok-Sayr}, \citenamefont
	{Lumsden}, \citenamefont {Sales}, \citenamefont {McGuire}, \citenamefont
	{Sefat},\ and\ \citenamefont {Mandrus}}]{sonier2011}%
\BibitemOpen
\bibfield  {author} {\bibinfo {author} {\bibfnamefont {J.~E.}\ \bibnamefont
		{Sonier}}, \bibinfo {author} {\bibfnamefont {W.}~\bibnamefont {Huang}},
	\bibinfo {author} {\bibfnamefont {C.~V.}\ \bibnamefont {Kaiser}}, \bibinfo
	{author} {\bibfnamefont {C.}~\bibnamefont {Cochrane}}, \bibinfo {author}
	{\bibfnamefont {V.}~\bibnamefont {Pacradouni}}, \bibinfo {author}
	{\bibfnamefont {S.~A.}\ \bibnamefont {Sabok-Sayr}}, \bibinfo {author}
	{\bibfnamefont {M.~D.}\ \bibnamefont {Lumsden}}, \bibinfo {author}
	{\bibfnamefont {B.~C.}\ \bibnamefont {Sales}}, \bibinfo {author}
	{\bibfnamefont {M.~A.}\ \bibnamefont {McGuire}}, \bibinfo {author}
	{\bibfnamefont {A.~S.}\ \bibnamefont {Sefat}}, \ and\ \bibinfo {author}
	{\bibfnamefont {D.}~\bibnamefont {Mandrus}},\ }\href {\doibase
	10.1103/PhysRevLett.106.127002} {\bibfield  {journal} {\bibinfo  {journal}
		{Phys. Rev. Lett.}\ }\textbf {\bibinfo {volume} {106}},\ \bibinfo {pages}
	{127002} (\bibinfo {year} {2011})}\BibitemShut {NoStop}%
\bibitem [{\citenamefont {Brandt}(1988)}]{brandt1988}%
\BibitemOpen
\bibfield  {author} {\bibinfo {author} {\bibfnamefont {E.~H.}\ \bibnamefont
		{Brandt}},\ }\href {\doibase 10.1103/PhysRevB.37.2349} {\bibfield  {journal}
	{\bibinfo  {journal} {Phys. Rev. B}\ }\textbf {\bibinfo {volume} {37}},\
	\bibinfo {pages} {2349} (\bibinfo {year} {1988})}\BibitemShut {NoStop}%
\bibitem [{\citenamefont {P\"umpin}\ \emph {et~al.}(1990)\citenamefont
	{P\"umpin}, \citenamefont {Keller}, \citenamefont {K\"undig}, \citenamefont
	{Odermatt}, \citenamefont {Savi\ifmmode~\acute{c}\else \'{c}\fi{}},
	\citenamefont {Schneider}, \citenamefont {Simmler}, \citenamefont
	{Zimmermann}, \citenamefont {Kaldis}, \citenamefont {Rusiecki}, \citenamefont
	{Maeno},\ and\ \citenamefont {Rossel}}]{pumpin1990}%
\BibitemOpen
\bibfield  {author} {\bibinfo {author} {\bibfnamefont {B.}~\bibnamefont
		{P\"umpin}}, \bibinfo {author} {\bibfnamefont {H.}~\bibnamefont {Keller}},
	\bibinfo {author} {\bibfnamefont {W.}~\bibnamefont {K\"undig}}, \bibinfo
	{author} {\bibfnamefont {W.}~\bibnamefont {Odermatt}}, \bibinfo {author}
	{\bibfnamefont {I.~M.}\ \bibnamefont {Savi\ifmmode~\acute{c}\else
			\'{c}\fi{}}}, \bibinfo {author} {\bibfnamefont {J.~W.}\ \bibnamefont
		{Schneider}}, \bibinfo {author} {\bibfnamefont {H.}~\bibnamefont {Simmler}},
	\bibinfo {author} {\bibfnamefont {P.}~\bibnamefont {Zimmermann}}, \bibinfo
	{author} {\bibfnamefont {E.}~\bibnamefont {Kaldis}}, \bibinfo {author}
	{\bibfnamefont {S.}~\bibnamefont {Rusiecki}}, \bibinfo {author}
	{\bibfnamefont {Y.}~\bibnamefont {Maeno}}, \ and\ \bibinfo {author}
	{\bibfnamefont {C.}~\bibnamefont {Rossel}},\ }\href {\doibase
	10.1103/PhysRevB.42.8019} {\bibfield  {journal} {\bibinfo  {journal} {Phys.
			Rev. B}\ }\textbf {\bibinfo {volume} {42}},\ \bibinfo {pages} {8019}
	(\bibinfo {year} {1990})}\BibitemShut {NoStop}%
\bibitem [{\citenamefont {Liu}\ \emph {et~al.}(2014)\citenamefont {Liu},
	\citenamefont {Tanatar}, \citenamefont {Straszheim}, \citenamefont {Jensen},
	\citenamefont {Dennis}, \citenamefont {McCallum}, \citenamefont {Kogan},
	\citenamefont {Prozorov},\ and\ \citenamefont {Lograsso}}]{liu2014}%
\BibitemOpen
\bibfield  {author} {\bibinfo {author} {\bibfnamefont {Y.}~\bibnamefont
		{Liu}}, \bibinfo {author} {\bibfnamefont {M.~A.}\ \bibnamefont {Tanatar}},
	\bibinfo {author} {\bibfnamefont {W.~E.}\ \bibnamefont {Straszheim}},
	\bibinfo {author} {\bibfnamefont {B.}~\bibnamefont {Jensen}}, \bibinfo
	{author} {\bibfnamefont {K.~W.}\ \bibnamefont {Dennis}}, \bibinfo {author}
	{\bibfnamefont {R.~W.}\ \bibnamefont {McCallum}}, \bibinfo {author}
	{\bibfnamefont {V.~G.}\ \bibnamefont {Kogan}}, \bibinfo {author}
	{\bibfnamefont {R.}~\bibnamefont {Prozorov}}, \ and\ \bibinfo {author}
	{\bibfnamefont {T.~A.}\ \bibnamefont {Lograsso}},\ }\href {\doibase
	10.1103/PhysRevB.89.134504} {\bibfield  {journal} {\bibinfo  {journal} {Phys.
			Rev. B}\ }\textbf {\bibinfo {volume} {89}},\ \bibinfo {pages} {134504}
	(\bibinfo {year} {2014})}\BibitemShut {NoStop}%
\bibitem [{\citenamefont {Charnukha}(2014)}]{charnukhareview2014}%
\BibitemOpen
\bibfield  {author} {\bibinfo {author} {\bibfnamefont {A.}~\bibnamefont
		{Charnukha}},\ }\href {http://stacks.iop.org/0953-8984/26/i=25/a=253203}
{\bibfield  {journal} {\bibinfo  {journal} {Journal of Physics: Condensed
			Matter}\ }\textbf {\bibinfo {volume} {26}},\ \bibinfo {pages} {253203}
	(\bibinfo {year} {2014})}\BibitemShut {NoStop}%
\bibitem [{\citenamefont {Calder\'on}\ \emph {et~al.}(2014)\citenamefont
	{Calder\'on}, \citenamefont {Medici}, \citenamefont {Valenzuela},\ and\
	\citenamefont {Bascones}}]{calderon2014}%
\BibitemOpen
\bibfield  {author} {\bibinfo {author} {\bibfnamefont {M.~J.}\ \bibnamefont
		{Calder\'on}}, \bibinfo {author} {\bibfnamefont {L.~d.}\ \bibnamefont
		{Medici}}, \bibinfo {author} {\bibfnamefont {B.}~\bibnamefont {Valenzuela}},
	\ and\ \bibinfo {author} {\bibfnamefont {E.}~\bibnamefont {Bascones}},\
}\href {\doibase 10.1103/PhysRevB.90.115128} {\bibfield  {journal} {\bibinfo
	{journal} {Phys. Rev. B}\ }\textbf {\bibinfo {volume} {90}},\ \bibinfo
{pages} {115128} (\bibinfo {year} {2014})}\BibitemShut {NoStop}%
\bibitem [{\citenamefont {Georges}\ \emph {et~al.}(2013)\citenamefont
	{Georges}, \citenamefont {de'Medici},\ and\ \citenamefont
	{Mravlje}}]{georges2013}%
\BibitemOpen
\bibfield  {author} {\bibinfo {author} {\bibfnamefont {A.}~\bibnamefont
		{Georges}}, \bibinfo {author} {\bibfnamefont {L.}~\bibnamefont {de'Medici}},
	\ and\ \bibinfo {author} {\bibfnamefont {J.}~\bibnamefont {Mravlje}},\ }\href
{\doibase 10.1146/annurev-conmatphys-020911-125045} {\bibfield  {journal}
	{\bibinfo  {journal} {Annual Review of Condensed Matter Physics}\ }\textbf
	{\bibinfo {volume} {4}},\ \bibinfo {pages} {137} (\bibinfo {year}
	{2013})}\BibitemShut {NoStop}%
\bibitem [{\citenamefont {Litvinchuk}\ \emph {et~al.}(2008)\citenamefont
	{Litvinchuk}, \citenamefont {Hadjiev}, \citenamefont {Iliev}, \citenamefont
	{Lv}, \citenamefont {Guloy},\ and\ \citenamefont {Chu}}]{litvinchuk2008}%
\BibitemOpen
\bibfield  {author} {\bibinfo {author} {\bibfnamefont {A.~P.}\ \bibnamefont
		{Litvinchuk}}, \bibinfo {author} {\bibfnamefont {V.~G.}\ \bibnamefont
		{Hadjiev}}, \bibinfo {author} {\bibfnamefont {M.~N.}\ \bibnamefont {Iliev}},
	\bibinfo {author} {\bibfnamefont {B.}~\bibnamefont {Lv}}, \bibinfo {author}
	{\bibfnamefont {A.~M.}\ \bibnamefont {Guloy}}, \ and\ \bibinfo {author}
	{\bibfnamefont {C.~W.}\ \bibnamefont {Chu}},\ }\href {\doibase
	10.1103/PhysRevB.78.060503} {\bibfield  {journal} {\bibinfo  {journal} {Phys.
			Rev. B}\ }\textbf {\bibinfo {volume} {78}},\ \bibinfo {pages} {060503}
	(\bibinfo {year} {2008})}\BibitemShut {NoStop}%
\bibitem [{\citenamefont {Akrap}\ \emph {et~al.}(2009)\citenamefont {Akrap},
	\citenamefont {Tu}, \citenamefont {Li}, \citenamefont {Cao}, \citenamefont
	{Xu},\ and\ \citenamefont {Homes}}]{akrap2009}%
\BibitemOpen
\bibfield  {author} {\bibinfo {author} {\bibfnamefont {A.}~\bibnamefont
		{Akrap}}, \bibinfo {author} {\bibfnamefont {J.~J.}\ \bibnamefont {Tu}},
	\bibinfo {author} {\bibfnamefont {L.~J.}\ \bibnamefont {Li}}, \bibinfo
	{author} {\bibfnamefont {G.~H.}\ \bibnamefont {Cao}}, \bibinfo {author}
	{\bibfnamefont {Z.~A.}\ \bibnamefont {Xu}}, \ and\ \bibinfo {author}
	{\bibfnamefont {C.~C.}\ \bibnamefont {Homes}},\ }\href {\doibase
	10.1103/PhysRevB.80.180502} {\bibfield  {journal} {\bibinfo  {journal} {Phys.
			Rev. B}\ }\textbf {\bibinfo {volume} {80}},\ \bibinfo {pages} {180502}
	(\bibinfo {year} {2009})}\BibitemShut {NoStop}%
\bibitem [{\citenamefont {Storey}\ \emph {et~al.}(2013)\citenamefont {Storey},
	\citenamefont {Loram}, \citenamefont {Cooper}, \citenamefont {Bukowski},\
	and\ \citenamefont {Karpinski}}]{storey2013}%
\BibitemOpen
\bibfield  {author} {\bibinfo {author} {\bibfnamefont {J.~G.}\ \bibnamefont
		{Storey}}, \bibinfo {author} {\bibfnamefont {J.~W.}\ \bibnamefont {Loram}},
	\bibinfo {author} {\bibfnamefont {J.~R.}\ \bibnamefont {Cooper}}, \bibinfo
	{author} {\bibfnamefont {Z.}~\bibnamefont {Bukowski}}, \ and\ \bibinfo
	{author} {\bibfnamefont {J.}~\bibnamefont {Karpinski}},\ }\href {\doibase
	10.1103/PhysRevB.88.144502} {\bibfield  {journal} {\bibinfo  {journal} {Phys.
			Rev. B}\ }\textbf {\bibinfo {volume} {88}},\ \bibinfo {pages} {144502}
	(\bibinfo {year} {2013})}\BibitemShut {NoStop}%
\bibitem [{\citenamefont {Xu}\ \emph {et~al.}(2016)\citenamefont {Xu},
	\citenamefont {Dai}, \citenamefont {Xiao}, \citenamefont {Shen},
	\citenamefont {Ye}, \citenamefont {Forget}, \citenamefont {Colson},
	\citenamefont {Feng}, \citenamefont {Wen}, \citenamefont {Qiu},\ and\
	\citenamefont {Lobo}}]{xu2016}%
\BibitemOpen
\bibfield  {author} {\bibinfo {author} {\bibfnamefont {B.}~\bibnamefont
		{Xu}}, \bibinfo {author} {\bibfnamefont {Y.~M.}\ \bibnamefont {Dai}},
	\bibinfo {author} {\bibfnamefont {H.}~\bibnamefont {Xiao}}, \bibinfo {author}
	{\bibfnamefont {B.}~\bibnamefont {Shen}}, \bibinfo {author} {\bibfnamefont
		{Z.~R.}\ \bibnamefont {Ye}}, \bibinfo {author} {\bibfnamefont
		{A.}~\bibnamefont {Forget}}, \bibinfo {author} {\bibfnamefont
		{D.}~\bibnamefont {Colson}}, \bibinfo {author} {\bibfnamefont {D.~L.}\
		\bibnamefont {Feng}}, \bibinfo {author} {\bibfnamefont {H.~H.}\ \bibnamefont
		{Wen}}, \bibinfo {author} {\bibfnamefont {X.~G.}\ \bibnamefont {Qiu}}, \ and\
	\bibinfo {author} {\bibfnamefont {R.~P. S.~M.}\ \bibnamefont {Lobo}},\ }\href
{\doibase 10.1103/PhysRevB.94.085147} {\bibfield  {journal} {\bibinfo
		{journal} {Phys. Rev. B}\ }\textbf {\bibinfo {volume} {94}},\ \bibinfo
	{pages} {085147} (\bibinfo {year} {2016})}\BibitemShut {NoStop}%
\bibitem [{\citenamefont {Hashimoto}\ \emph {et~al.}(2012)\citenamefont
	{Hashimoto}, \citenamefont {Cho}, \citenamefont {Shibauchi}, \citenamefont
	{Kasahara}, \citenamefont {Mizukami}, \citenamefont {Katsumata},
	\citenamefont {Tsuruhara}, \citenamefont {Terashima}, \citenamefont {Ikeda},
	\citenamefont {Tanatar}, \citenamefont {Kitano}, \citenamefont {Salovich},
	\citenamefont {Giannetta}, \citenamefont {Walmsley}, \citenamefont
	{Carrington}, \citenamefont {Prozorov},\ and\ \citenamefont
	{Matsuda}}]{hashimoto2012}%
\BibitemOpen
\bibfield  {author} {\bibinfo {author} {\bibfnamefont {K.}~\bibnamefont
		{Hashimoto}}, \bibinfo {author} {\bibfnamefont {K.}~\bibnamefont {Cho}},
	\bibinfo {author} {\bibfnamefont {T.}~\bibnamefont {Shibauchi}}, \bibinfo
	{author} {\bibfnamefont {S.}~\bibnamefont {Kasahara}}, \bibinfo {author}
	{\bibfnamefont {Y.}~\bibnamefont {Mizukami}}, \bibinfo {author}
	{\bibfnamefont {R.}~\bibnamefont {Katsumata}}, \bibinfo {author}
	{\bibfnamefont {Y.}~\bibnamefont {Tsuruhara}}, \bibinfo {author}
	{\bibfnamefont {T.}~\bibnamefont {Terashima}}, \bibinfo {author}
	{\bibfnamefont {H.}~\bibnamefont {Ikeda}}, \bibinfo {author} {\bibfnamefont
		{M.~A.}\ \bibnamefont {Tanatar}}, \bibinfo {author} {\bibfnamefont
		{H.}~\bibnamefont {Kitano}}, \bibinfo {author} {\bibfnamefont
		{N.}~\bibnamefont {Salovich}}, \bibinfo {author} {\bibfnamefont {R.~W.}\
		\bibnamefont {Giannetta}}, \bibinfo {author} {\bibfnamefont {P.}~\bibnamefont
		{Walmsley}}, \bibinfo {author} {\bibfnamefont {A.}~\bibnamefont
		{Carrington}}, \bibinfo {author} {\bibfnamefont {R.}~\bibnamefont
		{Prozorov}}, \ and\ \bibinfo {author} {\bibfnamefont {Y.}~\bibnamefont
		{Matsuda}},\ }\href {\doibase 10.1126/science.1219821} {\bibfield  {journal}
	{\bibinfo  {journal} {Science}\ }\textbf {\bibinfo {volume} {336}},\ \bibinfo
	{pages} {1554} (\bibinfo {year} {2012})}\BibitemShut {NoStop}%
\bibitem [{\citenamefont {Putzke}\ \emph {et~al.}(2014)\citenamefont {Putzke},
	\citenamefont {Walmsley}, \citenamefont {Fletcher}, \citenamefont {Malone},
	\citenamefont {Vignolles}, \citenamefont {Proust}, \citenamefont {Badoux},
	\citenamefont {See}, \citenamefont {Beere}, \citenamefont {Ritchie} \emph
	{et~al.}}]{putzke2014}%
\BibitemOpen
\bibfield  {author} {\bibinfo {author} {\bibfnamefont {C.}~\bibnamefont
		{Putzke}}, \bibinfo {author} {\bibfnamefont {P.}~\bibnamefont {Walmsley}},
	\bibinfo {author} {\bibfnamefont {J.}~\bibnamefont {Fletcher}}, \bibinfo
	{author} {\bibfnamefont {L.}~\bibnamefont {Malone}}, \bibinfo {author}
	{\bibfnamefont {D.}~\bibnamefont {Vignolles}}, \bibinfo {author}
	{\bibfnamefont {C.}~\bibnamefont {Proust}}, \bibinfo {author} {\bibfnamefont
		{S.}~\bibnamefont {Badoux}}, \bibinfo {author} {\bibfnamefont
		{P.}~\bibnamefont {See}}, \bibinfo {author} {\bibfnamefont {H.}~\bibnamefont
		{Beere}}, \bibinfo {author} {\bibfnamefont {D.}~\bibnamefont {Ritchie}},
	\emph {et~al.},\ }\href {\doibase 10.1038/ncomms6679} {\bibfield  {journal}
	{\bibinfo  {journal} {Nature communications}\ }\textbf {\bibinfo {volume}
		{5}} (\bibinfo {year} {2014}),\ 10.1038/ncomms6679}\BibitemShut {NoStop}%
\end{thebibliography}

%

\end{document}